\documentclass[aps,floats]{revtex4}
\usepackage{amsmath,amssymb}
\usepackage{graphicx,epsfig}
\usepackage[greek,english]{babel}
\usepackage{bbold}

\begin{document}
\bibliographystyle {plain}

\pdfoutput=1
\def\oppropto{\mathop{\propto}} 
\def\opsimeq{\mathop{\simeq}}
\def\opoverderline{\mathop{\overline}}
\def\operarrow{\mathop{\longrightarrow}}
\def\opsim{\mathop{\sim}}

\def\fig#1#2{\includegraphics[height=#1]{#2}}
\def\figx#1#2{\includegraphics[width=#1]{#2}}


\title{ Anomalous dynamical large deviations of local empirical densities and activities 
\\ in the pure and in the random kinetically-constrained East Model
  } 


\author{ C\'ecile Monthus }
 \affiliation{Institut de Physique Th\'{e}orique, 
Universit\'e Paris Saclay, CNRS, CEA,
91191 Gif-sur-Yvette, France}

\begin{abstract}
The East model is the simplest one-dimensional kinetically-constrained model of $N$ spins with a trivial equilibrium that displays anomalously large spatio-temporal fluctuations, with characteristic "space-time bubbles" in trajectory space, and with a discontinuity at the origin for the first derivative of the scaled cumulant generating function of the total activity. These striking dynamical properties are revisited via the large deviations at various levels for the relevant local empirical densities and activities that only involve two consecutive spins. This framework allows to characterize their anomalous rate functions and to analyze the consequences for all the time-additive observables that can be reconstructed from them, both for the pure and for the random East model. These singularities in dynamical large deviations properties disappear when the hard-constraint of the East model is replaced by the soft constraint.

\end{abstract}

\maketitle

\section{ Introduction }

Kinetically-Constrained-Models have attracted a lot of interest in the field of glassy dynamics
(see the reviews \cite{ritort2003,lecture,chapter,math2013,garrahan_lecture} and references therein).
Despite their trivial equilibrium properties, they can display 
very slow cooperative relaxation and anomalously large spatio-temporal fluctuations,
with characteristic "space-time bubbles" in trajectory space.
In order to understand these striking properties, 
the large deviations of the total activity have been much studied in various models \cite{garrahan_lecture,lecomte_glass,kristina1,kristina2,bodineau,LeeYang,jack_soft,jack_sol,banuls,noninter,XOR,MPS,garrahan2021}
and have pointed towards the general scenario 
of a discontinuity in the first derivative of the scaled cumulant generating function at the origin.

Indeed, the formalism of large deviations has become the unifying language to characterize Markov processes.
While the initial classification involved only three nested levels 
(see the reviews \cite{oono,ellis,review_touchette} and references therein),
 with Level 1 for empirical observables,
Level 2 for the empirical density,
and Level 3 for the empirical process, 
the introduction of the Level 2.5 has been a major progress
in order to characterize the joint distribution of the empirical density and of the empirical flows.
Its essential advantage is that the rate functions at Level 2.5 can be written explicitly for general Markov processes,
including discrete-time Markov chains
 \cite{fortelle_thesis,fortelle_chain,review_touchette,c_largedevdisorder,c_reset,c_inference},
continuous-time Markov jump processes
\cite{fortelle_thesis,fortelle_jump,maes_canonical,maes_onandbeyond,wynants_thesis,chetrite_formal,BFG1,BFG2,chetrite_HDR,c_ring,c_interactions,c_open,c_detailed,barato_periodic,chetrite_periodic,c_reset,c_inference,c_runandtumble,c_jumpdiff,c_skew,c_metastable,c_exclusion}
and Diffusion processes 
\cite{wynants_thesis,maes_diffusion,chetrite_formal,engel,chetrite_HDR,c_reset,c_lyapunov,c_inference,c_metastable}.
As a consequence, the explicit Level 2.5 can be considered as the central starting point
 from which many other large deviations properties can be derived
via the appropriate contraction. In particular, the Level 2 for the empirical density alone 
can be obtained via the optimization of the Level 2.5 over the empirical flows,
so that the Level 2 will be closed only if this contraction can be performed explicitly.
More generally, the Level 2.5 can be contracted
to obtain the large deviations properties of any time-additive observable
of the dynamical trajectory involving both the configuration and the flows.
The link with the studies of the generating function of time-additive observables 
via deformed Markov operators  \cite{derrida-lecture,sollich_review,lazarescu_companion,lazarescu_generic,jack_review,vivien_thesis,lecomte_chaotic,lecomte_thermo,lecomte_formalism,lecomte_glass,kristina1,kristina2,jack_ensemble,simon1,simon2,simon3,Gunter1,Gunter2,Gunter3,Gunter4,chetrite_canonical,chetrite_conditioned,chetrite_optimal,chetrite_HDR,touchette_circle,touchette_langevin,touchette_occ,touchette_occupation,derrida-conditioned,derrida-ring,bertin-conditioned,touchette-reflected,touchette-reflectedbis,c_lyapunov,previousquantum2.5doob,quantum2.5doob,quantum2.5dooblong,c_ruelle,lapolla}
 can be then understood via the corresponding conditioned process obtained from the generalization of Doob's h-transform.

However for many-body models of $N$ spins, these large deviations properties 
are formulated in the configuration space of size $2^N$ 
and the consequences for real-space properties can be difficult to extract.
When the dynamical rules are local in space, one would like to analyze the dynamics via $O(N)$
local empirical observables instead.
A famous example of this type of space-local analysis for many-body dynamics
 is the Macroscopic Fluctuation Theory (see the review \cite{mft} and references therein)
where interacting driven diffusive systems are renormalized in the hydrodynamic limit
into an action for dynamical trajectories based on an elementary space-time local Lagrangian containing
only the empirical local density and the local empirical current.
Similarly, for interacting random walkers in the continuous-time discrete-space framework, 
 space-time local Lagrangian have been analyzed in Refs \cite{c_interactions,chemical,chabane}.

In the present paper, we take advantage of the detailed-balance property of the dynamics
in order to write closed large deviations for the relevant local empirical densities and the local empirical activities
that only involve two consecutive spins for the East model. This approach allows to characterize the anomalous
large deviations of the pure and the random East model via simple analytical results.

The paper is organized as follows.
In section \ref{sec_models}, we recall the definitions of the pure and random East Model, with the hard or the soft kinetic constraint.
In Section \ref{sec_Erandomsoft}, the random East Model with the soft kinetic constraint $\epsilon>0$
is analyzed via the large deviations properties with respect to the time-window $T$ of the $O(N)$ 
relevant empirical time-averaged densities and activities involving only two consecutive spins.
In Section \ref{sec_Epuresoft}, the pure East Model with the soft kinetic constraint $\epsilon>0$
is studied via the large deviations properties with respect to the space-time volume $(TN)$ 
of the $O(1)$ relevant empirical time-space-averaged densities and activities involving only two consecutive spins.
We then discuss the anomalous large deviations properties that emerge in the hard-constraint limit $\epsilon=0$,
both for the pure East model in section \ref{sec_Epure} and for 
random East Model in section \ref{sec_Erandom}.
Our conclusions are summarized in section \ref{sec_conclusion}.
Five appendices include the following complementary material.
Appendix \ref{appendix_relevant} explains the general procedure to obtain large deviations for the relevant empirical observables of Markov trajectories.
Appendix \ref{app_LargeDevDB} describes how the detailed-balance condition allows to contract explicitly
the Level 2.5 towards lower levels. 
Appendix \ref{app_configEast} contains the large deviations at various Levels in the space of the $2^N$ configurations of the random soft East model.
Finally, we show how the Level 2.25 concerning the empirical density and the empirical activities in the space of the $2^N$ configurations of the soft East model can be contracted,
either towards the Level 2.25 for the $O(N)$ empirical local densities and local activities for the random model (Appendix \ref{app_GlobalLocal}), or towards the Level 2.25 
for the $O(1)$ empirical time-space-averaged densities and activities for the pure model (Appendix \ref{app_GlobalLocalPure}).


\section{Pure and random East Model, with hard or soft kinetic constraint }

\label{sec_models}

\subsection{ Detailed-Balance single-spin-flip Markov dynamics for spin chains}

For a chain of $N$ classical spins $S_i=\pm 1$ with periodic boundary conditions, 
the single-spin-flip Markov dynamics can be formulated as follows.
From the configuration $C=\{S_1,..,S_i,.,S_N\}$, the possible elementary moves are towards
the $N$ configurations $(\sigma_i^x C)=\{S_1,..,-S_i,.,S_N\} $
obtained by the flip of a single spin $S_i$ towards $(-S_i)$ and occur with rates $W(\sigma_i^x C, C ) $.
The master equation 
for the probability $ P_t( C )$ to be in configuration $C=\{S_1,..,S_i,.,S_N\}$ at time $t$ reads
\begin{eqnarray}
 \partial_t P_t( C )    =   \sum_{i=1 }^N \left[ W(C,\sigma_i^x C ) P_t(\sigma_i^x C) 
 -  W(\sigma_i^x C ,C) P_t(C) \right]
\label{master}
\end{eqnarray}
The convergence
towards the Boltzmann equilibrium $\Pi(C)$ in the potential $U(C)$ at inverse temperature $\beta$
\begin{eqnarray}
\Pi(C) = \frac{ e^{- \beta U(C)} }{Z}
\nonumber \\
Z = \sum_{ C}  e^{- \beta U(C)}
\label{partition}
\end{eqnarray}
can be ensured if the flip rates $W(\sigma_i^xC,C) $ satisfy the detailed-balance condition
\begin{eqnarray}
0= W(C,\sigma_i^x C ) e^{- \beta U(\sigma_i^x C)}  -  W(\sigma_i^x C,C)e^{- \beta U( C)}
\label{DB}
\end{eqnarray}


\subsection{ Kinetically-constrained East Model satisfying detailed-balance}

The East model is the simplest example of Kinetically-Constrained-Models 
and can be described as follows.

\subsubsection { Trivial equilibrium distribution  }

The equilibrium distribution $\Pi(C) $ of Eq. \ref{partition}
involves the trivial potential containing only fields $h_i$
\begin{eqnarray}
U(C) = -  \sum_{i=1}^N  h_i S_i 
\label{ekin}
\end{eqnarray}
and is thus factorized 
\begin{eqnarray}
 \Pi(S_1,..,S_N)  = \prod_{i=1}^N \Pi_i^{S_i}
\label{peqKCM}
\end{eqnarray}
into the equilibrium distributions $ \Pi_i^{S_i}$ of individual spins
\begin{eqnarray}
\Pi_i^{S_i} && \equiv c_i \delta_{S_i,+} + (1-c_i) \delta_{S_i,-}
\label{peqKCMi}
\end{eqnarray}
parametrized by the $N$ parameters   
\begin{eqnarray}
 c_i \equiv \Pi_i^{S_i=+} = \frac{ e^{   \beta h_i } }{ e^{   \beta h_i }+ e^{ -  \beta h_i }} \in ]0,1[
\label{defci}
\end{eqnarray}
for the random version of the model, or by the single parameter $c_i=c$ for the pure version of the model.

\subsubsection { The East Model, with its 'hard' kinetic-constraint   }

In the East Model, the rate to flip the spin $S_i$ towards $(-S_i)$
satisfies the detailed-balance condition with respect to the equilibrium of Eq. \ref{peqKCM}
\begin{eqnarray}
 W^{East}(\sigma_i^x C,C) = \delta_{S_{i-1},+ } \left[  (1-c_i) \delta_{S_i,+} + c_i \delta_{S_i,-}\right] 
\label{weasthard}
\end{eqnarray}
but contains the kinetic constraint $  \delta_{S_{i-1},+ }$, so that the rate vanishes when the 
neighboring spin $(S_{i-1})$ is in the $-$ state.

\subsubsection{ The 'soft' version of the East Model parametrized by $\epsilon>0$ }

For comparison, it will be useful to consider the 'soft' version of the East model of parameter $\epsilon>0$
(see \cite{jack_soft} and references therein)
\begin{eqnarray}
 W^{EastSoft}(\sigma_i^x C,C) = \left[ \delta_{S_{i-1},+ } + \epsilon \delta_{S_{i-1},- } \right] 
 \left[  (1-c_i) \delta_{S_i,+} + c_i \delta_{S_i,-}\right]
\label{weast}
\end{eqnarray}
where the 'hard constraint' $\epsilon=0$ of Eq. \ref{weasthard}
has been replaced by the 'soft-constraint' $\epsilon>0$.


\subsection{ Probability of a trajectory
 $C(t)=\{S_1(t),..,S_i(t),.,S_N(t) \}$ during the large time-window $0 \leq t \leq T$ }

The probability of a trajectory $[C(0 \leq t \leq T)] $
can be written for any continuous-time Markov chain as
\begin{eqnarray}
{\cal P}[C(0 \leq t \leq T)]   
=   e^{ \displaystyle    \sum_{t \in [0,T]: C(t^+) \ne C(t) } \ln (W(C(t^+),C(t))  ) +  \int_0^T dt W(C(t),C(t))   }
\label{pwtrajjump}
\end{eqnarray}
in terms of the jump rates $W(C',C) \geq 0 $ from configuration $C$ to configuration $C' \ne C$
and in terms of the diagonal element of the Markov Matrix $W$ fixed by the conservation of probability to be
\begin{eqnarray}
W(C,C) = - \sum_{C' \ne C} W(C',C) 
\label{Wdiag}
\end{eqnarray}
For the single-spin-flip dynamics of Eq. \ref{master} where the flip rate 
\begin{eqnarray}
 W^{KCM}(\sigma_i^x C,C) = w_i^{S_i}(S_{i-1}) 
\label{wKCM}
\end{eqnarray}
only involves the flipping spin $S_i$ and its left neighboring spin $S_{i-1}$ as in Eqs \ref{weasthard}
and \ref{weast},
the probability of Eq. \ref{pwtrajjump} for the trajectory 
$C(t)=\{S_1(t),..,S_i(t),.,S_N(t) \}$ during the large time-window $0 \leq t \leq T$
becomes
\begin{eqnarray}
\ln \left( {\cal P}[C(0 \leq t \leq T)]   \right)
=   
  \sum_{i=1}^N \sum_{t \in [0,T] : S_i(t^+) \ne S_i(t) }   \ln \left( w_i^{S_i(t)} (S_{i-1}(t)) \right)
  -  \sum_{i=1}^N \int_0^T dt \ w_i^{S_i(t)}(S_{i-1}(t))    
\label{pwtrajspin}
\end{eqnarray}
and can be rewritten in terms of empirical time-averaged densities and flows
involving only two consecutive spins. In the two following sections,
we consider both the random East model where the flip rate $w_i^{S_i}(S_{i-1})  $ depends on the location $i$
and the pure East model where the flip rate $w_i^{S_i}(S_{i-1}) = w^{S_i}(S_{i-1}) $ 
does not depend on the location $i$.


\section{ Large deviations for the random East model with the soft constraint $\epsilon>0$ }

\label{sec_Erandomsoft}

In this section, the goal is to analyze the dynamical large deviations properties of 
the random East model with the soft constraint of parameter $\epsilon>0$ in the flip rates of Eq. \ref{weast}
\begin{eqnarray}
 W^{RandomEast}_{\epsilon}(\sigma_i^x C,C) = \left[ \delta_{S_{i-1},+ } + \epsilon \delta_{S_{i-1},- } \right] 
 \left[  (1-c_i) \delta_{S_i,+} + c_i \delta_{S_i,-}\right] \equiv w_i^{S_i}(S_{i-1}) 
\label{weastrandomsoft}
\end{eqnarray}


\subsection{ Identification of the relevant empirical observables  }

As recalled in Appendix \ref{appendix_relevant}, the relevant empirical observables of a Markov process
are the time-averages of time-local operators computed over the trajectory $C(0 \leq t \leq T) $ 
that are sufficient to evaluate the probability of this dynamical trajectory $C(0 \leq t \leq T) $.
Here the trajectory probability of Eq. \ref{pwtrajspin} can be rewritten as
\begin{eqnarray}
\ln {\cal P}[C(0 \leq t \leq T)]   
=  
T   \sum_{i=1}^N \sum_{S_L=\pm,S=\pm}   \left[ q_i^S(S_L)    \ln \left( w_i^S(S_L) \right)
  -  w_i^S(S_L) \rho_{i-1,i}^{S_L,S}   \right]
\label{plogempi2}
\end{eqnarray}
where the empirical flow $ q_i^S(S_L) $ represents the density of flips of the spin $S_i$
 from the value $S$ to $(-S)$ when the left neighboring spin $S_{i-1}$ takes the value $S_L$ 
\begin{eqnarray}
 q_i^S(S_L)= \frac{1}{ T }  \sum_{t \in [0,T] : S_i(t^+) \ne S_i(t) } 
\ \delta_{S_{i-1}(t) ,S_L}  \delta_{S_i(t^+) ,-S} \delta_{S_i(t) ,S}
\label{q2}
\end{eqnarray}
while the empirical density $\rho_{i-1,i}^{S_L,S} $ represents the fraction of time
where of two consecutive spins $(S_{i-1},S_i)$ located at $(i-1,i)$ take the values $(S_L,S)$
\begin{eqnarray}
\rho_{i-1,i}^{S_L,S} \equiv \frac{1}{ T }    \int_0^T dt 
\ \delta_{S_{i-1}(t) ,S_L}  \delta_{S_i(t) ,S} 
\label{rho2}
\end{eqnarray}

With respect to the general formalism of Appendix \ref{appendix_relevant},
this means that the relevant empirical observables $E$ for the random East model
are the empirical flows $q_.^.(.) $ of Eq. \ref{q2}
and the empirical densities $\rho_{.,.}^{.,.}$ of Eq. \ref{rho2},
while the corresponding action introduced in Eq. \ref{ptrajectempi} reads 
\begin{eqnarray}
\Phi_{[w]} [  q_.^.(.) ; \rho_{.,.}^{.,.} ] && = 
 \sum_{i=1}^N \sum_{S_L=\pm,S=\pm}  
  \left[w_i^S(S_L) \rho_{i-1,i}^{S_L,S} - q_i^S(S_L)    \ln \left( w_i^S(S_L) \right)       \right]
\label{actionjump}
\end{eqnarray}


\subsection{ Rate function at Level 2.5 for the relevant empirical observables  }

As explained in Appendix \ref{appendix_relevant} around Eq. \ref{modeleeff}, 
it is useful to introduce the modified Markov rates $\hat w_i^S(S_L) $
that would make typical the relevant empirical observables  $[  q_.^.(.,.) ; \rho_{.,.}^{.,.} ]$
\begin{eqnarray}
\hat w_i^{S}(S_L) && = \frac{q_i^{S}(S_L) } { \rho_{i-1,i}^{S_L,S}  }
\label{weffeast}
\end{eqnarray}
in order to evaluate the action of Eq. \ref{actionjump} corresponding to this modified Markov generator $\hat w$ 
\begin{eqnarray}
\Phi_{[\hat w]} [  q_.^.(.,.) ; \rho_{.,.}^{.,.} ] && = 
 \sum_{i=1}^N \sum_{S_L=\pm,S=\pm}  
  \left[ \hat w_i^S(S_L) \rho_{i-1,i}^{S_L,S} - q_i^S(S_L)    \ln \left( \hat w_i^S(S_L) \right)       \right]
  \nonumber \\
&& =  \sum_{i=1}^N \sum_{S_L=\pm,S=\pm}  
  \left[ q_i^{S}(S_L)  - q_i^S(S_L)    \ln \left( \frac{q_i^{S}(S_L) } { \rho_{i-1,i}^{S_L,S}  } \right)       \right]  
\label{actionjumptyp}
\end{eqnarray}
The rate function $ I_{2.5} [  q_.^.(.) ; \rho_{.,.}^{.,.}]  $ can be then obtained
from the difference of Eq. \ref{rateempi}
between the actions of Eqs \ref{actionjump} and \ref{actionjumptyp} 
\begin{eqnarray}
 I_{2.5} [  q_.^.(.) ; \rho_{.,.}^{.,.}]  
&& = \Phi_{[w]} [  q_.^.(.,.);  \rho_{.,.}^{.,.} ] -  \Phi_{[\hat w]} [  q_.^.(.,.) ; \rho_{.,.}^{.,.} ]
\nonumber \\
&& =  \sum_{i=1}^N  \sum_{S_L=\pm,S=\pm}  
\left[ q_i^S(S_L)    \ln \left( \frac{  q_i^S(S_L)  }{w_i^S(S_L)  \rho_{i-1,i}^{S_L,S} } \right)
- q_i^S(S_L) +   w_i^S(S_L) \rho_{i-1,i}^{S_L,S} \right]
\label{rate2.5e}
\end{eqnarray}
where one recognizes the relative entropy cost of having empirical flows $q_i^S(S_L) $
different from the typical flows $w_i^S(S_L)  \rho_{i-1,i}^{S_L,S} $
 that would be produced by the empirical densities $ \rho_{i-1,i}^{S_L,S} $.

For each site $i=1,..,N$ and each value $S_L=\pm$ of the left neighboring spin,
 it is useful to parametrize the two empirical flows $q_i^+(S_L) $ and $q_i^-(S_L)$ concerning the flip of the spin $S_i$ when the left neighboring spin $S_{i-1}$ takes the value $S_L$
 \begin{eqnarray}
q_i^+(S_L)  \equiv \frac{a_i(S_L)  + j_i(S_L) }{2} 
\nonumber \\
q_i^-(S_L)  \equiv \frac{a_i(S_L)  - j_i(S_L) }{2} 
\label{ajrecieast}
\end{eqnarray}
 by their symmetric and antisymmetric parts called the empirical activity and the empirical current
\begin{eqnarray}
a_i(S_L)  \equiv q_i^+(S_L)  +q_i^-(S_L) 
\nonumber \\
j_i(S_L)  \equiv q_i^+(S_L)  -q_i^-(S_L) 
\label{ajeast}
\end{eqnarray}

Via this change of variables, the rate function of Eq. \ref{rate2.5e} translates into
\begin{eqnarray}
&& I_{2.5} [  a_.(.) ; j_.(.) ; \rho_{.,.}^{.,.}]  
 =  \sum_{i=1}^N  \sum_{S_L=\pm} 
 \bigg[  \frac{ j_i(S_L) }{2}     
 \ln \left( \frac{  [ a_i(S_L)  + j_i(S_L) ]  w_i^-(S_L)  \rho_{i-1,i}^{S_L,-} }
 { [a_i(S_L)  - j_i(S_L)  ]  w_i^+(S_L)  \rho_{i-1,i}^{S_L,+} } \right)
\nonumber \\
&& + \frac{a_i(S_L)  }{2}     \ln \left( \frac{  a_i^2(S_L)  - j_i^2(S_L)  }{ 4 w_i^+(S_L)  \rho_{i-1,i}^{S_L,+} w_i^-(S_L)  \rho_{i-1,i}^{S_L,-}} \right)
- a_i(S_L) +   w_i^+(S_L) \rho_{i-1,i}^{S_L,+}   +   w_i^-(S_L) \rho_{i-1,i}^{S_L,-} \bigg]
\label{rate2.5aj}
\end{eqnarray}


\subsection{ Constitutive constraints for the relevant empirical observables  }

\label{subsec_c2.5eastrandom}

As recalled in Appendix \ref{appendix_relevant}, the large deviations at Level 2.5
do not involve only the rate function $I_{2.5}$, but they also contain the constitutive constraints
that the empirical observables should satisfy (see the constraints $C(E)$ for the general formulation of Eq. \ref{probaaempi} and the explicit constraints of Eq. \ref{level2.5master}
for the special case of Markov jump processes in the full configuration space).

\subsubsection{ Closed constraints for the local empirical densities for one spin and for two consecutive spins}

The empirical 2-spin density of Eq. \ref{rho2} contains the information on 
the empirical 1-spin density $\rho_{i}^{S_i} $ that can be computed either from 
$\rho_{i-1,i}^{S_{i-1},S_{i}} $
 or from 
 $ \rho_{i,i+1}^{S_i,S_{i+1}} $
\begin{eqnarray}
\rho_{i}^{S_i} \equiv  \frac{1}{ T }    \int_0^T dt 
\   \delta_{S_i(t) ,S_i}= \sum_{S_{i-1}} \rho_{i-1,i}^{S_{i-1},S_{i}}= \sum_{S_{i+1}} \rho_{i,i+1}^{S_i,S_{i+1}}
\label{rho1from2}
\end{eqnarray}
and that should be normalized
\begin{eqnarray}
\rho_{i}^+ + \rho_i^- =1
\label{rho1norma}
\end{eqnarray}
In practice, it will be convenient to take into account these consistency constraints as follows :

(i) for each site $i$, one keeps the basic variable $\rho_i^+ \in [0,1]$ and one eliminates $\rho_i^-$ via
the normalization of Eq. \ref{rho1norma}
\begin{eqnarray}
\rho_i^-&& =1- \rho_i^+
\label{elimc1}
\end{eqnarray}

(ii) for each link $(i-1,i)$, one keeps the basic variable $\rho_{i-1,i}^{++}$ and 
one eliminates the three other link variables via the consistency constraints of Eq. \ref{rho1from2}
\begin{eqnarray}
\rho_{i-1,i}^{+-} && =\rho_{i-1}^+ - \rho_{i-1,i}^{++}
\nonumber \\
\rho_{i-1,i}^{-+} && =\rho_{i}^+ - \rho_{i-1,i}^{++}
\nonumber \\
\rho_{i-1,i}^{--} && = 1 - \rho_{i-1}^+ -\rho_{i}^+ + \rho_{i-1,i}^{++}
\label{elimc2}
\end{eqnarray}

In summary, the constitutive constraints satisfied by the empirical 1-spin density $\rho_.^. $ and 
by the empirical 2-spin density $\rho_{.,.}^{.,.} $ are closed
and can be summarized as a way to compute all the values in terms of the 
$(2N)$ basic variables $(\rho_i^+,\rho_{i-1,i}^{++}) $ discussed in (i) and (ii) above
\begin{eqnarray}
 C_{2} [ \rho_.^. ;  \rho_{.,.}^{.,.}] 
  && 
= 
 \prod_{i=1}^N [ 
\delta \left(   \rho_i^- -[1- \rho_i^+ ] \right)
 \nonumber \\ && 
\delta \left ( \rho_{i-1,i}^{+-} - [\rho_{i-1}^+ - \rho_{i-1,i}^{++}] \right) 
\delta \left ( \rho_{i-1,i}^{-+} -[\rho_{i}^+ - \rho_{i-1,i}^{++}]\right)  
\delta \left ( \rho_{i-1,i}^{--} -[ 1 - \rho_{i-1}^+ -\rho_{i}^+ + \rho_{i-1,i}^{++}] \right) 
]
\label{c2e} 
\end{eqnarray}

\subsubsection{ Closure problem in the stationary constraints for the local empirical currents }

For Markov jump processes, the empirical flows in the full configuration space 
have to satisfy the stationary constraints of Eq. \ref{contrainteq},
that can be rewritten as Eq. \ref{contraintej}
in terms of the empirical currents only, while the empirical activities disappear from these constraints.

In the present case however, it is not possible to write closed constraints for the local empirical currents $j_i(S_L)$
of Eq. \ref{ajeast} that would ensure the stationarity of the local 2-spin empirical density $\rho_{.,.}^{.,.} $.
More generally for many-body models, the projection of the explicit stationarity constraints in the full configuration space towards local empirical observables is not closed but gives rise to a whole hierarchy, as 
discussed recently on the example of disordered exclusion processes between two reservoirs \cite{c_exclusion}.

\subsubsection{ Conclusion : Closure problem in the Level 2.5 for the local empirical currents }

In conclusion, despite the explicit rate function at Level 2.5 of Eq. \ref{rate2.5e}, or equivalently Eq. \ref{rate2.5aj},
and despite the explicit constitutive constraints of Eq. \ref{c2e} for the local empirical densities,
the Level 2.5 cannot be written in closed form for the local empirical observables
as a consequence of the closure problem in the stationary constraints for the local empirical currents discussed above. 
However we have not yet used the detailed-balance property of the East model as discussed below.


\subsection{ Explicit large deviations at Level 2.25 for the local empirical densities and the local empirical activities  }

\label{subsec_c2.25eastrandom}

As explained in Appendix \ref{app_LargeDevDB}, the detailed-balance property of the rates
induce special properties for the dynamical large deviations of Markov jump processes.
In particular for given empirical density and given empirical activity, 
the rate function at Level 2.5 is minimized when all the empirical currents vanish (Eq. \ref{jopt0}),
while the stationary constraints are trivially satisfied for vanishing currents.

For the present random soft East model that satisfies the detailed-balance property,
the vanishing of all the empirical currents $j_i(S_L)$ of Eq. \ref{ajeast}
\begin{eqnarray}
j_i^{opt}(S_L)  =0
\label{ajeastoptzero}
\end{eqnarray}
 leads to the closed Level 2.25 for the local empirical densities and the local empirical activities
 \begin{eqnarray}
  P^{[2.25]}_T [  a_.(.) ; \rho_.^. ;  \rho_{.,.}^{.,.}] 
  \opsimeq_{T \to +\infty} 
C_{2} [ \rho_.^. ;  \rho_{.,.}^{.,.}]
e^{ \displaystyle - T  I_{2.25} [  a_.(.) ; \rho_{.,.}^{.,.}]
 }
\label{level2.25e}
\end{eqnarray}
where the constitutive constraints $C_{2} [ \rho_.^. ;  \rho_{.,.}^{.,.}] $ concerning the local empirical densities 
have been written in Eq. \ref{c2e},
while the rate function at Level 2.25 is obtained from the rate function at Level 2.5 when all the empirical currents vanish $j^{opt}_i(S_L)=0$
\begin{eqnarray}
I_{2.25} [  a_.(.) ;  \rho_{.,.}^{.,.}]  && = I_{2.5} [  a_.(.) ; j^{opt}_.(.)=0 ; \rho_{.,.}^{.,.}]  
\label{rate2.25e}
 \\
&&  =  \sum_{i=1}^N  \sum_{S_L=\pm} 
 \bigg[  \frac{a_i(S_L)  }{2}     \ln \left( \frac{  a_i^2(S_L)    }{ 4 w_i^+(S_L)  \rho_{i-1,i}^{S_L,+} w_i^-(S_L)  \rho_{i-1,i}^{S_L,-}} \right)
- a_i(S_L) +   w_i^+(S_L) \rho_{i-1,i}^{S_L,+}   +   w_i^-(S_L) \rho_{i-1,i}^{S_L,-} \bigg]
\nonumber
\end{eqnarray}
An alternative derivation of this rate function $I_{2.25} [  a_.(.) ;  \rho_{.,.}^{.,.}]   $
 at Level 2.25 for the local activities $a_{i} (\pm)  $ 
and the local densities $\rho_{i-1,i}^{\pm,\pm} $
is given in Appendix \ref{app_GlobalLocal} via the explicit contraction from the Level 2.25 
in the space of the $2^N$ configurations.

For the random East model with the soft constraint of parameter $\epsilon$ in the flip rates of Eq. \ref{weastrandomsoft},
the rate function at Level 2.25 of Eq. \ref{rate2.25e} reads more explicitly
\begin{eqnarray}
 I_{2.25}^{RandomEastSoft} [  a_.(.) ; \rho_{.,.}^{.,.}]
=  \sum_{i=1}^N  
\left[ \frac{a_i(+)  }{2}    \ln \left( \frac{  a_i^2(+)  }
{ 4 c_i (1-c_i)  \rho_{i-1,i}^{++}   \rho_{i-1,i}^{+-} } \right)
-  a_i(+) + (1-c_i)  \rho_{i-1,i}^{++} 
+ c_i  \rho_{i-1,i}^{+-}  \right]
\nonumber \\
+
\sum_{i=1}^N  
\left[ \frac{a_i(-)  }{2}    \ln \left( \frac{  a_i^2(-)  }
{ 4 \epsilon^2 c_i (1-c_i)  \rho_{i-1,i}^{-+}   \rho_{i-1,i}^{--} } \right)
-  a_i(-) + \epsilon (1-c_i)  \rho_{i-1,i}^{-+} 
+  \epsilon c_i  \rho_{i-1,i}^{--}  \right]
 \label{rate2.25rEs} 
\end{eqnarray}


\subsection{ Explicit contraction over the local activities towards the Level 2 for the local empirical densities alone  }

The optimization of the rate function at Level 2.25 of Eq. \ref{rate2.25e} over the local empirical activities $a_i(S_L) $
\begin{eqnarray}
0 && = \frac{ \partial I_{2.25} [  a_.(.,.) ; \rho_{.,.}^{.,.}] } { \partial a_i(S_L)} 
=     \ln \left( \frac{  a_i^2(S_L)  }
{4 w_i^+(S_L)  \rho_{i-1,i}^{S_L,+} w_i^-(S_L)  \rho_{i-1,i}^{S_L,-} } \right)
\label{optimizeq}
\end{eqnarray}
leads to the optimal values 
\begin{eqnarray}
  a_i^{opt}(S_L) = 2 \sqrt{ w_i^+(S_L)  \rho_{i-1,i}^{S_L,+} w_i^-(S_L)  \rho_{i-1,i}^{S_L,-}}  
\label{qiopt}
\end{eqnarray}
that can be plugged into Eq. \ref{rate2.25e}
to obtain the rate function at Level 2
\begin{eqnarray}
 I_2 [   \rho_{.,.}^{.,.}]   && =  I_{2.25} [  a_.^{opt}(.) ; \rho_{.,.}^{.,.}]
   = \sum_{i=1}^N \sum_{S_L=\pm}     \left[   
   \sqrt{ w_i^+(S_L)  \rho_{i-1,i}^{S_L,+} }
   - \sqrt{ w_i^-(S_L)  \rho_{i-1,i}^{S_L,-}}     \right]^2
\label{rate2e}
\end{eqnarray}
that will govern the large deviations at Level 2 for the empirical densities alone
\begin{eqnarray}
  P^{[2]}_T [  \rho_.^. ;  \rho_{.,.}^{.,.}] \opsimeq_{T \to +\infty} 
C_2 [   \rho_.^. ;  \rho_{.,.}^{.,.}]
e^{ \displaystyle - T  I_{2}  [  \rho_{.,.}^{.,.}]  }
\label{level2e}
\end{eqnarray}

For the random East model with the soft constraint of parameter $\epsilon$ in the flip rates of Eq. \ref{weastrandomsoft},
the rate function at Level 2 of Eq. \ref{rate2e} reads more explicitly
\begin{eqnarray}
 I_{2}^{RandomEastSoft} [   \rho_{.,.}^{.,.}]
=  \sum_{i=1}^N  
\left( \left[ \sqrt{ (1-c_i)  \rho_{i-1,i}^{++} } - \sqrt{ c_i  \rho_{i-1,i}^{+-} } \right]^2
+ \epsilon \left[ \sqrt{  (1-c_i)  \rho_{i-1,i}^{-+} } - \sqrt{ c_i  \rho_{i-1,i}^{--} } \right]^2 \right)
 \label{rate2rEs} 
\end{eqnarray}


\subsection{ Typical fluctuations of order $\frac{1}{\sqrt{T} }$ for the empirical densities and activities around their equilibrium values }

The equilibrium 2-spin density of Eq. \ref{peqKCM}
\begin{eqnarray}
 \Pi_{i-1,i}^{S_{i-1},S_i} = \Pi_{i-1}^{S_{i-1} }\Pi_i^{S_i} =
\left[ c_{i-1} \delta_{S_{i-1},+} + (1-c_{i-1}) \delta_{S_{i-1},-} \right]
\left[ c_i \delta_{S_i,+} + (1-c_i) \delta_{S_i,-} \right]
\label{rhoeqe}
\end{eqnarray}
and the corresponding equilibrium activities
\begin{eqnarray}
 A_i(+)  && = 2 w_i^{+}(+) \Pi_{i-1,i}^{++} = 2 w_i^{-}(+) \Pi_{i-1,i}^{+-} = 2 c_{i-1} c_i (1-c_i)
 \nonumber \\
 A_i(-)  && = 2 w_i^{+}(-) \Pi_{i-1,i}^{-+} = 2 w_i^{-}(-) \Pi_{i-1,i}^{--} = 2 \epsilon (1-c_{i-1}) c_i (1-c_i)
\label{Aeqe}
\end{eqnarray}
are the only values of the local empirical densities and local empirical activities that satisfy the constraints $C_2 [   \rho_.^. ;  \rho_{.,.}^{.,.}] $ and that make  
the rate function $I_{2.25}^{RandomEastSoft} [  a_.(.) ; \rho_{.,.}^{.,.}] $ of Eq. \ref{rate2.25rEs} vanish.
The Level 2.25 of Eq. \ref{level2.25e} characterizes how rare it is for large $T$  to see 
empirical densities $\rho_{.,.}^{.,.} $ and activities $a_.(.)  $ 
that are different from their equilibrium values $\Pi_{.,.,.}^{.,.,.} $ and $A_.(.)  $.
If one is interested only in the small typical fluctuations of order $\frac{1}{\sqrt{T} }$ around these equilibrium values
\begin{eqnarray}
 \rho_{i-1,i}^{S_L,S} && = \Pi_{i-1,i}^{S_L,S} + \frac{{\hat \rho}_{i-1,i}^{S_L,S}}{\sqrt{T} }
\nonumber \\
 a_i(S_L) && =  A_i(S_L)+ \frac{{\hat a}_i(S_L) }{\sqrt{T} }
\label{hatrhoq}
\end{eqnarray}
one just needs to expand
the rate function $I_{2.25}^{RandomEastSoft} [  a_.(.) ; \rho_{.,.}^{.,.}] $ of Eq. \ref{rate2.25rEs}
at second order in the perturbations to obtain
the rescaled Gaussian rate function for the rescaled densities ${\hat \rho}_{.,.,.}^{.,.,.} $ and 
the rescaled activities ${\hat a}_.(.,.)  $
\begin{eqnarray}
 {\hat I}_{2.25}^{small} [  {\hat a}_.(.) ; {\hat \rho}_{.,.}^{.,.}] 
 && \equiv  \lim_{T \to + \infty}
 \left( T I_{2.25}^{RandomEastSoft} [  a_.(.)= A_.(.) +\frac{{\hat a}_.(.) }{\sqrt{T}} ; 
\rho_{.,.}^{.,.}= \Pi_{.,.}^{.,.}+ \frac{{\hat \rho}_{.,.}^{.,.}}{\sqrt{T}} ] \right)
\nonumber \\
&& =   \sum_{i=1}^N  \sum_{S_L=\pm}  
    \frac{  \left[ \frac{ {\hat a}_i(S_L)}{2}  - w_i^{+}(S_L){\hat \rho}_{i-1,i}^{S_L,+}\right]^2
    + \left[ \frac{ {\hat a}_i(S_L)}{2} - w_i^{-}(S_L){\hat \rho}_{i-1,i}^{S_L,-}\right]^2  }
{  A_i(S_L)  } 
 \nonumber \\
 && 
 = \sum_{i=1}^N  \sum_{S_L=\pm}  
    \frac{   \left[  {\hat a}_i(S_L) -  \left( w_i^{+}(S_L){\hat \rho}_{i-1,i}^{S_L,+}
    +w_i^{-}(S_L){\hat \rho}_{i-1,i,}^{S_L,-} \right)\right]^2
    +  \left[ w_i^{+}(S_L){\hat \rho}_{i-1,i}^{S_L,+}  - w_i^{-}(S_L){\hat \rho}_{i-1,i}^{S_L,-}  \right]^2  }
{   2 A_i(S_L)  } 
 \nonumber \\
 && 
 = \sum_{i=1}^N  
    \frac{   \left[ {\hat a}_i(+) - 
    \left( (1-c_i) {\hat \rho}_{i-1,i}^{++}
    + c_i {\hat \rho}_{i-1,i,}^{+-} \right) \right]^2
    +  \left[  (1-c_i){\hat \rho}_{i-1,i}^{++}  - c_i{\hat \rho}_{i-1,i}^{+-}  \right]^2  }
{   4 c_{i-1} c_i (1-c_i)  } 
 \nonumber \\ && 
 + \sum_{i=1}^N 
    \frac{   \left[ {\hat a}_i(-) - \epsilon
    \left( (1-c_i) {\hat \rho}_{i-1,i}^{-+}
    +c_i{\hat \rho}_{i-1,i,}^{--} \right)\right]^2
    + \epsilon^2 \left[  (1-c_i){\hat \rho}_{i-1,i}^{-+}  - c_i{\hat \rho}_{i-1,i}^{--}  \right]^2  }
{   4 \epsilon (1-c_{i-1}) c_i (1-c_i)  } 
 \label{rate2.5egauss}
\end{eqnarray}
that will govern the probability of the rescaled fluctuations $ [  {\hat a}_.(.) ; {\hat \rho}_.^. ;  {\hat \rho}_{.,.}^{.,.}] $
of the empirical local observables
\begin{eqnarray}
  {\hat P}^{[2.25]}_T [  {\hat a}_.(.) ; {\hat \rho}_.^. ;  {\hat \rho}_{.,.}^{.,.}] 
  \opsimeq_{T \to +\infty} 
 {\hat C}^{small}_{2} [ {\hat \rho}_.^. ;  {\hat \rho}_{.,.}^{.,.}]
e^{ \displaystyle -  {\hat I}_{2.25}^{small} [  {\hat a}_.(.) ; {\hat \rho}_{.,.}^{.,.}] 
 }
\label{level2.25egauss}
\end{eqnarray}
together with the constraints inherited from Eq. \ref{c2e} 
\begin{eqnarray}
 {\hat C}^{small}_{2} [ {\hat \rho}_.^. ;  {\hat \rho}_{.,.}^{.,.}]
 = 
 \prod_{i=1}^N \left[ 
\delta \left(   {\hat \rho}_i^- + {\hat \rho}_i^+  \right)
\delta \left ( {\hat \rho}_{i-1,i}^{+-} - [{\hat \rho}_{i-1}^+ - {\hat \rho}_{i-1,i}^{++}] \right) 
\delta \left ( {\hat \rho}_{i-1,i}^{-+} -[{\hat \rho}_{i}^+ - {\hat \rho}_{i-1,i}^{++}]\right)  
\delta \left ( {\hat \rho}_{i-1,i}^{--} -[  {\hat \rho}_{i-1,i}^{++} - {\hat \rho}_{i-1}^+ -{\hat \rho}_{i}^+ ] \right) 
\right] \ \ 
\label{c2hat} 
\end{eqnarray}

If one is interested into the rescaled fluctuations of the local empirical densities alone,
the rescaled Gaussian rate function reads
\begin{eqnarray}
 {\hat I}_2^{small} [ {\hat \rho}_{.,.}^{.,.}]    && \equiv \lim_{T \to + \infty}
 \left(  T I_{2}^{RandomEastSoft} [ 
\rho_{.,.}^{.,.,.}= \Pi_{.,.}^{.,.}+ \frac{{\hat \rho}_{.,.}^{.,.,.}}{\sqrt{T}} ] \right)
 \nonumber \\ &&
 =   \sum_{i=1}^N  \sum_{S_L=\pm}  
    \frac{   \left[  w_i^{+}(S_L){\hat \rho}_{i-1,i}^{S_L,+}  - w_i^{-}(S_L){\hat \rho}_{i-1,i}^{S_L,-}  \right]^2  }
{   2 A_i(S_L)  } 
 \nonumber \\ &&
       = \sum_{i=1}^N  
\left(    \frac{    \left[ (1-c_i){\hat \rho}_{i-1,i}^{++}  - c_i{\hat \rho}_{i-1,i}^{+-}  \right]^2  }
{   4 c_{i-1} c_i (1-c_i)  } 
+\epsilon   \frac{  
     \left[  (1-c_i){\hat \rho}_{i-1,i}^{-+}  - c_i{\hat \rho}_{i-1,i}^{--}  \right]^2  }
{    4 (1-c_{i-1}) c_i (1-c_i)  } 
\right)
 \label{rate2egauss}
\end{eqnarray}
Since the rescaled empirical densities $ [{\hat \rho}_{.}^{.}, {\hat \rho}_{.,.}^{.,.}]   $ belong to $]-\infty,+\infty[$,
one can use the constraints of Eq. \ref{c2hat} to keep only the $(2N)$ rescaled empirical densities 
$ [{\hat \rho}_{.}^+, {\hat \rho}_{.,.}^{++}]   $ 
and one obtains the rescaled rate function
\begin{eqnarray}
 {\hat I}_2^{small} [ {\hat \rho}_{.}^+, {\hat \rho}_{.,.}^{++} ]  
       = \sum_{i=1}^N  
\left(    \frac{    \left[ {\hat \rho}_{i-1,i}^{++} 
 - c_i {\hat \rho}_{i-1}^+  \right]^2  }
{   4 c_{i-1} c_i (1-c_i)  } 
+\epsilon   \frac{  
     \left[  {\hat \rho}_{i-1,i}^{++}  - c_i {\hat \rho}_{i-1}^+    - {\hat \rho}_{i}^+   
      \right]^2  }
{    4 (1-c_{i-1}) c_i (1-c_i)  } 
\right)
 \label{rate2egausspositive}
\end{eqnarray}
that will govern their joint Gaussian probability without any remaining constraint
\begin{eqnarray}
  {\hat P}^{[2]}_T [ {\hat \rho}_{.}^+, {\hat \rho}_{.,.}^{++}   ] 
  \opsimeq_{T \to +\infty} 
e^{ \displaystyle -  {\hat I}_{2}^{small} [ {\hat \rho}_{.}^+, {\hat \rho}_{.,.}^{++} ] 
 }
\label{level2egausspositive}
\end{eqnarray}
The Gaussian integration over the rescaled 2-spin density ${\hat \rho}_{.,.}^{++} $ yields 
that the probability of the rescaled 1-spin density $ {\hat \rho}_{.}^+ $ alone 
\begin{eqnarray}
  {\hat P}^{[2]}_T [ {\hat \rho}_{.}^+  ] 
  \opsimeq_{T \to +\infty} 
e^{ \displaystyle -  \sum_{i=1}^N  \frac{ [{\hat \rho}_{i}^+ ]^2}{ 2 v_i}   }
\label{level2egausspositive1spin}
\end{eqnarray}
reduces to independent Gaussian distributions for the $\rho_i^+$ with variances
\begin{eqnarray}
 v_i = 2 c_i (1-c_i) \left[ c_{i-1} + \frac{ (1-c_{i-1} ) }{\epsilon} \right]
 \label{variancei}
\end{eqnarray}
that are finite for the soft version $\epsilon>0$ of the East model, but that diverge for 
the true East model corresponding to the hard-constraint $\epsilon=0$ that will be analyzed in section \ref{sec_Erandom}.


\subsection{ Time-additive observables that can be reconstructed from the local empirical densities and activities}

 The local empirical activity of Eq. \ref{ajeast} with Eq. \ref{q2}
\begin{eqnarray}
a_i(S_L)  = \sum_{S=\pm} q_i^S(S_L)  =  \frac{1}{ T }  \sum_{t \in [0,T] : S_i(t^+) \ne S_i(t) } 
\ \delta_{S_{i-1}(t) ,S_L} 
\label{ajeastexpli}
\end{eqnarray}
represents the density of the total number of flips of the spin $S_i$ 
when the left neighboring spin $S_{i-1}$ takes the value $S_L$.

These local empirical activities and the local empirical densities $\rho_{i-1,i}^{S_L,S} $ of Eq. \ref{rho2} 
allow to reconstruct any time-additive space-local observable $O_T $
that can be parametrized by some functions $\alpha_i^{S}(S_L) $ and $\beta_i(S_L) $
\begin{eqnarray}
O_T 
&& =  \sum_{i=1}^N  \sum_{S_L=\pm}  \left[\sum_{S=\pm}\alpha_i^{S}(S_L) \rho_{i-1,i}^{S_L,S} 
+ \beta_i(S_L)  a_i(S_L)\right]
\nonumber \\
&& = 
 \sum_{i=1}^N  \left[ \frac{1}{T} \int_0^T dt 
   \alpha_i^{S_i(t)}(S_{i-1}(t) )   
+   \frac{1}{T}   \sum_{t \in [0,T] : S_i(t^+) \ne S_i(t) }  \beta_i(S_{i-1}(t)) \right]
\label{additive2}
\end{eqnarray}

 \subsubsection{ Averaged value and rescaled variance }

The first cumulant $G_1$ corresponding to the averaged value $\langle O_T \rangle $ 
coincides with the equilibrium value $ O_{eq} $ computed from the equilibrium density $\Pi_{i-1,i}^{S_L,S} $ of Eq. \ref{rhoeqe}
and from the corresponding equilibrium activities $A_i(S_L) $ of Eq. \ref{Aeqe}
\begin{eqnarray}
G_1 \equiv \langle O_T \rangle = O_{eq} 
\equiv  \sum_{i=1}^N  \sum_{S_L=\pm}  \left[\sum_{S=\pm}\alpha_i^{S}(S_L) \Pi_{i-1,i}^{S_L,S} 
+ \beta_i(S_L)  A_i(S_L)\right]
\label{g1}
\end{eqnarray}
The small typical fluctuations of order $\frac{1}{\sqrt{T} } $ around this equilibrium value 
can be rewritten in terms of the rescaled fluctuations 
${\hat \rho}_{...}^{...} $ and ${\hat a}_.(.,.) $ of Eq. \ref{hatrhoq}
\begin{eqnarray}
O_T - \langle O_T \rangle
&& = \frac{1}{\sqrt{T} } \sum_{i=1}^N  \sum_{S_L=\pm,S=\pm}  \left[\alpha_i^{S}(S_L) {\hat \rho}_{i-1,i,i+1}^{S_L,S}
+ \beta_i^S(S_L)  {\hat a}_i(S_L)  \right]
\label{additivesmallfluct}
\end{eqnarray}
so that the rescaled variance
\begin{eqnarray}
G_2  \equiv  T   \langle  \left( O_T  - \langle  O_T  \rangle \right)^2  \rangle
= \bigg\langle \left( \sum_{i=1}^N  \sum_{S_L=\pm} 
 \left[\sum_{S=\pm}\alpha_i^{S}(S_L) {\hat \rho}_{i-1,i}^{S_L,S,}
+ \beta_i(S_L)  {\hat a}_i(S_L)  \right] \right)^2  \bigg\rangle
\label{g2}
\end{eqnarray}
can be evaluated via the average over the probability ${\hat P}^{[2.25]}_T [  {\hat a}_.(.) ; {\hat \rho}_.^. ;  {\hat \rho}_{.,.}^{.,.}] $ of the rescaled fluctuations of Eq. \ref{level2.25egauss}.


\subsubsection{ Large deviations governed by the rate function $I(O)$ and the scaled cumulant generating function $G(k)$  }

The large deviations properties for large $T$
 \begin{eqnarray}
 P_T( O ) \opsimeq_{T \to +\infty} e^{- T I ( O )}
\label{level1def}
\end{eqnarray} 
are governed by the rate function $I(O)$ that vanishes only for the equilibrium value $O_{eq}$ of Eq. \ref{g1}
 \begin{eqnarray}
 I ( O_{eq} ) =0
\label{iaeqvanish}
\end{eqnarray}
The generating function
\begin{eqnarray}
\langle e^{k T O_T } \rangle \equiv \int dO  e^{k T O} P_T( O ) \opsimeq_{T \to +\infty} 
\int dO e^{ T \left[ k O - I ( O ) \right] }\opsimeq_{T \to +\infty} e^{ T G(k) }
\label{level1gen}
\end{eqnarray} 
involves the generating function $G(k)$ of the whole series of the scaled cumulants $G_n$ (beyond 
the two first scaled cumulants $G_1$ and $G_2$ discussed in Eqs \ref{g1} and \ref{g2} )
\begin{eqnarray}
G(k) && = \sum_{n=1}^{+\infty} G_n \frac{k^n}{n!} =   G_1 k + G_2 \frac{k^2}{2}  + O(k^3)
\label{gkper}
\end{eqnarray}
The saddle-point evaluation in $T$ of the integral of Eq. \ref{level1gen}
yields that $G(k)$
 is the Legendre transform of the rate function $ I ( O ) $
 \begin{eqnarray}
 k O - I ( O ) && =  G(k) 
 \nonumber \\
 k - I'(O) && =0
\label{legendre}
\end{eqnarray}

The generating function of Eq. \ref{level1gen}
 can be evaluated from the Level 2.25 of Eq. \ref{level2.25e} 
via the integral over the empirical activities $a_.(.) $ and the empirical densities $ [   \rho_.^. ;  \rho_{.,.}^{.,.}]  $
\begin{eqnarray}
 \langle e^{ k T O_T} \rangle 
 && =  \int  d a_.(.) \int d\rho_.^. \int d\rho_{.,.}^{.,.}
 P^{[2.25]}_T [  a_.(.) ; \rho_.^. ;  \rho_{.,.}^{.,.}]  
 \ e^{\displaystyle  k T \sum_{i=1}^N  \sum_{S_L=\pm,}  \left[\sum_{S=\pm}\alpha_i^S(S_L) \rho_{i-1,i}^{S_L,S} 
+ \beta_i(S_L)  a_i(S_L)\right]}
\nonumber \\
&&
 \opsimeq_{T \to +\infty} 
 \int d\rho_.^. \int d\rho_{.,.}^{.,.}
 C_2 [  \rho_.^. ;  \rho_{.,.}^{.,.}] 
 \int  d a_.(.) e^{ T L_{2.25}^{[k]}  [  a_.(.) ; \rho_{.,.}^{.,.}]  }
\label{gene}
\end{eqnarray}
where the function
\begin{eqnarray}
 L_{2.25}^{[k]} [  a_.(.) ; \rho_{.,.}^{.,.}]  && =  - I_{2.25} [  a_.(.) ; \rho_{.,.}^{.,.}]
+ k  \sum_{i=1}^N  \sum_{S_L=\pm}  
\left[\sum_{S=\pm}\alpha_i^S(S_L) \rho_{i-1,i}^{S_L,S} 
+ \beta_i(S_L)  a_i(S_L)\right] 
\nonumber \\
&& = - \sum_{i=1}^N  \sum_{S_L=\pm}  
\frac{a_i(S_L) }{2}   \ln \left( \frac{  a_i^2(S_L)  }
{4 e^{2 k \beta_i(S_L)} w_i^+(S_L)  \rho_{i-1,i}^{S_L,+} 
w_i^-(S_L)  \rho_{i-1,i}^{S_L,-} } \right)
\nonumber \\
&& + \sum_{i=1}^N  \sum_{S_L=\pm}  
\left[  a_i(S_L) 
+ \left( k \alpha_i^+(S_L)- w_i^+(S_L) \right) \rho_{i-1,i}^{S_L,+} 
+ \left( k \alpha_i^-(S_L)- w_i^-(S_L) \right) \rho_{i-1,i}^{S_L,-}  \right]
\label{Lk}
\end{eqnarray}
involves the standard $k$-deformation rules for the Markov Matrix adapted to the time-additive observable of Eq. \ref{additive2}.
The optimization of Eq. \ref{Lk}
 over the activities $a_i(S_L) $
 \begin{eqnarray}
0 && = \frac{ \partial L_{2.25}^{[k]} [  a_.(.) ; \rho_{.,.}^{.,.}] } { \partial a_i(S_L)} 
=   - \frac{1 }{2}   \ln \left( \frac{  a_i^2(S_L)  }
{4 e^{2 k \beta_i(S_L)} w_i^+(S_L)  \rho_{i-1,i}^{S_L,+} 
w_i^-(S_L)  \rho_{i-1,i}^{S_L,-} } \right)
\label{optimizeqlk}
\end{eqnarray}
leads to the optimal values 
\begin{eqnarray}
  a_i^{opt}(S_L) = 2 e^{ k \beta_i(S_L)}
  \sqrt{ w_i^+(S_L)    \rho_{i-1,i,i+1}^{S_L,+} w_i^-(S_L)  \rho_{i-1,i,i+1}^{S_L,-}}  
\label{qioptk}
\end{eqnarray}
that can be plugged into Eq. \ref{Lk}
to obtain the function of the empirical densities alone
\begin{eqnarray}
 L_{2}^{[k]} [   \rho_{.,.}^{.,.}]  &&  = 
 \sum_{i=1}^N  \sum_{S_L=\pm}  
\left[  2  e^{k \beta_i(S_L)  }  \sqrt{ w_i^+(S_L)  
  \rho_{i-1,i}^{S_L,+} w_i^-(S_L)  \rho_{i-1,i}^{S_L,-}}  
 \right]
\nonumber \\
&& + 
 \sum_{i=1}^N  \sum_{S_L=\pm}  
\left[   \left( k \alpha_i^+(S_L)- w_i^+(S_L) \right) \rho_{i-1,i}^{S_L,+} 
+ \left( k \alpha_i^-(S_L)- w_i^-(S_L) \right) \rho_{i-1,i}^{S_L,-}  \right]
\label{Lk2bis}
\end{eqnarray}
that governs the generating function of Eq. \ref{gene}
\begin{eqnarray}
 \langle e^{ k T A_T} \rangle 
 &&
 \opsimeq_{T \to +\infty} 
 \int d\rho_.^. \int d\rho_{.,.}^{.,.}
 C_2 [  \rho_.^. ;  \rho_{.,.}^{.,.}] 
 e^{ T L_{2}^{[k]}  [  \rho_{.,.}^{.,.}]  }
\label{gene2bis}
\end{eqnarray}

For the random East model with the soft constraint of parameter $\epsilon$ in the  flip rates of Eq. \ref{weastrandomsoft},
the function of Eq. \ref{Lk2bis} reads
 \begin{eqnarray}
&& L_{2}^{RandomEastSoft[k]} [   \rho_{.,.}^{.,.}]  
 \nonumber \\
&&   = 
 \sum_{i=1}^N   
\left[  2  e^{k \beta_i(+)  }  \sqrt{ c_i (1-c_i)   \rho_{i-1,i}^{++}   \rho_{i-1,i}^{+-}}  
+   \left( k \alpha_i^+(+)- (1-c_i) \right) \rho_{i-1,i}^{++} 
+ \left( k \alpha_i^-(+)- c_i \right) \rho_{i-1,i}^{+-}  \right]
\nonumber \\
&&+  \sum_{i=1}^N   
\left[  2  e^{k \beta_i(-)  } \epsilon \sqrt{ c_i (1-c_i)   \rho_{i-1,i}^{-+}   \rho_{i-1,i}^{--}}  
+   \left( k \alpha_i^+(-)- \epsilon (1-c_i) \right) \rho_{i-1,i}^{-+} 
+ \left( k \alpha_i^-(-)- \epsilon c_i \right) \rho_{i-1,i}^{--}  \right]
\label{Lk2bisres}
\end{eqnarray}


\section{ Large deviations for the pure East model with the soft constraint $\epsilon>0$ }

\label{sec_Epuresoft}

In this section, the goal is to analyze the dynamical large deviations properties of 
 the pure East model with the soft constraint of parameter $\epsilon>0$ in the flip rates of Eq. \ref{weast}
\begin{eqnarray}
 W^{PureEast}_{\epsilon}(\sigma_i^x C,C) = \left[ \delta_{S_{i-1},+ } + \epsilon \delta_{S_{i-1},- } \right] 
 \left[  (1-c) \delta_{S_i,+} + c \delta_{S_i,-}\right] \equiv w^{S_i}(S_{i-1}) 
\label{weastpuresoft}
\end{eqnarray}


\subsection{ Identification of the relevant empirical time-space-averaged observables }

For flip rates of the form $w^{S_i}(S_{i-1})  $, the trajectory probability of Eq. \ref{pwtrajspin}
\begin{eqnarray}
\ln \left( {\cal P}[C(0 \leq t \leq T)]   \right)
=   
  \sum_{i=1}^N \sum_{t \in [0,T] : S_i(t^+) \ne S_i(t) }   \ln \left( w^{S_i(t)} (S_{i-1}(t)) \right)
  -  \sum_{i=1}^N \int_0^T dt \ w^{S_i(t)}(S_{i-1}(t))    
\label{pwtrajspin2pure}
\end{eqnarray}
 can be written as
\begin{eqnarray}
\ln {\cal P}[C(0 \leq t \leq T)]   
=  
T  N \sum_{S_L=\pm,S=\pm}   \left[ q^S(S_L)    \ln \left( w^S(S_L) \right)
  -  \rho^{S_L,S} w^S(S_L)   \right]
\label{plogempipureeast}
\end{eqnarray}
in terms of the empirical time-space-averaged density of flips from $S$ to $(-S)$ when the 
left neighboring spin is $S_L$
\begin{eqnarray}
q^S(S_L) \equiv 
\frac{1}{N T } \sum_{i=1}^N  \sum_{t \in [0,T] : S_i(t^+) \ne S_i(t) } 
\ \delta_{S_{i-1}(t) ,S_L}  \delta_{S_i(t^+) ,-S} \delta_{S_i(t) ,S}
\label{q2pure}
\end{eqnarray}
and in terms of the empirical time-space-averaged density of two consecutive spins $(S_L,S)$
\begin{eqnarray}
\rho^{S_L,S}  \equiv 
\frac{1}{N T } \sum_{i=1}^N    \int_0^T dt 
\ \delta_{S_{i-1}(t) ,S_L}  \delta_{S_i(t) ,S}
\label{rho2pure}
\end{eqnarray}

With respect to the general formalism of Appendix \ref{appendix_relevant},
this means that the relevant empirical observables $E$ for the pure East model
are the empirical flows $q^.(.) $ of Eq. \ref{q2pure}
and the empirical density $\rho^{.,.}$ of Eq. \ref{rho2pure},
while the corresponding action introduced in Eq. \ref{ptrajectempi} reads 
\begin{eqnarray}
\Phi_{[w]} [  q^.(.) ; \rho^{.,.} ]  = 
N \sum_{S_L=\pm,S=\pm}   \left[  \rho^{S_L,S} w^S(S_L) - q^S(S_L)    \ln \left( w^S(S_L) \right)      \right]
\label{actionjumppure}
\end{eqnarray}


\subsection{ Rate function at Level 2.5 for the relevant empirical observables  }

As explained in Appendix \ref{appendix_relevant} around Eq. \ref{modeleeff}, 
it is useful to introduce the modified Markov rates $\hat w^S(S_L) $
that would make typical the relevant empirical observables  $[  q^.(.,.) ; \rho^{.,.} ]$
\begin{eqnarray}
\hat w^{S}(S_L) && = \frac{q^{S}(S_L) } { \rho^{S_L,S}  }
\label{weffeastpure}
\end{eqnarray}
in order to evaluate the action of Eq. \ref{actionjumppure} corresponding to this modified Markov generator $\hat w$ 
\begin{eqnarray}
\Phi_{[\hat w]} [  q^.(.) ; \rho^{.,.} ]  && = 
N \sum_{S_L=\pm,S=\pm}   \left[  \rho^{S_L,S} \hat w^S(S_L) - q^S(S_L)    \ln \left( \hat w^S(S_L) \right)      \right]
\nonumber \\
&& =N \sum_{S_L=\pm,S=\pm}   \left[ q^{S}(S_L)  - q^S(S_L)    \ln \left(  \frac{q^{S}(S_L) } { \rho^{S_L,S}  } \right)      \right]
\label{actionjumptyppure}
\end{eqnarray}

The rate function ${\cal I}_{2.5} [  q^.(.) ; \rho^{.,.}] $ with respect to the space-time volume $(TN)$ 
can be then obtained
from the difference of Eq. \ref{rateempi}
between the actions of Eqs \ref{actionjumppure} and \ref{actionjumptyppure} 
\begin{eqnarray}
{\cal I}_{2.5} [  q^.(.) ; \rho^{.,.}] 
&& =\frac{ \Phi_{[w]} [  q^.(.,.);  \rho^{.,.} ] -  \Phi_{[\hat w]} [  q^.(.,.) ; \rho^{.,.} ] }{N}
\nonumber \\
&& =     \sum_{S_L=\pm,S=\pm}  
\left[ q^S(S_L)    \ln \left( \frac{  q^S(S_L)  }{w^S(S_L)  \rho^{S_L,S} } \right)
- q^S(S_L) +   w^S(S_L) \rho^{S_L,S} \right]
\label{rate2.5pure}
\end{eqnarray}
where one recognizes the relative entropy cost of having empirical flows $q^S(S_L) $
different from the typical flows $w^S(S_L)  \rho^{S_L,S} $
 that would be produced by the empirical densities $ \rho^{S_L,S} $.

The parametrization of the empirical flows 
 \begin{eqnarray}
q^+(S_L)  \equiv \frac{a(S_L)  + j(S_L) }{2} 
\nonumber \\
q^-(S_L)  \equiv \frac{a(S_L)  - j(S_L) }{2} 
\label{ajrecieastpure}
\end{eqnarray}
 in terms of the activities and the currents
\begin{eqnarray}
a(S_L)  \equiv q^+(S_L)  +q^-(S_L) 
\nonumber \\
j(S_L)  \equiv q^+(S_L)  -q^-(S_L) 
\label{ajeastpure}
\end{eqnarray}
yields that the rate function of Eq. \ref{rate2.5pure}
becomes
\begin{eqnarray}
{\cal I}_{2.5} [  q(.) ; j(.) ; \rho^{.,.}] 
&&  =     \sum_{S_L=\pm} 
 \bigg[  \frac{ j(S_L) }{2}     
 \ln \left( \frac{  [ a(S_L)  + j(S_L) ]  w^-(S_L)  \rho^{S_L,-} }
 { [a(S_L)  - j(S_L)  ]  w^+(S_L)  \rho^{S_L,+} } \right)
\nonumber \\
&& + \frac{a(S_L)  }{2}     \ln \left( \frac{  a^2(S_L)  - j^2(S_L)  }{ 4 w^+(S_L)  \rho^{S_L,+} w_i^-(S_L)  \rho_{i-1,i}^{S_L,-}} \right)
- a(S_L) +   w^+(S_L) \rho^{S_L,+}   +   w^-(S_L) \rho^{S_L,-} \bigg]
\label{rate2.5pureaj}
\end{eqnarray}


\subsection{ Constitutive constraints of the empirical observables  }

The constraints concerning the empirical densities of Eq. \ref{c2e} simplify into
\begin{eqnarray}
&& {\cal C}_{2} [ \rho^. ;  \rho^{.,.}]
 = 
\delta \left(   \rho^- -[1- \rho^+ ] \right)
\delta \left ( \rho^{+-} - [\rho^+ - \rho^{++}] \right) 
\delta \left ( \rho^{-+} -[\rho^+ - \rho^{++}]\right)  
\delta \left ( \rho^{--} -[ 1 - 2 \rho^+  + \rho^{++}] \right) 
\label{c2pureeast} 
\end{eqnarray}
and allow to compute all the values of the empirical densities $[ \rho^. ;  \rho^{.,.}] $
 in terms of the two basic variables $(\rho^+,\rho^{++}) $.

As discussed in detail in subsection \ref{subsec_c2.5eastrandom},
despite the explicit rate function at Level 2.5 of Eq. \ref{rate2.5pure}, or equivalently Eq. \ref{rate2.5pureaj},
and despite the explicit constitutive constraints of Eq. \ref{c2pureeast} for the empirical density,
the Level 2.5 cannot be written in closed form 
as a consequence of the closure problem in the stationary constraints for the empirical currents.


\subsection{ Explicit large deviations at Level 2.25 for the local empirical densities and the local empirical activities  }

As discussed in subsection \ref{subsec_c2.25eastrandom},
the detailed-balance property ensures that the vanishing of the empirical currents $j(S_L)$ of Eq. \ref{ajeastpure}
\begin{eqnarray}
j^{opt}(S_L)  =0
\label{ajeastoptzeropure}
\end{eqnarray}
 leads to the closed Level 2.25 for the empirical density and the empirical activity
\begin{eqnarray}
P^{[2.25]}_T [  a(.) ; \rho^. ;  \rho^{.,.}] \opsimeq_{T \to +\infty} 
 {\cal C}_{2} [  \rho^. ;   \rho^{.,.}] 
e^{ \displaystyle - T N {\cal I}_{2.25} [  a(.) ; \rho^{.,.}]   }
\label{level2.25pureEastsimplia}
\end{eqnarray}
where the constitutive constraints ${\cal C}_{2} [  \rho^. ;   \rho^{.,.}] $ 
have been written in Eq. \ref{c2pureeast},
while the rate function at Level 2.25 is obtained from the rate function at Level 2.5 
of Eq. \ref{rate2.5pureaj} when all the empirical currents vanish $j^{opt}(S_L)=0$
\begin{eqnarray}
 {\cal I}_{2.25} [  a(.) ; \rho^{.,.}] && =  {\cal I}_{2.5} [  a(.) ; j(.)=0 ; \rho^{.,.}]
 \nonumber \\
&&  =  
   \sum_{S_L=\pm} 
 \bigg[  \frac{a(S_L)  }{2}     \ln \left( \frac{  a^2(S_L)    }{ 4 w^+(S_L)  \rho^{S_L,+} w^-(S_L)  \rho^{S_L,-}} \right)
- a(S_L) +   w^+(S_L) \rho^{S_L,+}   +   w^-(S_L) \rho^{S_L,-} \bigg]
 \label{rate2.25pureEastsimpli}
\end{eqnarray}
An alternative derivation of this rate function $  {\cal I}_{2.25} [  a(.) ; \rho^{.,.}]  $
 at Level 2.25 for empirical time-space-averaged activities $a (S_L)  $ 
and the local time-space-averaged densities $\rho^{S_L,S} $
is given in Appendix \ref{app_GlobalLocalPure} via the explicit contraction from the Level 2.25 
in the space of the $2^N$ configurations.

For the pure East model with the soft constraint of parameter $\epsilon$ in the flip rates of Eq. \ref{weastpuresoft},
the rate function at Level 2.25 reads more explicitly
 \begin{eqnarray}
 {\cal I}_{2.25}^{PureEastSoft} [  a(.) ;  ; \rho^{.,.}]
&&  =    
\frac{a(+)  }{2}     \ln \left( \frac{  a^2(+)  }{4 c (1-c)  \rho^{++}   \rho^{+-}} \right)
- a(+) +  (1-c)  \rho^{++} + c  \rho^{+-}  
 \nonumber \\  &&
 +   
\frac{a(-)  }{2}   \ln \left( \frac{  a^2(-)  }{ 4 \epsilon^2 c (1-c)  \rho^{-+}   \rho^{--}} \right)
- a(-) +  \epsilon (1-c)  \rho^{-+} + \epsilon c    \rho^{--}   
 \label{rate2.25pEs}
\end{eqnarray}


\subsection{ Explicit contraction towards the Level 2 for the empirical density alone  }

The optimization of the rate function of Eq. \ref{rate2.25pureEastsimpli} over the activities $a(S_L) $
\begin{eqnarray}
0 && = \frac{ \partial {\cal I}_{2.25} [  a(.) ; \rho^{.,.}] } { \partial a(S_L)} 
=  \frac{1}{2}  \ln \left( \frac{  a^2(S_L)  }{4w^+(S_L)  \rho^{S_L,+} w^-(S_L)  \rho^{S_L,-}} \right)
\label{optimizeqpure}
\end{eqnarray}
leads to the optimal values 
\begin{eqnarray}
  a^{opt}(S_L) = 2 \sqrt{ w^+(S_L)  \rho^{S_L,+} w^-(S_L)  \rho^{S_L,-}}  
\label{qioptpure}
\end{eqnarray}
that can be plugged into Eq. \ref{rate2.25pureEastsimpli}
to obtain the rate function at Level 2
\begin{eqnarray}
 {\cal I}_{2} [ \rho^{.,.}] && =   {\cal I}_{2.25} [  a^{opt}(.) ; \rho^{.,.}]
= \sum_{S_L=\pm}  \left[  \sqrt{ w^+(S_L)  \rho^{S_L,+} } - \sqrt{ w^-(S_L)  \rho^{S_L,-} } \right]^2
\label{rate2pure}
\end{eqnarray}
that governs the probability of the empirical density alone
\begin{eqnarray}
P^{[2]}_T [   \rho^. ;  \rho^{.,.}] \opsimeq_{T \to +\infty} 
 {\cal C}_{2} [  \rho^. ;   \rho^{.,.}] 
e^{ \displaystyle - T N {\cal I}_{2} [  \rho^. ; \rho^{.,.}]   }
\label{level2pureEastsimplip}
\end{eqnarray}

For the pure East model with the soft constraint of parameter $\epsilon$ in the flip rates of Eq. \ref{weastpuresoft},
the rate function at Level 2 reads more explicitly
\begin{eqnarray}
 {\cal I}^{PureEastSoft}_{2} [ \rho^{.,.} ] 
 =      \left[  \sqrt{ (1-c)  \rho^{++} } - \sqrt{ c  \rho^{+-} } \right]^2 
   +      \epsilon
   \left[   \sqrt{ (1-c)\rho^{-+}  }   - \sqrt{ c  \rho^{--}   }     \right]^2
\label{rate2pEs}
\end{eqnarray}


\subsection{ Typical fluctuations of order $\frac{1}{\sqrt{TN} }$ for the empirical densities and activities around  equilibrium values }

The equilibrium 2-spin density of Eq. \ref{peqKCM}
\begin{eqnarray}
\Pi^{S_{i-1},S_i} = \Pi^{S_{i-1} }\Pi^{S_i} =
\left[ c \delta_{S_{i-1},+} + (1-c) \delta_{S_{i-1},-} \right]
\left[ c \delta_{S_i,+} + (1-c) \delta_{S_i,-} \right]
\label{rhoeqepure}
\end{eqnarray}
and the corresponding equilibrium activities 
\begin{eqnarray}
 A(+)  && = 2w^{+}(+) \Pi^{++} = 2w^{-}(+) \Pi^{+-} = 2c^2 (1-c)
 \nonumber \\
 A(-)  && = 2w^{+}(-) \Pi^{-+} = 2w^{-}(-) \Pi^{--} = 2\epsilon c (1-c)^2
\label{Aeqepure}
\end{eqnarray}
are the only values of the empirical densities and activities that satisfy the constraints and that make  
the rate function ${\cal I}_{2.25}^{PureEastSoft} [  a(.) ; \rho^{.,.}] $ of Eq. \ref{rate2.25pEs} vanish.

If one is interested only in the small fluctuations of order $\frac{1}{\sqrt{TN} }$ around these equilibrium values
\begin{eqnarray}
 \rho^{S_L,S} && = \Pi^{S_L,S} + \frac{{\hat \rho}^{S_L,S}}{\sqrt{TN} }
\nonumber \\
 a(S_L) && =  A(S_L)+ \frac{{\hat a}(S_L) }{\sqrt{TN} }
\label{hatrhoqpure}
\end{eqnarray}
one just needs to expand
the rate function ${\cal I}_{2.25}^{PureEastSoft} [  a(.) ; \rho^{.,.}] $ of Eq. \ref{rate2.25pEs}
at second order in the perturbations to obtain
the rescaled Gaussian rate function for the rescaled densities ${\hat \rho}^{.,.,.} $ and activities ${\hat a}(.,.)  $ 
\begin{eqnarray}
&& {\hat I}_{2.25}^{small} [  {\hat a}(.) ; {\hat \rho}^{.,.}]  \equiv  \lim_{TN \to + \infty}
 \left( TN {\cal I}_{2.25} [  a(.)= A(.,.) +\frac{{\hat a}(.) }{\sqrt{TN}} ; 
\rho^{.,.}= \Pi^{.,.}+ \frac{{\hat \rho}^{.,.}}{\sqrt{TN}} ] \right)
\nonumber \\
&& =    \sum_{S_L=\pm}  
    \frac{  \left[ \frac{{\hat a}(S_L)}{2} - w^{+}(S_L){\hat \rho}^{S_L,+}\right]^2
    + \left[ \frac{{\hat a}(S_L) }{2} - w^{-}(S_L){\hat \rho}^{S_L,-}\right]^2  }
{  A(S_L)  } 
 \nonumber \\
 && 
 =   \sum_{S_L=\pm}  
    \frac{   \left[ {\hat a}(S_L) -     ( w^{+}(S_L){\hat \rho}^{S_L,+}    +w^{-}(S_L){\hat \rho}^{S_L,-} )\right]^2
    +  \left[  w^{+}(S_L){\hat \rho}^{S_L,+}  - w^{-}(S_L){\hat \rho}^{S_L,-}  \right]^2  }
{   2 A(S_L)  } 
 \nonumber \\
 && 
 =   
    \frac{   \left[ {\hat a}(+) -     ( (1-c){\hat \rho}^{++}    +c{\hat \rho}^{+-} )\right]^2
    +  \left[  (1-c){\hat \rho}^{++}  - c {\hat \rho}^{+-}  \right]^2  }
{    4 c^2 (1-c)  } 
 \nonumber \\
 &&  +       \frac{   \left[ {\hat a}(-) -    \epsilon ( (1-c){\hat \rho}^{S_L,+}    +c{\hat \rho}^{S_L,-} )\right]^2
   +  \epsilon^2 \left[  (1-c) {\hat \rho}^{-+}  - c {\hat \rho}^{--}  \right]^2  }
{  4  \epsilon c (1-c)^2  } 
 \label{rate2.25puregauss}
\end{eqnarray}
that will govern the probability of the rescaled fluctuations 
\begin{eqnarray}
  {\hat P}^{[2.25]}_T [  {\hat a}(.) ;{\hat \rho}^{.} ; {\hat \rho}^{.,.}] 
  \opsimeq_{TN \to +\infty} 
 {\hat {\cal C}}_{2} [ {\hat \rho}^{.} ; {\hat \rho}^{.,.}]
e^{ \displaystyle -  {\hat I}_{2.25}^{small} [  {\hat a}(.) ; {\hat \rho}^{.,.}] 
 }
\label{level2.25simplihatpure}
\end{eqnarray}
together with the constraints inherited from Eq. \ref{c2pureeast} 
\begin{eqnarray}
{\hat {\cal C}}_{2} [  {\hat \rho}^{.} ; {\hat \rho}^{.,.}] 
 =  \delta \left(   {\hat \rho}^- -[- {\hat \rho}^+ ] \right)
\delta \left ( {\hat \rho}^{+-} - [{\hat \rho}^+ - {\hat \rho}^{++}] \right) 
\delta \left ( {\hat \rho}^{-+} -[{\hat \rho}^+ - {\hat \rho}^{++}]\right)  
\delta \left ( {\hat \rho}^{--} -[  - 2 {\hat \rho}^+  + {\hat \rho}^{++}] \right)  
\label{c2purerho2hat}  
\end{eqnarray}

If one is interested into the small fluctuations ${\hat \rho}^{.,.}  $ of the empirical density alone,
the rescaled rate function reduces to
\begin{eqnarray}
 {\hat I}_2^{small} [ {\hat \rho}^{.,.}]    && \equiv \lim_{TN \to + \infty}
 \left(  TN {\cal I}_{2} [ 
\rho^{.,.}= \Pi^{.,.}+ \frac{{\hat \rho}^{.,.}}{\sqrt{N T}} ] \right)
     = 
      \frac{   \left[  (1-c){\hat \rho}^{++}  - c {\hat \rho}^{+-} \right]^2  }
{    4 c^2 (1-c)  } 
 +   \epsilon   \frac{     \left[  (1-c) {\hat \rho}^{-+}  - c {\hat \rho}^{--}  \right]^2  }
{     4 c (1-c)^2  } 
 \label{rate2gausspure}
\end{eqnarray}
Since the rescaled empirical densities $ [{\hat \rho}_{.}^{.}, {\hat \rho}_{.,.}^{.,.}]   $ belong to $]-\infty,+\infty[$,
one can use the constraints of Eq. \ref{c2purerho2hat} to keep only the two rescaled empirical densities 
$ [{\hat \rho}^+, {\hat \rho}^{++}]   $ 
and one obtains the rescaled rate function
\begin{eqnarray}
 {\hat I}_2^{small} [ {\hat \rho}_{.}^+, {\hat \rho}_{.,.}^{++} ]  
       =
\left(    \frac{    \left[ {\hat \rho}^{++} 
 - c {\hat \rho}^+  \right]^2  }
{   4 c^2 (1-c)  } 
+\epsilon   \frac{  
     \left[  {\hat \rho}^{++}  - c {\hat \rho}^+    - {\hat \rho}^+   
      \right]^2  }
{    4 c (1-c)^2  } 
\right)
 \label{rate2egausspositivepure}
\end{eqnarray}
that will govern their joint Gaussian probability without any remaining constraint
\begin{eqnarray}
  {\hat P}^{[2]}_T [ {\hat \rho}^+, {\hat \rho}^{++}   ] 
  \opsimeq_{T \to +\infty} 
e^{ \displaystyle -  {\hat I}_{2}^{small} [ {\hat \rho}^+, {\hat \rho}^{++} ] 
 }
\label{level2egausspositivepure}
\end{eqnarray}
The Gaussian integration over ${\hat \rho}^{++} $ yields 
that the probability of the 1-spin rescaled density $ {\hat \rho}^+ $ alone reduces to the Gaussian
\begin{eqnarray}
  {\hat P}^{[2]}_T [ {\hat \rho}^+  ] 
  \opsimeq_{T \to +\infty} 
e^{ \displaystyle -    \frac{ [{\hat \rho}^+ ]^2}{ 2 v}   }
\label{level2egausspositive1spinpure}
\end{eqnarray}
of variance
\begin{eqnarray}
 v = 2 c (1-c) \left[ c + \frac{ (1-c ) }{\epsilon} \right]
 \label{variancepure}
\end{eqnarray}
that is finite for the soft version $\epsilon>0$ of the pure East model, but that diverges for 
the true pure East model corresponding to the hard-constraint $\epsilon=0$ that will be analyzed in section \ref{sec_Epure}.



\subsection{ Time-additive observables that can be reconstructed from the empirical densities and activities}

The empirical activity of Eq. \ref{ajeastpure} with Eq. \ref{q2pure}
\begin{eqnarray}
a(S_L)  =   \frac{1}{ T N } \sum_{i=1}^N  \sum_{t \in [0,T] : S_i(t^+) \ne S_i(t) } 
\ \delta_{S_{i-1}(t) ,S_L} 
\label{ajeastexplipure}
\end{eqnarray}
represents the density of the flips in the whole system 
when the left neighboring spin takes the value $S_L$.

These empirical activities and the empirical densities of Eq. \ref{rho2pure}
allow to reconstruct any time-additive space-averaged observable ${\cal O}_T  $
that can be parametrized by some functions $\alpha^S(S_L) $ and $\beta(S_L) $
\begin{eqnarray}
{\cal O}_T 
&& =    \sum_{S_L=\pm}  \left[\sum_{S=\pm}\alpha^S(S_L) \rho^{S_L,S} 
+ \beta(S_L)  a(S_L)\right]
\nonumber \\
&& = \frac{1}{NT}
 \sum_{i=1}^N  \left[  \int_0^T dt 
   \alpha^{S_i(t)}(S_{i-1}(t) )   
+     \sum_{t \in [0,T] : S_i(t^+) \ne S_i(t) }  \beta(S_{i-1}(t)) \right]
\label{additivepureEast}
\end{eqnarray}

 \subsubsection{ Averaged value and rescaled variance }

The first cumulant ${\cal G}_1$ corresponding to the averaged value $\langle {\cal O}_T \rangle $
coincides with the equilibrium value $ {\cal O}_{eq} $ computed
 from the equilibrium density $\Pi^{.,.} $ of Eq. \ref{rhoeqepure}
and from the corresponding equilibrium activities $A(.) $ of Eq. \ref{Aeqepure}
\begin{eqnarray}
{\cal G}_1 \equiv \langle {\cal O}_T \rangle = {\cal O}_{eq} 
 \equiv    \sum_{S_L=\pm}  \left[ \sum_{S=\pm} \alpha^{S}(S_L) \Pi^{S_L,S} 
+ \beta(S_L)  A(S_L)\right]
\label{g1pure}
\end{eqnarray}
The small fluctuations of order $\frac{1}{\sqrt{TN} } $ around this equilibrium value 
can be rewritten in terms of the rescaled fluctuations 
${\hat \rho}^{...} $ and ${\hat a}(.,.) $ of Eq. \ref{hatrhoqpure}
\begin{eqnarray}
{\cal O}_T - \langle {\cal O}_T \rangle
&& = \frac{1}{\sqrt{TN } }   \sum_{S_L=\pm}  \left[\sum_{S=\pm} \alpha^{S}(S_L) {\hat \rho}^{S_L,S}
+ \beta(S_L)  {\hat a}(S_L)  \right]
\label{additivesmallfluctpure}
\end{eqnarray}
so that the rescaled variance
\begin{eqnarray}
{\cal G}_2  \equiv  T N  \langle  \left( {\cal O}_T  - \langle  {\cal O}_T  \rangle \right)^2  \rangle
= \bigg\langle \left( \sum_{S_L=\pm}  \left[\sum_{S=\pm}\alpha^{S}(S_L) {\hat \rho}^{S_L,S}
+ \beta(S_L)  {\hat a}(S_L)  \right] \right)^2  \bigg\rangle
\label{g2pure}
\end{eqnarray}
can be evaluated from the probability $ {\hat P}^{[2.25]}_T [  {\hat a}(.)  ;  {\hat \rho}^{.};  {\hat \rho}^{.,.}] $ of the rescaled fluctuations of Eq. \ref{level2.25simplihatpure}.


\subsubsection{ Large deviations governed by the ate function ${\cal I}({\cal O} )$ and the scaled cumulant generating function ${\cal G}(k)$  }

The large deviations properties for large $(TN)$  
 \begin{eqnarray}
 P_T( {\cal O} ) \opsimeq_{TN \to +\infty} e^{- T N {\cal I}({\cal O})}
\label{level1defpure}
\end{eqnarray} 
are governed by the rate function ${\cal I}({\cal O} )$
that vanishes only for the equilibrium value ${\cal O}_{eq}$ of Eq. \ref{g1pure}
 \begin{eqnarray}
{\cal I}({\cal O}_{eq} ) =0
\label{iaeqvanishpure}
\end{eqnarray}
The generating function ${\cal G}(k)$ of the scaled cumulants ${\cal G}_n$ 
\begin{eqnarray}
{\cal G}(k) && = \sum_{n=1}^{+\infty} {\cal G}_n \frac{k^n}{n!} =   {\cal G}_1 k + {\cal G}_2 \frac{k^2}{2}  + O(k^3)
\label{gkperpure}
\end{eqnarray}
corresponds to the Legendre transform of the rate function ${\cal I}({\cal O} )$
via the saddle-point evaluation of the generating function
\begin{eqnarray}
\langle e^{k TN {\cal O}_T } \rangle \equiv \int d{\cal O}  e^{k TN {\cal O}} P_T( {\cal O} ) \opsimeq_{TN \to +\infty} 
\int d{\cal O} e^{ TN \left[ k {\cal O} - {\cal I}({\cal O} ) \right] }\opsimeq_{T \to +\infty} e^{ T N {\cal G}(k) }
\label{level1genpure}
\end{eqnarray} 
The adaptation of the computation of Eq. \ref{gene} to the present case 
yields that the generating function 
\begin{eqnarray}
 \langle e^{ k TN {\cal O}_T } \rangle 
 &&
 \opsimeq_{TN \to +\infty} 
  \int d\rho^{.,.}
 {\cal C}_2 [ \rho^. ;    \rho^{.,.}] 
 e^{ T N {\cal L}_{2}^{[k]}  [  \rho^{.,.}]  }
\label{gene2pureEast}
\end{eqnarray}
involves the function
\begin{eqnarray}
 {\cal L}_{2}^{[k]} [   \rho^{.,.}]  
 && = 
  \sum_{S_L=\pm}  
\left[  2  e^{k \beta(S_L)  } 
\sqrt{ w^+(S_L)   \rho^{S_L,+} w^-(S_L)  \rho^{S_L,-}}  
 \right]
 \nonumber \\&&
 + 
  \sum_{S_L=\pm}  
\left[   \left( k \alpha^+(S_L)- w^+(S_L) \right) \rho^{S_L,+} 
+ \left( k \alpha^-(S_L)- w^-(S_L) \right) \rho^{S_L,-}  \right]
\label{Lk2pureEast}
\end{eqnarray}

For the pure East model with the soft constraint of parameter $\epsilon$
in the flip rates of Eq. \ref{weastpuresoft},
the function of Eq. \ref{Lk2pureEast} reads
 \begin{eqnarray} 
&& {\cal L}_{2}^{PureEastSoft[k]} [   \rho^{.,.} ] 
   = 
 2  e^{k \beta(+)  }  \sqrt{ c (1-c)   \rho^{++}   \rho^{+-}}  
+   \left( k \alpha^+(+)- (1-c) \right) \rho^{++} 
+ \left( k \alpha^-(+)- c \right) \rho^{+-}  
\nonumber \\
&&+   2  e^{k \beta(-)  } \epsilon \sqrt{ c (1-c)   \rho^{-+}  \rho^{--}  }  
+   \left( k \alpha^+(-)- \epsilon (1-c) \right) \rho^{-+} 
+ \left( k \alpha^-(-)- \epsilon c \right)   \rho^{--}  
\ \ \ \ \ \ \ \ 
\label{Lk2bispes}
\end{eqnarray}


\section{ Anomalous large deviations properties of the pure East model }

\label{sec_Epure}

In this section, we  consider the pure East model with the flip rates of Eq. \ref{weasthard}
\begin{eqnarray}
 W^{PureEast}(\sigma_i^x C,C) = \delta_{S_{i-1},+ } 
 \left[  (1-c) \delta_{S_i,+} + c \delta_{S_i,-}\right] \equiv w^{S_i}(S_{i-1}) 
\label{weastpure}
\end{eqnarray}
in order to analyze the anomalous dynamical large deviations properties that emerge for 
this hard-constrained model $\epsilon=0$ with respect to the soft-constrained model $\epsilon>0$ discussed
 in section \ref{sec_Epuresoft}.


\subsection{ Anomalous large deviations properties at Level 2.25 for the empirical densities and activities  }

\label{subsec_vanish2.5pure}

In the limit $\epsilon \to 0$, the large deviations at Level 2.25 of Eq. \ref{level2.25pureEastsimplia}
become
\begin{eqnarray}
P^{[2.25]}_T [  a(+) ; \rho^. ;  \rho^{.,.}] \opsimeq_{T \to +\infty} 
 {\cal C}_{2} [  \rho^. ;   \rho^{.,.}] \theta \left( \rho^+>0 \right)
e^{ \displaystyle - T N {\cal I}_{2.25}^{PureEast} [  a(+) ; \rho^{.,.}]   }
\label{level2.25pureEastsimplia0}
\end{eqnarray}
with the rate function of Eq. \ref{rate2.25pEs} for $\epsilon=0$
 \begin{eqnarray}
 {\cal I}_{2.25}^{PureEast} [  a(+)  ; \rho^{..}]
  =    
 \frac{a(+)}{2}    \ln \left( \frac{  a^2(+)  }{4c (1-c)  \rho^{++}   \rho^{+-}} \right)
- a(+) +  (1-c)  \rho^{++} + c  \rho^{+-}  
 \label{rate2.25pureEast}
\end{eqnarray}
Note that besides the constitutive constraints ${\cal C}_{2} [  \rho^. ;   \rho^{.,.}]  $ of Eq. \ref{c2pureeast},
we have added the Heaviside constraint $\theta \left( \rho^+>0 \right) $ in order to exclude explicitly 
the configuration $\{S_i=-1 ; i=1,2,..,N\}$ that is disconnected from the set of the other $(2^N-1)$ configurations 
for the hard-constrained model $\epsilon=0$.

We are interested into the region of empirical activities $a_*(+) $ and empirical densities $  [\rho_*^. ;  \rho_*^{.,.}]$
that make vanish the rate function $  {\cal I}_{2.25}^{PureEast} [  a_*(+) ; \rho_*^{.,.}] =0 $ of Eq. \ref{rate2.25pureEast}  
\begin{eqnarray}
  \rho_*^{++} && = \frac{a_*(+)}{2(1-c)}
  \nonumber \\
  \rho_*^{+-}  && = \frac{a_*(+)}{2 c}
 \label{rate2.5pEzero} 
\end{eqnarray}
and that satisfy the constraints ${\cal C}_{2} [  \rho^. ;   \rho^{.,.}]  $ of Eq. \ref{c2pureeast} 
and the supplementary constraint $\rho^+>0$ of Eq. \ref{level2.25pureEastsimplia0}
\begin{eqnarray}
0<\rho_*^+  && = \rho_*^{++}  + \rho_*^{+-} = \frac{a_*(+)}{ 2c(1-c)}
\nonumber \\
\rho_*^{-+} && =\rho_*^+ - \rho_*^{++} =  \frac{a_*(+)}{2c}
\nonumber \\
\rho_*^{--} && = 1- \rho_*^{++} - \rho_*^{+-} - \rho_*^{-+}= 1 -  a_*(+) \frac{2-c}{2 c(1-c)} 
\label{elimc2zeropure}
\end{eqnarray}
Since all these empirical densities  
have to be in $[0,1]$,
one obtains that the remaining constraint for the positive empirical activity $a_*(+)$ reduces to
\begin{eqnarray}
0 < a_*(+) \leq  \frac{2 c (1-c) }{2-c} \equiv a_*^{max}(+)
\label{qinterval}
\end{eqnarray}

In summary, for any empirical activity $a_*(+) $ in the interval $]0,a_*^{max}(+) =  \frac{2c (1-c) }{2-c}]$,
the empirical densities $  [\rho_*^. ;  \rho_*^{.,.}]$ computed from Eqs \ref{rate2.5pEzero}  and \ref{elimc2zeropure} satisfy the constraints of Eq. \ref{c2pureeast}
and make the rate function of Eq. \ref{rate2.25pureEast}
  vanish $  {\cal I}_{2.25}^{PureEast} [  a_*(+) ; \rho_*^{.,.}] =0 $.


\subsection{ Anomalous large deviations properties at Level 2 for the empirical densities alone }

The large deviations at Level 2 of Eq. \ref{level2pureEastsimplip}
\begin{eqnarray}
P^{[2]}_T [   \rho^. ;  \rho^{.,.}] \opsimeq_{T \to +\infty} 
 {\cal C}_{2} [  \rho^. ;   \rho^{.,.}] \theta \left( \rho^+>0 \right)
e^{ \displaystyle - T N {\cal I}_{2}^{PureEast} [  \rho^. ; \rho^{.,.}]   }
\label{level2pureEastsimplipappli}
\end{eqnarray}
involve the rate function of Eq. \ref{rate2pEs} for $\epsilon=0$
\begin{eqnarray}
 {\cal I}^{PureEast}_{2} [ \rho^{..} ] 
 =      \left[  \sqrt{ (1-c)  \rho^{++} } - \sqrt{ c  \rho^{+-} } \right]^2 
\label{rate2pE}
\end{eqnarray}

To obtain the region of empirical densities $  [\rho_*^. ;  \rho_*^{.,.}]$
that make the rate function ${\cal I}^{PureEast}_{2} [ \rho^{..} ]  $ 
of Eq. \ref{rate2pE}  vanish and that satisfy the constraints $C_{2} [ \rho_.^. ;  \rho_{.,.}^{.,.}] $ of Eq. \ref{c2e},
we can use the previous analysis of subsection \ref{subsec_vanish2.5pure} concerning the Level 2.25 :
the first equation of Eq. \ref{elimc2zeropure} allows to replace the empirical activity in terms of the empirical 1-spin density
\begin{eqnarray}
a_*(+) =   2 c (1-c)  \rho_*^+
\label{elimqversrhopure}
\end{eqnarray}
into all the other equations of Eqs \ref{rate2.5pEzero}  and \ref{elimc2zeropure}
to obtain the empirical 2-spin density
\begin{eqnarray}
  \rho_*^{++} && =\rho_*^+   c 
  \nonumber \\
  \rho_*^{+-}  && = \rho_*^+ (1-  c )
\nonumber \\
\rho_*^{-+} && = \rho_*^+ (1-  c )
\nonumber \\
\rho_*^{--} && = 1 - \rho_*^+  (2- c ) 
\label{elimc2zeroelimqpure}
\end{eqnarray}
while the inequality constraint of Eq. \ref{qinterval}
translates into 
\begin{eqnarray}
0 < \rho_*^+    \leq  \frac{ 1 }{2-c} 
\label{rhointerval}
\end{eqnarray}

In summary, for any empirical density $\rho_*^+ $ in the interval $]0,  \frac{1 }{2-c}]$,
the empirical densities $    \rho_*^{.,.}$ computed from Eq \ref{elimc2zeroelimqpure} satisfy the constraints of Eq. \ref{c2pureeast}
and make the rate function of Eq. \ref{rate2pE}
  vanish $  {\cal I}_{2}^{PureEast} [   \rho_*^{.,.}] =0 $.


\subsection{ Anomalous large deviations properties for the activity $a(+)$ alone   }

The probability of the empirical activity $a(+)$
can be obtained via the integration over the empirical densities
of the Level 2.25 of Eq.  \ref{level2.25pureEastsimplia0}
\begin{eqnarray}
P_T [  a(+) ]  
 \opsimeq_{T \to +\infty} 
\int d \rho^. \int d  \rho^{.,.}
 {\cal C}_{2} [  \rho^. ;   \rho^{.,.}]  \theta \left( \rho^+>0 \right)
e^{ \displaystyle - T N {\cal I}_{2.25}^{PureEast} [  a(+) ; \rho^{.,.}]   }
 \opsimeq_{T \to +\infty} e^{ \displaystyle - T N  {\cal I}^{PureEast} [ a(+) ]  }
\label{levelpureEastqonly}
\end{eqnarray}

From the analysis of subsection \ref{subsec_vanish2.5} concerning the Level 2.25,
one obtains that the rate function ${\cal I}^{PureEast} [ a(+) ]   $ vanishes on the interval of Eq. \ref{qinterval}
\begin{eqnarray}
{\cal I}^{PureEast} [ a_*(+) ]  =0 \ \ \ {\rm for } \ \ \ \ 0 < a_*(+) \leq  \frac{2c (1-c) }{2-c} \equiv a_*^{max}(+)
\label{qintervalvanish}
\end{eqnarray}

For bigger activities $a(+) > a_*^{max}(+) =\frac{2c (1-c) }{2-c} $, the rate function will be positive $ {\cal I}^{PureEast} [ a(+) ] >0 $.
Its value can be then computed via the optimization of  
\begin{eqnarray}
{\cal I}_{2.25}^{PureEast} [ a(+)  ; \rho^{++}; \rho^{+-}=\frac{1- \rho^{++}}{2}] 
=\frac{a(+)}{2}    \ln \left( \frac{  a^2(+)  }{2 c (1-c)  \rho^{++}  (1- \rho^{++}) } \right)
- 2a(+) +  \frac{c}{2} + \left( 1 -  \frac{3 }{2} c \right)  \rho^{++} 
 \label{rate2.5pureEastexplieps0deribordvalue}
\end{eqnarray}
over the remaining density $\rho^{++} $
\begin{eqnarray}
0 = \frac{ \partial  {\cal I}_{2.25}^{PureEast} [ a(+)  ; \rho^{++}; \rho^{+-}=
\frac{1- \rho^{++}}{2}] } {\partial \rho^{++} }
  = \frac{a(+)}{2}  \left( \frac{2 \rho^{++} -1 }{ \rho^{++} (1-\rho^{++}) } \right) + 1 -  \frac{3 }{2} c
 \label{rate2.5pureEastexplieps0deribord}
\end{eqnarray}
leading to the second order equation for $\rho^{++}$
\begin{eqnarray}
0   = (\rho^{++})^2 - \rho^{++} \left[ 1 +    \frac{  a(+)  }{ 1 - \frac{3}{2} c} \right]  +     \frac{  a(+) }{ 2-3 c }  
 \label{2dorder}
\end{eqnarray}
Let us now focus on the region $0<c<\frac{2}{3}$
where the solution of Eq. \ref{2dorder}
in the interval $[0,1]$ yields the optimal value
\begin{eqnarray}
\rho^{++}_{opt} = \frac{ \left[ 1 +    \frac{  a(+)  }{ 1 - \frac{3}{2} c} \right] 
 -  \sqrt{1 +  \left(  \frac{  a(+)  }{ 1 - \frac{3}{2} c}  \right)^2 }  }{2}  
 \label{2dordersol}
\end{eqnarray}
that can be plugged into Eq. \ref{rate2.5pureEastexplieps0deribordvalue}
in order to obtain the rate function for the activity $a(+)$ alone in the region $a(+)>a_*^{max}(+) =\frac{2c (1-c) }{2-c} $
\begin{eqnarray}
&& {\cal I}^{PureEast} [ a(+)]  =
{\cal I}_{2.25}^{PureEast} [ a(+)  ; \rho^{++}_{opt}; \rho^{+-}=\frac{1- \rho^{++}_{opt}}{2}] 
\nonumber \\
&& =\frac{a(+)}{2}     \ln \left( \frac{ a(+)  }{   \frac{ c (1-c)   }{ 1 - \frac{3}{2} c} 
 \left( \sqrt{1 +  \left(  \frac{  a(+)  }{ 1 - \frac{3}{2} c}  \right)^2 }
 - \frac{  a(+)  }{ 1 - \frac{3}{2} c} \right) 
 } \right)
- a(+) +  \frac{c}{2} + \left( 1 -  \frac{3 }{2} c \right)  
\frac{ \left[ 1 +    \frac{  a(+)  }{ 1 - \frac{3}{2} c} \right] 
 -  \sqrt{1 +  \left(  \frac{  a(+)  }{ 1 - \frac{3}{2} c}  \right)^2 }  }{2}   
 \nonumber \\
&& =\frac{a(+)}{2}     \ln \left( \frac{a(+)}{c(1-c)} \left[a(+) +  \sqrt{     a^2(+) +  \left( 1 -  \frac{3 }{2} c \right)^2  }\right] \right) +    \frac{2- c }{4} 
- \frac{a(+)}{2}   -  \frac{1}{2}\sqrt{     a^2(+) +  \left( 1 -  \frac{3 }{2} c \right)^2  }  
 \label{rateqalone}
\end{eqnarray}


\subsection{ Anomalous large deviations properties of time-additive space-averaged observables  }

 In the pure East model $\epsilon=0$, using the constraints of Eq. \ref{c2pureeast}
 to eliminate $\rho^{-+}=\rho^{+-}$ and $\rho^{--}=1-\rho^{++}-2 \rho^{+-}$,
 one obtains that 
 the 
 time-additive space-averaged observable of Eq. \ref{additivepureEast}
 can be rewritten in terms of three parameters $(\beta, \alpha^{++},  \alpha^{+-}) $ only
\begin{eqnarray}
{\cal O}  =  \beta  a(+) +  \alpha^{++} \rho^{++} +  \alpha^{+-} \rho^{+-}
\label{additivepureEastappli}
\end{eqnarray}


\subsubsection{ Anomalous vanishing of the rate function ${\cal I}({\cal O} )$ on a finite interval  }

The rate function ${\cal I}({\cal O} )$ will vanish ${\cal I}({\cal O}_* )=0$
for any value $O_* $ that can be reconstructed from the region $[  a_*(+) ;  \rho_*^{.} ; \rho_*^{.,.}]  $ discussed in subsection \ref{subsec_vanish2.5pure}
\begin{eqnarray}
{\cal O}_*  =  \beta  a_*(+) +  \alpha^{++} \rho_*^{++} +  \alpha^{+-} \rho_*^{+-}
=   \left[  \beta + \frac{\alpha^{++}}{2(1-c)} +  \frac{\alpha^{+-} }{2c}  \right] a_*(+)
 \ \ \ \  { \rm for } \ \ \  0 <  a_*(+) \leq a_c(+)  = \frac{2c (1-c) }{2-c} \ \ 
\label{additivepureEastapplistar}
\end{eqnarray}

To be more concrete, let us now focus on the cases where the three parameters of Eq. \ref{additivepureEastappli} are positive  
\begin{eqnarray}
\alpha^{++} \geq 0 \ \ \ ; \ \  \alpha^{+-} \geq 0 \ \ \  ; \ \ \beta  \geq 0
 \label{positiveab}
\end{eqnarray}
Then Eq. \ref{additivepureEastapplistar} means that the rate function 
vanishes on the interval $]0,  {\cal O}_*^{max}]$
\begin{eqnarray}
 {\cal I} ( {\cal O}_*  )    
  =     0 \ \ \ \ \ \ { \rm for } \ \ \ \ \ \ \ 0 <  {\cal O}_* \leq  {\cal O}_*^{max}
  =  \left[  \beta + \frac{\alpha^{++}}{2(1-c)} +  \frac{\alpha^{+-} }{2c}  \right] \frac{2c (1-c) }{2-c}
 \label{rate1vanishintervalpurea}
\end{eqnarray}


\subsubsection{ Corresponding singularity in the scaled cumulant generating function ${\cal G}(k ) $ 
at the origin $k \to 0^{\pm}$  }

For a positive observable ${\cal O} \geq 0$ parametrized by the the three positive parameters of Eq. \ref{positiveab},
the generating function will be dominated for small $k$ by the interval 
${\cal O}_* \in [0,{\cal O}_*^{max}]$
where the rate function ${\cal I} ( {\cal O}_*  )   =0$ vanishes (Eq. \ref{rate1vanishintervalpurea}) 
\begin{eqnarray}
 \langle e^{ k TN {\cal O}_T } \rangle 
 \opsimeq_{TN \to +\infty}  e^{ T N {\cal G}(k)  }
 \opsimeq_{TN \to +\infty} \int_0^{+\infty} d {\cal O}  e^{ T N \left[ k {\cal O} -  {\cal I} ( {\cal O} )    \right]  }
 \opsimeq_{k \to 0}   \int_0^{{\cal O}_*^{max}} d {\cal O}  e^{ T N \left[ k {\cal O}     \right]  }
\label{saddlevanish}
\end{eqnarray}
This integral will be dominated by the maximal value ${\cal O}_*^{max} $ for $k >0$
and by the minimal value ${\cal O}_*^{min} =0$ for $k <0$,
so that the scaled cumulant generating function ${\cal G}(k) $ will present the different behaviors for $k \to 0^{\pm}$
\begin{eqnarray}
{\cal G}(k) && \opsimeq_{k \to 0^-}   k {\cal O}_*^{min}
\nonumber \\
{\cal G}(k) && \opsimeq_{k \to 0^+}   k {\cal O}_*^{max}
\label{gminmax}
\end{eqnarray}
i.e. its first derivative $  {\cal G}'(k)$ will display the following discontinuity at the origin $k \to 0^{\pm}$
\begin{eqnarray}
{\cal G}'(k=0^-) && = {\cal O}_*^{min} =0
\nonumber \\
 {\cal G}'(k=0^+) && = {\cal O}_*^{max} =  \left[  \beta + \frac{\alpha^{++}}{2(1-c)} +  \frac{\alpha^{+-} }{2c}  \right] \frac{2c (1-c) }{2-c}
  \label{gprimejump}
\end{eqnarray}


\section{ Anomalous large deviations properties of the random East model }

\label{sec_Erandom}

In this section, we consider the random East model with the flip rates of Eq. \ref{weasthard}
\begin{eqnarray}
 W^{RandomEast}(\sigma_i^x C,C) =  \delta_{S_{i-1},+ }  
 \left[  (1-c_i) \delta_{S_i,+} + c_i \delta_{S_i,-}\right] \equiv w_i^{S_i}(S_{i-1}) 
\label{weastrandom}
\end{eqnarray}
in order to analyze the anomalous dynamical large deviations properties that emerge for 
this hard-constrained model $\epsilon=0$ with respect to the soft-constrained model $\epsilon>0$ discussed in section
\ref{sec_Erandomsoft}.


\subsection{ Anomalous large deviations at Level 2.25 for the empirical densities and activities  }

\label{subsec_vanish2.5}

In the limit $\epsilon \to 0$, the large deviations at Level 2.25 of Eq. \ref{level2.25e} become
 \begin{eqnarray}
  P^{[2.25]}_T [  a_.(+) ; \rho_.^. ;  \rho_{.,.}^{.,.}] 
  \opsimeq_{T \to +\infty} 
C_{2} [ \rho_.^. ;  \rho_{.,.}^{.,.}]\theta \left( \sum_{i=1}^N \rho_i^+>0 \right) 
e^{ \displaystyle - T  I_{2.25}^{RandomEast} [  a_.(+) ; \rho_{.,.}^{.,.}]
 }
\label{level2.5esimpliappli}
\end{eqnarray}
with the rate function of Eq. \ref{rate2.25rEs} for  $\epsilon=0$
\begin{eqnarray}
 I_{2.25}^{RandomEast} [  a_.(+) ; \rho_{.,.}^{.,.}]
=  \sum_{i=1}^N  
\left[ \frac{a_i(+)  }{2}    \ln \left( \frac{  a_i^2(+)  }
{ 4 c_i (1-c_i)  \rho_{i-1,i}^{++}   \rho_{i-1,i}^{+-} } \right)
-  a_i(+) + (1-c_i)  \rho_{i-1,i}^{++} 
+ c_i  \rho_{i-1,i}^{+-}  \right]
 \label{rate2.5rE} 
\end{eqnarray}
Note that besides the constitutive constraints $C_{2} [ \rho_.^. ;  \rho_{.,.}^{.,.}] $ of Eq. \ref{c2e},
we have added the Heaviside constraint $\theta \left( \sum_{i=1}^N \rho_i^+>0 \right) $ in order to exclude explicitly 
the configuration $\{S_i=-1 ; i=1,2,..,N\}$ that is disconnected from the set of the other $(2^N-1)$ configurations 
for the hard-constrained model $\epsilon=0$.

We are interested into the region of empirical activities $a_.(+) $ and empirical densities $  [\rho_.^. ;  \rho_{.,.}^{.,.}]$
that make the rate function $ I_{2.25}^{RandomEast} [  a_.(+) ; \rho_{.,.}^{.,.}] $ of Eq. \ref{rate2.5rE}  vanish
\begin{eqnarray}
  \rho_{i-1,i}^{++} && = \frac{a_i(+)}{2(1-c_i)}
  \nonumber \\
  \rho_{i-1,i}^{+-}  && = \frac{a_i(+)}{2 c_i}
 \label{rate2.5rEzero} 
\end{eqnarray}
and that satisfy the constraints $C_{2} [ \rho_.^. ;  \rho_{.,.}^{.,.}] $ of Eq. \ref{c2e}
\begin{eqnarray}
\rho_{i-1}^+  && = \rho_{i-1,i}^{++}  + \rho_{i-1,i}^{+-} = \frac{a_i(+)}{ 2 c_i (1-c_i)}
\nonumber \\
\rho_{i-1,i}^{-+} && =\rho_{i}^+ - \rho_{i-1,i}^{++} = \frac{a _{i+1}(+)}{ 2 c_{i+1} (1-c_{i+1})} - \frac{a _i(+)}{2(1-c_i)}
\nonumber \\
\rho_{i-1,i}^{--} && = 1- \rho_{i-1,i}^{++} - \rho_{i-1,i}^{+-} - \rho_{i-1,i}^{-+}
= 1 - \frac{a_i(+)}{2 c_i} - \frac{a_{i+1}(+)}{ 2 c_{i+1} (1-c_{i+1})}
\label{elimc2zero}
\end{eqnarray}
Since all these empirical densities have to be in $[0,1]$,
one obtains that the remaining constraints for the $N$ positive empirical activities $a_i(+) \geq 0$ can be summarized by
the following inequalities for any $i=1,..,N$
\begin{eqnarray}
0 \leq \frac{a_i(+)}{2(1-c_i)}  \leq \frac{a_{i+1}(+)}{ 2 c_{i+1} (1-c_{i+1})}  \leq 1 - \frac{a_i(+)}{2 c_i} 
\label{elimc2zeropositivity}
\end{eqnarray}

In summary, whenever the $N$ empirical activities $a_i(+) $ satisfy the inequality constraints of Eq. \ref{elimc2zeropositivity}
 for any $i=1,..,N$, and the supplementary constraint  $\theta \left( \sum_{i=1}^N a_i^+>0 \right) $,
the empirical densities computed from Eqs \ref{rate2.5rEzero} and \ref{elimc2zero} satisfy the constraints of Eq. \ref{c2e}
and make the rate function of Eq. \ref{rate2.5rE}  vanish $ I_{2.25}^{RandomEast} [  a_.(+) ; \rho_{.,.}^{.,.}] =0$.


\subsection{ Anomalous large deviations at Level 2 for the empirical densities alone  }

The large deviations at Level 2 of Eq. \ref{level2e}
\begin{eqnarray}
  P^{[2]}_T [  \rho_.^. ;  \rho_{.,.}^{.,.}] \opsimeq_{T \to +\infty} 
C_2 [   \rho_.^. ;  \rho_{.,.}^{.,.}]\theta \left( \sum_{i=1}^N \rho_i^+>0 \right) 
e^{ \displaystyle - T  I_{2}^{RandomEast}  [  \rho_{.,.}^{.,.}]  }
\label{level2eappli}
\end{eqnarray}
involve the 
the rate function at Level 2 of Eq. \ref{rate2rEs}
for  $\epsilon=0$
\begin{eqnarray}
 I_{2}^{RandomEast} [   \rho_{.,.}^{.,.}]
=  \sum_{i=1}^N  
 \left[ \sqrt{ (1-c_i)  \rho_{i-1,i}^{++} } - \sqrt{ c_i  \rho_{i-1,i}^{+-} } \right]^2
 \label{rate2rE} 
\end{eqnarray}

To obtain the region of empirical densities $  [\rho_.^. ;  \rho_{.,.}^{.,.}]$
that make the rate function $ I_{2}^{RandomEast} [   \rho_{.,.}^{.,.}] $ 
of Eq. \ref{rate2rE}  vanish and that satisfy the constraints $C_{2} [ \rho_.^. ;  \rho_{.,.}^{.,.}] $ of Eq. \ref{c2e},
we can use the previous analysis of subsection \ref{subsec_vanish2.5} concerning the Level 2.25 :
the first equation of Eq. \ref{elimc2zero} allows to replace the $N$ empirical activities in terms of the empirical 1-spin density
\begin{eqnarray}
a_i(+) = 2  \rho_{i-1}^+   c_i (1-c_i)
\label{elimc2zeroelimq}
\end{eqnarray}
into all the other equations of Eqs \ref{rate2.5rEzero}  and \ref{elimc2zero}
to obtain the empirical 2-spin density
\begin{eqnarray}
  \rho_{i-1,i}^{++} && =\rho_{i-1}^+   c_i 
  \nonumber \\
  \rho_{i-1,i}^{+-}  && = \rho_{i-1}^+ (1-  c_i )
\nonumber \\
\rho_{i-1,i}^{-+} && = \rho_i^+ - \rho_{i-1}^+   c_i 
\nonumber \\
\rho_{i-1,i}^{--} && = 1 - \rho_{i-1}^+  (1- c_i ) -  \rho_i^+
\label{elimc2zeroelimq2}
\end{eqnarray}
while the inequality constraints of Eq. \ref{elimc2zeropositivity}
translate into 
\begin{eqnarray}
0 \leq \rho_{i-1}^+   c_i   \leq \rho_i^+  \leq 1 - \rho_{i-1}^+  (1- c_i )
\label{elimc2zeropositivityrho}
\end{eqnarray}

In summary, whenever the $N$ empirical 1-spin density $\rho_i^+ \in [0,1]$  satisfy the inequality constraints of Eq. \ref{elimc2zeropositivityrho} for any $i=1,..,N$ and the supplementary constraint  $\theta \left( \sum_{i=1}^N \rho_i^+>0 \right) $ : 
together with the empirical 2-spin density computed via Eq. \ref{elimc2zeroelimq2},
they satisfy the constraints of Eq. \ref{c2e} and
they make the rate function of Eq. \ref{rate2.5rE}  vanish $ I_{2}^{RandomEast} [   \rho_{.,.}^{.,.}] =0$.


\subsection{ Anomalous large deviations properties of time-additive space-local observables  }

 In the Random East model with the hard constraint $\epsilon=0$, the time-additive space-local observable of Eq. \ref{additive2}
can be parametrized by some functions $\beta_i $ and $\alpha_i^{S_L,S} $ 
\begin{eqnarray}
O_T 
 =  \sum_{i=1}^N  \left(  \beta_i  a_i(+) + \sum_{S_L=\pm,S=\pm}  \alpha_i^{S_L,S} \rho_{i-1,i}^{S_L,S}  \right)
\label{additive2zero}
\end{eqnarray}


\subsubsection{ Anomalous vanishing of the rate function $I(O)$ on a finite interval  }

The rate function $I(O)$ will vanish $I(O_*) =0$
for any value $O_* $ that can be reconstructed from the region discussed in subsection \ref{subsec_vanish2.5} 
of empirical activities $a_.(+) $ and empirical densities $  [\rho_.^. ;  \rho_{.,.}^{.,.}]$
that make the rate function $ I_{2.25}^{RandomEast} [  a_.(+) ; \rho_{.,.}^{.,.}] $ of Eq. \ref{rate2.5rE}  vanish
and that satisfy the constraints $C_{2} [ \rho_.^. ;  \rho_{.,.}^{.,.}] $ of Eq. \ref{c2e}
\begin{small}
\begin{eqnarray}
&& O_* 
 =  \sum_{i=1}^N  \left( \beta_i  a_i(+)
 +   \alpha_i^{++} \rho_{i-1,i}^{++} 
   + \alpha_i^{+-} \rho_{i-1,i}^{+-}
   +   \alpha_i^{-+} \rho_{i-1,i}^{-+} 
   + \alpha_i^{--} \rho_{i-1,i}^{--} \right)
 \label{additive2zerovanish} \\
&& =  \sum_{i=1}^N  \left( \beta_i  a_i(+)
 +   \alpha_i^{++} \frac{a_i(+)}{2(1-c_i)}
   + \alpha_i^{+-}  \frac{a_i(+)}{2c_i}
   +   \alpha_i^{-+} \left[ \frac{a_{i+1}(+)}{ 2c_{i+1} (1-c_{i+1})} - \frac{a_i(+)}{2(1-c_i)} \right] 
   + \alpha_i^{--} \left[1 - \frac{a_i(+)}{2c_i} - \frac{a_{i+1}(+)}{ 2 c_{i+1} (1-c_{i+1})} \right] \right)   
\nonumber 
\end{eqnarray}
\end{small}
where the $N$ empirical activities $a_i(+)$ have to satisfy the inequality constraints of Eq. \ref{elimc2zeropositivity}
and the supplementary constraint  $\theta \left( \sum_{i=1}^N a_i^+>0 \right) $.

As a consequence, the rate function $I(O)$ will vanish on the finite interval $[O_*^{min},O_*^{max}]$
\begin{eqnarray}
 I ( O_*  )    
  =     0 \ \ \ \ \ \ { \rm for } \ \ \ \ \ \ \ O_*^{min} \leq O_* \leq O_*^{max}
  \label{rate1vanishinterval}
\end{eqnarray}
where $O_*^{min} $ and $O_*^{max} $ are the minimal and the maximal values that can be reconstructed via Eq. \ref{additive2zerovanish}.


\subsubsection{ Corresponding singularity in the scaled cumulant generating function $G(k ) $ 
at the origin $k \to 0^{\pm}$  }

For small $k$, the generating function will be dominated by the interval $O_* \in [O_*^{min},O_*^{max}]$
where the rate function $I(O_*)=0$ vanishes (Eq. \ref{rate1vanishinterval})
\begin{eqnarray}
 \langle e^{ k T O_T } \rangle  \opsimeq_{T \to +\infty}  e^{ T  G(k)  }
 \opsimeq_{T \to +\infty} \int_{-\infty}^{+\infty} d O  e^{ T  \left[ k O -  I(O)   \right]  }
 \opsimeq_{k \to 0}  \int_{O_*^{min}}^{O_*^{max}} d O_*  e^{ T  k O_*  }
 \label{saddlevanishrE}
\end{eqnarray}
This integral will be dominated by the maximal value $O_*^{max} $ for $k >0$
and by the minimal value $O_*^{min} $ for $k <0$,
so that the scaled cumulant generating function $G(k ) $ will present the different behaviors for $k \to 0^{\pm}$
\begin{eqnarray}
G(k) && \opsimeq_{k \to 0^-}   k O_*^{min}
\nonumber \\
G(k) && \opsimeq_{k \to 0^+}   k O_*^{max}
\label{gminmaxrE}
\end{eqnarray}
i.e. its first derivative $  G'(k)$ will display the following discontinuity at the origin $k \to 0^{\pm}$
\begin{eqnarray}
G'(k=0^-) && = O_*^{min}
\nonumber \\
 G'(k=0^+) && = O_*^{max}
 \label{gprimejumprE}
\end{eqnarray}


\section{ Conclusion }

\label{sec_conclusion}

In this paper, we have revisited the East model via the large deviations for the relevant local empirical densities and the relevant local empirical activities that only involve two consecutive spins.
We have first considered the random East Model with the soft kinetic constraint $\epsilon>0$
in order to derive the regular large deviations properties with respect to the time-window $T$ of the $O(N)$ 
relevant empirical time-averaged densities and activities. 
We have then turned to the pure East Model with the soft kinetic constraint $\epsilon>0$
in order to derive the regular large deviations properties with respect to the space-time volume $(TN)$ 
of the $O(1)$ relevant empirical time-space-averaged densities and activities.
Finally, we have analyzed in detail the anomalous large deviations properties 
that emerge in the hard-constraint limit $\epsilon=0$
both for the pure East model and for random East Model.

In the future, it will be interesting to study the dynamical properties of other spin models with local rates
satisfying detailed-balance
via the closed large deviations at Level 2.25 for the relevant local empirical densities and activities involving only a few spins.

\appendix


\section{ Reminder on large deviations for the relevant empirical observables  }

\label{appendix_relevant}

In this Appendix, we summarize the general procedure to derive the large deviations properties for
the relevant empirical observables of Markov trajectories.

\subsection{ Identification of the relevant time-empirical observables that determine the trajectories probabilities }

For the Markov model defined by the Markov generator $W$, 
the first step consists in rewriting the probability of a long dynamical trajectory $C(0 \leq t \leq T)$ 
\begin{eqnarray}
 {\cal P}[C(0 \leq t \leq T)] \opsimeq_{T \to +\infty}   e^{\displaystyle  -T  \Phi_{[W]} \left(E [C(0 \leq t \leq T)] \right) }
\label{ptrajectempi}
\end{eqnarray}
in terms of an intensive action $\Phi_{[W]} \left( E [C(0 \leq t \leq T)] \right) $ that depends on the Markov generator $W$,
and that only involves a few relevant time-empirical observables $E [C(0 \leq t \leq T)] $ of the dynamical trajectory $C(0 \leq t \leq T) $.

\subsection{ Number of dynamical trajectories of length $T$ with the same value of the time-empirical observables }

Since all the individual dynamical trajectories $ C(0 \leq t \leq T)  $
 that have the same empirical observables $E=E [C(0 \leq t \leq T)] $
  have the same probability given by Eq. \ref{ptrajectempi},
 one can rewrite the normalization over all possible trajectories 
as a sum over these empirical observables
\begin{eqnarray}
1= \sum_{C(0 \leq t \leq T)}  {\cal P}[C(0 \leq t \leq T)] 
\opsimeq_{T \to +\infty}  
 \sum_{E} \Omega_T(E ) e^{\displaystyle  -T  \Phi_{[W]} \left( E  \right) }
\label{normaempi}
\end{eqnarray}
where the number of dynamical trajectories of length $T$ associated to given values $E $ of these 
empirical observables
\begin{eqnarray}
\Omega_T ( E ) \equiv \sum_{C(0 \leq t \leq T)} \delta \left(E [C(0 \leq t \leq T)] - E \right)
\label{omegaaempi}
\end{eqnarray}
 grows exponentially with respect to the length $T$ of the trajectories
\begin{eqnarray}
 \Omega_T( E) \opsimeq_{T \to +\infty} C(E) \ e^{\displaystyle T S( E )  }
\label{omegat}
\end{eqnarray}
The prefactor $C(E)$ denotes the appropriate constitutive constraints for the empirical observables $E$.
The factor $S( E )  = \frac{\ln \Omega_T(E) }{ T }  $ represents the 
Boltzmann intensive entropy of the set of trajectories of length $T$ with given empirical observables $E $.
Let us now recall how it can be evaluated without any actual computation (i.e. one does not need 
to use combinatorial methods to count the appropriate configurations).

The normalization of Eq. \ref{normaempi} becomes for large $T$
\begin{eqnarray}
1  \opsimeq_{T \to +\infty} \sum_{E} C(E) \  e^{\displaystyle  T \left[ S( E  ) - \Phi_{[W]} \left( E \right)   \right] }
\label{normaempit}
\end{eqnarray}
When the empirical variables $E $ take their typical values $E_{[W]}^{typ}$ for the Markov generator $W$,
the exponential behavior in $T$ of Eq. \ref{normaempit}
should exactly vanish,
i.e. the entropy $S( E_{[W]}^{typ}  ) $ should exactly compensate the action $\Phi_{[W]} \left( E_{[W]}^{typ}\right)   $ 
\begin{eqnarray}
S( E_{[W]}^{typ}  )=    \Phi_{[W]} \left( E_{[W]}^{typ}\right) 
\label{compensation}
\end{eqnarray}
To obtain the intensive entropy $S( E  ) $ for any other given value $E  $ of the empirical observables,
one just needs to introduce the modified Markov generator $\hat W_E$ that would make 
the empirical values $E $ typical for this modified model
\begin{eqnarray}
   E = E_{[\hat W_E]}^{typ}
\label{modeleeff}
\end{eqnarray}
and to use Eq. \ref{compensation} for this modified model to obtain
\begin{eqnarray}
S( E) = S( E_{[\hat W (E)]}^{typ}  )=    \Phi_{[{\hat W}_E]} \left(E_{\hat W_E}^{typ}  \right) 
= \Phi_{[\hat W_E]} \left(E \right) 
\label{entropyempi}
\end{eqnarray}
Here one should stress that the modified generator ${\hat W}_E $  and thus $S(E)$ 
depends only on the empirical observables $E$
and do not involve the initial generator $W$.
Plugging Eq. \ref{entropyempi} into Eq. \ref{omegat}
yields that the number $ \Omega_T ( E )$ of dynamical trajectories of length $T$ associated to given values $E $ of these empirical observables of Eq \ref{omegaaempi} 
\begin{eqnarray}
\Omega_T ( E )  \opsimeq_{T \to +\infty} C(E) \ e^{\displaystyle T S( E )  }
\opsimeq_{T \to +\infty} C(E) \ e^{\displaystyle T \Phi_{[\hat W_E]} \left(E \right)   }
\label{omegafinal}
\end{eqnarray}
only involve the action $\Phi_{[\hat W_E]} \left(E \right) $ of the empirical observables $E$ evaluated for the 
 modified generator ${\hat W}_E $ defined by Eq. \ref{modeleeff}.


\subsection{ Large deviations for the relevant time-empirical observables $E$}

The normalization over trajectories of Eq \ref{normaempi} can be rewritten as the normalization
\begin{eqnarray}
 1  = \sum_{ E }  P_T^{[2.5]}  (E)
\label{normaprobaempi}
\end{eqnarray}
for the probability
\begin{eqnarray}
P_T^{[2.5]}  (E) \opsimeq_{T \to +\infty}  \Omega_T(E ) e^{\displaystyle  -T  \Phi_{[W]} (E) }
\label{probempi2.5}
\end{eqnarray}
to see the empirical observables $E$ when
the dynamical trajectories of length $T$ are governed by the Markov generator $W$.
Plugging Eq. \ref{omegafinal} into Eq. \ref{normaempi}
yields the large deviation form
\begin{eqnarray}
 P^{[2.5]}_T  (E) \opsimeq_{T \to +\infty}  C(E) \ e^{\displaystyle  - T  I_{2.5} (E)    }
\label{probaaempi}
\end{eqnarray}
where the rate function at Level 2.5
\begin{eqnarray}
I_{2.5}  (E)  
= \Phi_{[W]}  (E) - \Phi_{[\hat W_E]}  (E) 
\label{rateempi}
\end{eqnarray}
is simply given by the difference between the intensive action $\Phi_{[W]}  (E) $ associated to the true generator $W$
and the intensive action $\Phi_{[\hat W_E]}  (E) $ associated to the modified generator $\hat W_E$ that would make the empirical value $E$ typical (see Eq. \ref{modeleeff}).
It is positive $I_{2.5}  (E) \geq 0 $ and vanishes when $E$ takes the typical value $E_{[W]}^{typ}$
\begin{eqnarray}
0=I_{2.5}  (E_{[W]}^{typ}) 
\label{rateempizerotyp}
\end{eqnarray}
i.e. only when the modified generator $\hat W_E$ coincides with the true generator $W$.

\subsection{ Example : derivation of the large deviations at Level 2.5 for Markov jump processes  }

\label{subsec_2.5config}

Let us now describe how the general formalism described above
can be applied to the continuous-time Markov jump process described by the Master Equation
\begin{eqnarray}
\frac{\partial P_t(C)}{\partial t} =    \sum_{C' }   W(C,C')  P_t(C') 
\label{mastereq}
\end{eqnarray}

The trajectory probability of Eq. \ref{pwtrajjump}
can be rewritten as
\begin{eqnarray}
{\cal P}[x(0 \leq t \leq T)]   
=  e^{ \displaystyle  T  \sum_{C  }\sum_{ C' \ne C  } \left[ q(C',C) \ln ( W(C',C) ) -     \rho(C) W(C',C)      \right]}
\label{pwtrajjumpempi}
\end{eqnarray}
in terms of the empirical time-averaged density
\begin{eqnarray}
 \rho(C)  \equiv \frac{1}{T} \int_0^T dt \  \delta_{C(t),C}  
 \label{rhoc}
\end{eqnarray}
satisfying the normalization
\begin{eqnarray}
\sum_C \rho(C)  = 1
\label{rhocnorma}
\end{eqnarray}
and on the empirical flows from $C$ to $C' \ne C$
\begin{eqnarray}
q(C',C) \equiv  \frac{1}{T} \sum_{t : C(t) \ne C(t^+)} \delta_{C(t^+),C'} \delta_{C(t),C} 
\label{jumpempiricaldensity}
\end{eqnarray}
satisfying the following stationarity constraints.
For any configurations $C$, the total incoming flow into $C$
\begin{eqnarray}
q_{in}(C) \equiv \sum_{C' \ne C} q(C,C') = \frac{1}{T} \sum_{t : C(t) \ne C(t^+)} \delta_{C(t^+),C} 
\label{qin}
\end{eqnarray}
and the total outgoing flow from $C$
\begin{eqnarray}
q_{out}(C) \equiv \sum_{C' \ne C} q(C',C) = \frac{1}{T} \sum_{t : C(t) \ne C(t^+)}  \delta_{C(t),C} 
\label{qout}
\end{eqnarray}
should be equal up to boundary terms of order $1/T$ (involving the initial configuration at time $t=0$
and the final configuration at time $T$)
that can be neglected for large time-window $T \to +\infty$
\begin{eqnarray}
0=q_{out}(C) - q_{in}(C)  = \sum_{C' \ne C} \left(q(C',C)-  q(C,C') \right)
\label{contrainteq}
\end{eqnarray}

With respect to the general formalism summarized in Appendix \ref{appendix_relevant},
this means that the relevant empirical observables $E$ are the empirical density $\rho(.)$ and the 
empirical flows $q(.,.)$, while
the corresponding action introduced in Eq. \ref{ptrajectempi} reads 
\begin{eqnarray}
\Phi_{[W]} [ \rho(.) ; q(.,.)]  = \sum_{C  }\sum_{ C' \ne C  } \left[ \rho(C) W(C',C)   - q(C',C) \ln ( W(C',C) )    \right]
\label{actionjumpconfig}
\end{eqnarray}

For the modified rates $\hat W(C',C) $ that would make typical the empirical variables $[ \rho(.) ; q(.,.)] $ 
\begin{eqnarray}
 \hat W(C',C) = \frac{ q(C',C) }{\rho(C) }
\label{hatwconfig}
\end{eqnarray}
the action of Eq. \ref{actionjumpconfig} becomes
\begin{eqnarray}
\Phi_{[\hat W]} [ \rho(.) ; q(.,.)] && = \sum_{C  }\sum_{ C' \ne C  } \left[ \rho(C) \hat W(C',C)   - q(C',C) \ln ( \hat W(C',C) )    \right]
\nonumber \\
&& = \sum_{C  }\sum_{ C' \ne C  } \left[  q(C',C)    - q(C',C) \ln ( \frac{ q(C',C) }{\rho(C) } )    \right]
\label{actionjumptypconfig}
\end{eqnarray}
The difference of Eq. \ref{rateempi}
between the actions of Eqs \ref{actionjumpconfig} and \ref{actionjumptypconfig} 
allows to recover the well-known rate function at Level 2.5
\begin{eqnarray}
 I_{2.5}( \rho_. ; q_{.,.}  ) && = \Phi_{[W]} [ \rho(.) ; q(.,.)]   - \Phi_{[\hat W]} [ \rho(.) ; q(.,.)]
 \nonumber \\
 && =  \sum_{C } \sum_{C' \ne C} 
\left[ q(C',C)  \ln \left( \frac{ q(C',C)  }{  W(C',C)  \rho(C) }  \right) 
 - q(C',C)  + W(C',C)  \rho(C)  \right]
\label{rate2.5master}
\end{eqnarray}
that governs the probability of the empirical observables $ [ \rho(.) ; q(.,.)] $ for large  $T$ 
\cite{fortelle_thesis,fortelle_jump,maes_canonical,maes_onandbeyond,wynants_thesis,chetrite_formal,BFG1,BFG2,chetrite_HDR,c_ring,c_interactions,c_open,c_detailed,barato_periodic,chetrite_periodic,c_reset,c_inference,c_runandtumble,c_jumpdiff,c_skew,c_metastable,c_exclusion}
\begin{eqnarray}
P^{[2.5]}_{T}[ \rho(.) ; q(.,.) ] \oppropto_{T \to +\infty}  e^{- T I_{2.5}[ \rho(.) ; q(.,.) ] }
 \delta \left( \sum_C \rho(C) - 1 \right) 
 \prod_C \delta \left[  \sum_{C' \ne C} (q(C,C') - q(C',C) ) \right]
\label{level2.5master}
\end{eqnarray}
where the constitutive constraints of the empirical observables 
have been discussed in Eqs \ref{rhocnorma} and \ref{contrainteq}.


\section{Detailed-Balance Markov Jump processes : explicit contractions of the Level 2.5 }

\label{app_LargeDevDB}

In this Appendix, we describe how the Level 2.5 of Markov Jump processes described in the subsection \ref{subsec_2.5config} of the previous Appendix can be contracted explicitly towards lower levels
 when the rates satisfy the Detailed-Balance condition.


\subsection{ Level 2.5 when the empirical flows are replaced by the empirical activities and the empirical currents  }

It is convenient to order the configurations.
For each link $(C'>C)$, it is useful to parametrize the two empirical flows $q(C',C)$ and $q(C,C')$ of Eq. \ref{jumpempiricaldensity}
\begin{eqnarray}
q(C',C)  \equiv \frac{a(C',C)  +j(C',C) }{2} 
\nonumber \\
q(C,C')  \equiv \frac{a(C',C)  -j(C',C) }{2} 
\label{ajreci}
\end{eqnarray}
by their symmetric and antisymmetric parts called the activity and the current
\begin{eqnarray}
a(C',C)  \equiv q(C',C)  +q(C,C') = a(C,C')
\nonumber \\
j(C',C)  \equiv q(C',C)  -q(C,C')  = - j(C,C')
\label{aj}
\end{eqnarray}
Since the stationarity constraints of Eq. \ref{contrainteq}
only involve the currents $j(.,.) $ and not the activities $a(.,.) $
\begin{eqnarray}
 \sum_{C' \ne C} j(C',C) =0
\label{contraintej}
\end{eqnarray}
the Level 2.5 of Eq. \ref{level2.5master} becomes
\begin{eqnarray}
P^{[2.5]}_{T}[ \rho(.) ; a(.,.) ; j(.,.) ] \oppropto_{T \to +\infty} e^{- T I_{2.5}[ \rho(.) ; a(.,.) ; j(.,.) ] }
 \delta \left( \sum_C \rho(C) - 1 \right) 
 \prod_C  \delta \left[  \sum_{C' \ne C} j(C',C)  \right]
\label{level2.5masteraj}
\end{eqnarray}
The rate function translated from Eq. \ref{rate2.5master} via the change of variables of Eq \ref{ajreci}
\begin{eqnarray}
I_{2.5}[ \rho(.) ; a(.,.) ; j(.,.) ]=  \sum_{C } \sum_{C' > C} I^{[C',C]}_{2.5}[ \rho(C') ; \rho(C) ; a(C',C) ; j(C',C) ]
\label{rate2.5masteraj}
\end{eqnarray}
involves the following contribution for the link $[C',C]$ 
\begin{eqnarray}
&& I^{[C',C]}_{2.5}[ \rho(C') ; \rho(C) ; a(C',C) ; j(C',C) ]  \equiv 
  \frac{j(C',C) }{2}   \ln \left( \frac{ (a(C',C)  +j(C',C))  W(C,C')  \rho(C') }{ (a(C',C)  -j(C',C))  W(C',C)  \rho(C) }  \right) 
\nonumber \\
&&    + \frac{a(C',C) }{2}   \ln \left( \frac{ a^2(C',C)  - j^2(C',C)  }{  4 W(C',C)  \rho(C) W(C,C')  \rho(C')}  \right) 
 - a(C',C)     + W(C',C)  \rho(C)    + W(C,C')  \rho(C') 
\label{rate2.5masterajlink}
\end{eqnarray}
It is useful to separate the even and the odd parts with respect to the link current $j(C',C)$ 
\begin{small}
\begin{eqnarray}
 I^{[C',C]}_{2.5}[ \rho(C') ; \rho(C) ; a(C',C) ; j(C',C) ] =   I^{[C',C]Even}_{2.5}[ \rho(C') ; \rho(C) ; a(C',C) ; j(C',C) ]
 +   I^{[C',C]Odd}_{2.5}[ \rho(C') ; \rho(C) ; a(C',C) ; j(C',C) ] \ \ 
\label{rate2.5masterajlinkeo}
\end{eqnarray}
\end{small}
The even contribution reads
\begin{footnotesize}
\begin{eqnarray}
&& I^{[C',C]Even}_{2.5}[ \rho(C') ; \rho(C) ; a(C',C) ; j(C',C) ]  \equiv \frac{I^{[C',C]}_{2.5}[ \rho(C') ; \rho(C) ; a(C',C) ; j(C',C) ]+I^{[C',C]}_{2.5}[ \rho(C') ; \rho(C) ; a(C',C) ; -j(C',C) ]}{2}
\nonumber \\
&&  = \frac{j(C',C) }{2}   \ln \left( \frac{ a(C',C)  +j(C',C)    }{ a(C',C)  -j(C',C)     }  \right) 
   + \frac{a(C',C) }{2}   \ln \left( \frac{ a^2(C',C)  - j^2(C',C)  }{  4 W(C',C)  \rho(C) W(C,C')  \rho(C')}  \right) 
 - a(C',C)     + W(C',C)  \rho(C)    + W(C,C')  \rho(C') \ \ \ \ \ \
\label{rate2.5masterajlinkeven}
\end{eqnarray}
\end{footnotesize}
while the odd contribution
\begin{small}
\begin{eqnarray}
 I^{[C',C]Odd}_{2.5}[ \rho(C') ; \rho(C) ; a(C',C) ; j(C',C) ] && \equiv \frac{I^{[C',C]}_{2.5}[ \rho(C') ; \rho(C) ; a(C',C) ; j(C',C) ]-I^{[C',C]}_{2.5}[ \rho(C') ; \rho(C) ; a(C',C) ; -j(C',C) ]}{2}
\nonumber \\
&& 
=  \frac{j(C',C) }{2}   \ln \left( \frac{  W(C,C')  \rho(C')   } {   W(C',C)  \rho(C) }  \right) 
\label{rate2.5masterajlinkodd}
\end{eqnarray}
\end{small}
is simply linear in the current $j(C',C)$. The factor $  \ln \left( \frac{  W(C,C')  \rho(C')   } {   W(C',C)  \rho(C) }  \right)$ measures the irreversibility associated
to the two link flows $W(C,C')  \rho(C') $ and $W(C',C)  \rho(C) $ that would be typically
 produced by the empirical densities $ \rho(C') $ and $\rho(C) $.
This can be considered as an example of the Gallavotti-Cohen fluctuation relations
(see \cite{galla,kurchan_langevin,Leb_spo,maes1999,jepps,derrida-lecture,harris_Schu,kurchan,searles,zia,chetrite_thesis,maes2009,maes2017,chetrite_HDR} and references therein).


\subsection{Detailed-Balance : the sum of the current-odd contributions vanishes in the rate function at Level 2.5 }

When the rates satisfy the detailed balance condition on each link $C \ne C'$
\begin{eqnarray}
0= W(C,C') P_{eq} (C') - W(C',C) P_{eq}(C) 
\label{detailed}
\end{eqnarray}
on can plug the ratio of the two rates 
\begin{eqnarray}
\frac{ W(C,C') }{ W(C',C)}  = \frac{ P_{eq}(C) }{ P_{eq} (C') }
\label{detailedratio}
\end{eqnarray}
into the odd contribution of Eq. \ref{rate2.5masterajlinkodd} to obtain
\begin{eqnarray}
 I^{[C',C]Odd}_{2.5}[ \rho(C') ; \rho(C) ; a(C',C) ; j(C',C) ] 
=  \frac{j(C',C) }{2} \left[  \ln \left( \frac{    \rho(C')   } {   P_{eq}(C')   }  \right) 
-  \ln \left( \frac{    \rho(C)   } {   P_{eq}(C)  }  \right) 
\right]
\label{rate2.5masterajlinkodddb}
\end{eqnarray}
The sum of all the odd contributions of Eq. \ref{rate2.5masterajlinkodddb} 
then vanishes as a consequences 
of the antisymmetry of the current $j(C',C)  = - j(C,C') $ of Eq. \ref{aj}
and of the stationarity constraint of Eq. \ref{contraintej}
\begin{eqnarray}
&& \sum_{C } \sum_{C' > C}  I^{[C',C]Odd}_{2.5}[ \rho(C') ; \rho(C) ; a(C',C) ; j(C',C) ] 
\nonumber \\
&& = \sum_{C } \sum_{C' > C}  \frac{j(C',C) }{2}   \ln \left( \frac{    \rho(C')   } {   P_{eq}(C')   }  \right) 
- \sum_{C } \sum_{C' > C}  \frac{j(C',C) }{2} \ln \left( \frac{    \rho(C)   } {   P_{eq}(C)  }  \right) 
\nonumber \\
&& 
= - \frac{1}{2} \sum_{C } \ln \left( \frac{    \rho(C)   } {   P_{eq}(C)   }  \right) 
\left[ \sum_{C'<C}  j(C',C)   +  \sum_{C' > C} j(C',C)  \right] 
= - \frac{1}{2} \sum_{C } \ln \left( \frac{    \rho(C)   } {   P_{eq}(C)   }  \right) 
\left[ \sum_{C'\ne C}  j(C',C)   \right] 
=0
\label{rate2.5masterajlinkodddbsum}
\end{eqnarray}
 As recalled after Eq. \ref{rate2.5masterajlinkodd}, 
the odd contributions of the links to the rate function at Level 2.5
are linear in the link currents and are directly related to the irreversibility of the dynamics.
 The physical interpretation of the vanishing of the sum of all these odd contributions (Eq. \ref{rate2.5masterajlinkodddbsum}) is that a detailed-balance dynamics
 cannot have a global irreversible property.

So the total rate function of Eq. \ref{rate2.5masteraj}
reduces to the sum of the even contributions of the links
\begin{eqnarray}
I_{2.5}[ \rho(.) ; a(.,.) ; j(.,.) ] =  \sum_{C } \sum_{C' > C} I^{[C',C]Even}_{2.5}[ \rho(C') ; \rho(C) ; a(C',C) ; j(C',C) ]
\label{rate2.5masterajeonly}
\end{eqnarray}
As a consequence, when the empirical density $\rho(.) $ and the empirical activity $a(.,.)$ are given,
any configuration of the empirical currents $j(C',C)$ that satisfies the stationary constraints of Eq. \ref{contraintej}
has the same rate function as the configuration with the reversed empirical currents $(-j(C',C))$ that also satisfies the stationary constraints 
\begin{eqnarray}
I_{2.5}[ \rho(.) ; a(.,.) ; j(.,.) ] =  I_{2.5}[ \rho(.) ; a(.,.) ; -j(.,.) ]
\label{rate2.5masterajeonlyREVERSED}
\end{eqnarray}


\subsection{Explicit contraction over the currents towards Level 2.25 for the density $\rho(C)$ and the activity $a(C',C)$  }

The behavior of the even contribution of Eq. \ref{rate2.5masterajlinkeven}
 with respect to the current $j(C',C) \in ]- a(C',C),+a(C',C)[$ 
 can be analyzed as follows : 
 the first partial derivative with respect to the current $j(C',C)$ 
\begin{eqnarray}
 \frac{ \partial  I^{[C',C]Even}_{2.5}[ \rho(C') ; \rho(C) ; a(C',C) ; j(C',C) ]  }{ \partial j (C',C)}
 = \frac{ 1 }{2}   \ln \left( \frac{ a(C',C)  +j(C',C)    }{ a(C',C)  -j(C',C)     }  \right) 
\label{rate2.5masterajlinkevenderi1}
\end{eqnarray}
is of the sign of the current $j(C',C)$, while the second partial derivative remains positive
\begin{eqnarray}
 \frac{ \partial^2  I^{[C',C]Even}_{2.5}[ \rho(C') ; \rho(C) ; a(C',C) ; j(C',C) ]  }{ \partial j^2 (C',C)}
 = \frac{ a(C',C)   }{ a^2(C',C)  -j^2(C',C)     }  >0
\label{rate2.5masterajlinkevenderi2}
\end{eqnarray}

So the vanishing of the empirical currents on all the links $(C',C)$
\begin{eqnarray}
j^{opt}(C',C)=0
\label{jopt0}
\end{eqnarray} 
allows to minimize the rate function $I_{2.5}[ \rho(.) ; a(.,.) ; j(.,.) ] $ of Eq. \ref{rate2.5masterajeonly},
while the stationary constraints of Eq. \ref{contraintej} are trivially satisfied.
The physical meaning is that for any given empirical density $\rho(.) $ and any given empirical activity $a(.,.)$,
a detailed-balance dynamics prefers to remain detailed-balance even at the empirical level via the 
vanishing of all the link empirical currents (Eq. \ref{jopt0}).

As a consequence, this optimization over the current of
the Level 2.5 of Eq. \ref{level2.5masteraj} 
yields the large deviations for the joint probability of the empirical density $\rho(.)$ and of the activity $a(.,.)$,
that we will call the Level 2.25 in the present paper (just to mean that it is between the Level 2.5 
described above and the Level 2 that will be described in the next subsection)
\begin{eqnarray}
P^{[2.25]}_{T}[ \rho(.) ; a(.,.)  ]  \oppropto_{T \to +\infty} 
 \delta \left( \sum_C \rho(C) - 1 \right) 
 e^{- T I_{2.25}[ \rho(.) ; a(.,.)  ] }
\label{level2.25master}
\end{eqnarray}
where the rate function $ I_{2.25}[ \rho(.) ; a(.,.)  ] $ at Level 2.25 
is obtained from the rate function at Level 2.5 of Eqs \ref{rate2.5masteraj} and \ref{rate2.5masterajlink}
when all the empirical currents vanish $j^{opt}(C',C)=0=0 $ (Eq. \ref{jopt0})
\begin{eqnarray}
&& I_{2.25}[ \rho(.) ; a(.,.)  ]  =  I_{2.5}[ \rho(.) ; a(.,.) ; j^{opt}(.,.)=0 ] 
\nonumber \\
&& =  \sum_{C } \sum_{C' > C}
\left[ \frac{a(C',C) }{2}   \ln \left( \frac{ a^2(C',C)   }{  4 W(C',C)  \rho(C) W(C,C')  \rho(C')}  \right) 
 - a(C',C)     + W(C',C)  \rho(C)    + W(C,C')  \rho(C') \right] \ \ 
\label{rate2.25master}
\end{eqnarray}


\subsection{Explicit contraction of the Level 2.25 over the activity $ a(.,.)$ towards the Level 2 for the density $\rho(.)$   }

The optimization of the rate function at Level 2.25 of Eq. \ref{rate2.25master}
over the empirical activities $ a(C',C)$
\begin{eqnarray}
0 = \frac{\partial  I_{2.25}[ \rho(.) ; a(.,.)  ] }{ \partial a(C',C) } 
= \frac{1 }{2}   \ln \left( \frac{ a^2(C',C)   }{  4 W(C',C)  \rho(C) W(C,C')  \rho(C')}  \right) 
\label{rate2.5masterajeonly0deria}
\end{eqnarray}
lead to the optimal values
\begin{eqnarray}
a^{opt}(C',C)  = 2 \sqrt{  W(C',C)  \rho(C) W(C,C')  \rho(C')}  
\label{rate2.5masterajeonly0deriaopr}
\end{eqnarray}
that can be plugged into Eq. \ref{rate2.25master} to obtain the rate function at Level 2
\begin{eqnarray}
 I_{2}[ \rho(.)   ] = I_{2.25}[ \rho(.) ; a^{opt}(.,.)  ]  
  && =  \sum_{C } \sum_{C' > C}
\left[  W(C',C)  \rho(C)    + W(C,C')  \rho(C') - 2 \sqrt{  W(C',C)  \rho(C) W(C,C')  \rho(C')}  \right]
\nonumber \\
&& = \sum_{C } \sum_{C' > C}
\left[  \sqrt{ W(C',C)  \rho(C)  }   - \sqrt{  W(C,C')  \rho(C') }  \right]^2
\label{rate2master}
\end{eqnarray}
that will govern the large deviations properties of the probability of the empirical density $\rho(.)$ alone
\begin{eqnarray}
P^{[2]}_{T}[ \rho(.)  ]  \oppropto_{T \to +\infty} 
 \delta \left( \sum_C \rho(C) - 1 \right) 
 e^{- T I_{2}[ \rho(.)   ] }
\label{level2master}
\end{eqnarray}
The fact that the Level 2 is closed and explicit for detailed-balance Markov dynamics
is well-known since the works of Donsker and Varadhan \cite{DonskerV}.


\subsection{Contraction of the Level 2.25 over the density $\rho(.)$ to obtain the rate function for the activity $ a(.,.)$  }

A natural question is now whether on can contract the Level 2.25 of Eq. \ref{level2.25master}
over the empirical density $\rho(.)$ in order to obtain the rate function $I [   a(.,.) ]   $
 that governs the large deviations properties of the activities $a(.,.)$ alone
\begin{eqnarray}
P_{T}[  a(.,.)  ]  \oppropto_{T \to +\infty}  e^{- T I[  a(.,.)  ] }
\label{level2.25masteractivity}
\end{eqnarray}
It is useful to introduce the following notations for the total rate 
$W^{out}(C)$ out of the configuration $C$
\begin{eqnarray}
W^{out}(C) \equiv \sum_{C' \ne C} W(C',C) 
\label{WoutC}
\end{eqnarray}
and for the total activity of the links connected to the configuration $C$
\begin{eqnarray}
a^{tot}(C) \equiv \sum_{C' \ne C} a(C',C) 
\label{atotC}
\end{eqnarray}
Using the symmetry in $(C,C')$ of the activity $a(C',C)=a(C,C')$,
the rate function at Level 2.25 of Eq. \ref{rate2.25master}
can be rewritten using the notations of Eqs \ref{WoutC} and \ref{atotC}
as
\begin{eqnarray}
&& I_{2.25}[ \rho(.) ; a(.,.)  ]  
 = \frac{1}{2} \sum_{C } \sum_{C' \ne C}
 \frac{a(C',C) }{2} \left[  \ln \left( \frac{ a^2(C',C)   }{  4 W(C',C)W(C,C') }  \right) 
- \ln (\rho(C) ) - \ln (\rho(C') )  \right]
\nonumber \\
&& + \frac{1}{2} \sum_{C } \sum_{C' \ne C}
\left[  - a(C',C)     + W(C',C)  \rho(C)    + W(C,C')  \rho(C') \right] 
\nonumber \\
&&
=  \sum_{C } \sum_{C' \ne C}
 \frac{a(C',C) }{4}   \ln \left( \frac{ a^2(C',C)   }{  4 W(C',C)W(C,C') }  \right) 
-  \sum_{C }  \frac{a^{tot}(C) }{2} \ln (\rho(C) ) 
 -  \sum_{C } \frac{a^{tot}(C) }{2}  
   +  \sum_{C }  W^{out}(C)  \rho(C)   \ \ \ 
\label{rate2.25mastersym}
\end{eqnarray}

In order to optimize this rate function at Level 2.25 in the presence of the normalization constraint for the density
(Eq. \ref{level2.25master}), we consider the following Lagrangian involving the Lagrange multiplier $\omega$
\begin{eqnarray}
 \Upsilon[ \rho(.) ; a_.(.)  ]  &&  \equiv I_{2.25}[ \rho(.) ; a_.(.)  ]    +  \omega \left( 1- \sum_C \rho(C) \right)
  \nonumber \\
&& =  \sum_{C } \sum_{C' \ne C}
 \frac{a(C',C) }{4}   \ln \left( \frac{ a^2(C',C)   }{  4 W(C',C)W(C,C') }  \right) 
\nonumber \\
&&  +  \sum_{C } \left( - \frac{a^{tot}(C) }{2}  
  -    \frac{a^{tot}(C) }{2} \ln (\rho(C) ) 
   +   \left[ W^{out}(C) -\omega \right]   \rho(C) \right) +  \omega 
\label{lagrange2.25config}
\end{eqnarray}
The optimization of this Lagrangian over the empirical densities $\rho(C) $
\begin{eqnarray}
 0= \frac{ \partial  \Upsilon[ \rho(.) ; a_.(.)  ] }{ \partial \rho(C)}  
=  -    \frac{a^{tot}(C) }{2 \rho(C) }    +   \left[ W^{out}(C) -\omega \right]   
   \label{lagrange2.25configderi}
\end{eqnarray}
yields the optimal values
\begin{eqnarray}
\rho^{opt}(C)  =    \frac{a^{tot}(C) }{2  \left[ W^{out}(C) -\omega \right]   }
   \label{lagrange2.25configopti}
\end{eqnarray}
where the Lagrange multiplier $\omega$ has to be smaller than the total rate $ W^{out}(C)$ out of any configuration $C$ and has to be chosen in order to satisfy the normalization constraint
\begin{eqnarray}
1= \sum_C \rho^{opt}(C)  =  \sum_C  \frac{a^{tot}(C) }{2  \left[ W^{out}(C) -\omega \right]   }
   \label{lagrange2.25configoptinorma}
\end{eqnarray}

The rate function $I [   a(.,.) ]   $ that governs the large deviations properties of Eq. \ref{level2.25masteractivity}
for the activities $a(.,.)$ alone
corresponds to the value of the Lagrangian of Eq. \ref{lagrange2.25config}
when the empirical density $\rho(.)$ takes its optimal value $\rho^{opt}(.)  $ of Eq \ref{lagrange2.25configopti}
satisfying the normalization constraint of Eq. \ref{lagrange2.25configoptinorma} 
\begin{eqnarray}
 I[  a(.,.)  ]  && \equiv  \Upsilon[ \rho^{opt}(.) ; a(.,.)  ] 
  \nonumber \\
&& =  \sum_{C } \sum_{C' \ne C}
 \frac{a(C',C) }{4}   \ln \left( \frac{ a^2(C',C)   }{  4 W(C',C)W(C,C') }  \right) 
\nonumber \\
&&  +  \sum_{C } \left( - \frac{a^{tot}(C) }{2}  
  -    \frac{a^{tot}(C) }{2} \ln (\rho^{opt}(C) ) 
   +   \left[ W^{out}(C) -\omega \right]   \rho^{opt}(C) \right) +  \omega 
    \nonumber \\
&& =  \sum_{C } \sum_{C' \ne C}
 \frac{a(C',C) }{4}   \ln \left( \frac{ a^2(C',C)   }{  4 W(C',C)W(C,C') }  \right) 
  -  \sum_{C }     \frac{a^{tot}(C) }{2} \ln \left(  \frac{a^{tot}(C) }{2  \left[ W^{out}(C) -\omega \right]   } \right) 
  +  \omega 
\label{lagrange2.25configoptimal}
\end{eqnarray}
with the notations of Eqs \ref{WoutC} and \ref{atotC}.
However this rate function remains somewhat implicit since the Lagrange multiplier $\omega $
is defined via Eq. \ref{lagrange2.25configoptinorma}.
To see more clearly the physical meaning,
one can use the equilibrium state $P_{eq}(C)$ and the equilibrium activities
\begin{eqnarray}
A_{eq}(C',C) = 2 W(C',C) P_{eq}(C) = 2 W(C,C') P_{eq}(C') = A_{eq}(C,'C)
\label{AeqDB}
\end{eqnarray}
with the corresponding total equilibrium activities of Eq. \ref{atotC}
\begin{eqnarray}
A^{tot}_{eq}(C) = \sum_{C' \ne C} A_{eq}(C',C) = 2 W^{out}(C) P_{eq}(C)
\label{atotCeq}
\end{eqnarray}
in order to rewrite the rates as
\begin{eqnarray}
W(C',C) && = \frac{A_{eq}(C',C) }{ 2 P_{eq}(C) }
\nonumber \\
W^{out}(C) && = \frac{A^{tot}_{eq}(C) }{ 2 P_{eq}(C) }
\label{WAeqDB}
\end{eqnarray}
Then Eq. \ref{lagrange2.25configoptinorma} for the Lagrange multiplier $\omega $ reads
\begin{eqnarray}
1=   \sum_C P_{eq}(C) \  \frac{a^{tot}(C) }{ A^{tot}_{eq}(C) - 2 \omega P_{eq}(C)   }
   \label{lagrange2.25configoptinormast}
\end{eqnarray}
while the rate function of Eq. \ref{lagrange2.25configoptimal}
becomes
\begin{eqnarray}
 I[  a(.,.)  ]  && =  \sum_{C } \sum_{C' \ne C}
 \frac{a(C',C) }{4}   \ln \left( \frac{ a^2(C',C)   }{  A^2_{eq}(C',C)  }  \right) 
  -  \sum_{C }     \frac{a^{tot}(C) }{2} \ln \left( \frac{a^{tot}(C) }{  \left[ A^{tot}_{eq}(C) - 2 \omega P_{eq}(C) \right]   } \right) 
  +  \omega 
\label{lagrange2.25configoptimaleq}
\end{eqnarray}
where it is now obvious that the value $\omega=0$ is associated to the equilibrium where the rate function vanishes.


\subsection{ Comparison with the simplifications for the large deviations of detailed-balance diffusion processes }

As a final remark, it is useful to mention the similar simplifications for the large deviations properties
of detailed-balance diffusion processes, despite some technical differences.

\subsubsection{ Reminder on the Level 2.5 for diffusion processes  }

For diffusion processes described by the Fokker-Planck Equation
in the force field $\vec F(\vec x)$ in dimension $d$, with diffusion coefficient $D(\vec x)$
\begin{eqnarray}
\frac{ \partial P_t(\vec x)   }{\partial t}   =  -   \vec \nabla .  \left[ P_t(\vec x )   \vec F(\vec x ) 
-D (\vec x) \vec \nabla   P_t(\vec x)  \right]
\label{fokkerplanck}
\end{eqnarray}
the large deviations at Level 2.5 involve 
the empirical density
\begin{eqnarray}
 \rho(\vec x) && \equiv \frac{1}{T} \int_0^T dt \  \delta^{(d)} ( \vec x(t)- \vec x)  
\label{rhodiff}
\end{eqnarray}
satisfying the normalization
\begin{eqnarray}
\int d^d \vec x \ \rho (\vec x) && = 1
\label{rho1ptnormadiff}
\end{eqnarray}
and the empirical current $\vec j(\vec x)$ 
\begin{eqnarray} 
\vec j(\vec x) \equiv   \frac{1}{T} \int_0^T dt \ \frac{d \vec x(t)}{dt}   \delta^{(d)}( \vec x(t)- \vec x)  
\label{diffjlocal}
\end{eqnarray}
that should be divergence-free 
\begin{eqnarray}
 \vec \nabla . \vec j(\vec x) =0
\label{i2.5diffusion}
\end{eqnarray}
in order to ensure the stationarity.

The joint distribution of the empirical density $\rho(.)$ and the empirical current $\vec j(\vec x)$
follows the following large deviation form \cite{wynants_thesis,maes_diffusion,chetrite_formal,engel,chetrite_HDR,c_reset,c_lyapunov,c_inference,c_metastable}
\begin{eqnarray}
 P^{[2.5]}_T[ \rho(.), \vec j(.)]   \opsimeq_{T \to +\infty}  \delta \left(\int d^d \vec x \rho(\vec x) -1  \right)
\left[ \prod_{\vec x }  \delta \left(  \vec \nabla . \vec j(\vec x) \right) \right] 
e^{- \displaystyle I_{2.5}[ \rho(.), \vec j(.)]
 }
\label{ld2.5diff}
\end{eqnarray}
where the constitutive constraints have been discussed in Eqs \ref{rho1ptnormadiff}
and \ref{i2.5diffusion},
while the rate function is simply Gaussian with respect to the empirical current $\vec j(\vec x) $
\begin{eqnarray}
 I_{2.5}[ \rho(.), \vec j(.)]
=
\int \frac{d^d \vec x}{ 4 D (\vec x) \rho(\vec x) } \left[ \vec j(\vec x) - \rho(\vec x) \vec F(\vec x)+D (\vec x) \vec \nabla \rho(\vec x) \right]^2
\label{rate2.5diff}
\end{eqnarray}
As a consequence, the decomposition into even and odd contributions with respect to the current $\vec j(\vec x)$
\begin{eqnarray}
 I_{2.5}[ \rho(.), \vec j(.)] = I^{Even}_{2.5}[ \rho(.), \vec j(.)] + I^{Odd}_{2.5}[ \rho(.), \vec j(.)]
\label{rate2.5diffevenodd}
\end{eqnarray}
involves the even contribution
\begin{eqnarray}
 I^{Even}_{2.5}[ \rho(.), \vec j(.)] = \int \frac{d^d \vec x}{ 4 D (\vec x) \rho(\vec x) } 
 \left( \vec j^2(\vec x) + \left[  - \rho(\vec x) \vec F(\vec x)+D (\vec x) \vec \nabla \rho(\vec x) \right]^2 \right)
\label{rate2.5diffeven}
\end{eqnarray}
while the odd contribution is linear with respect to the current $\vec j(\vec x)$
\begin{eqnarray}
 I^{Odd}_{2.5}[ \rho(.), \vec j(.)] = \frac{1}{2} \int d^d \vec x \ 
  \vec j(\vec x) . \left[  -  \frac{ \vec F(\vec x)}{ D (\vec x)} + \vec \nabla \ln (\rho(\vec x) ) \right]
\label{rate2.5diffodd}
\end{eqnarray}

\subsubsection{ Simplifications when diffusion processes satisfy Detailed-Balance  }

For the Fokker-Planck of Eq. \ref{fokkerplanck},
the detailed-balance condition corresponds to the vanishing of the steady current
\begin{eqnarray}
0= P_{eq}(\vec x )   \vec F(\vec x ) -D (\vec x) \vec \nabla   P_{eq}(\vec x)  
\label{fokkerplanckDB}
\end{eqnarray}
 in the equilibrium state $P_{eq}(\vec x )$ in the potential $U(\vec x)$ at inverse temperature $\beta$
\begin{eqnarray}
 P_{eq}(\vec x )  = \frac{e^{- \beta U(\vec x)}}{Z}
\label{fokkerplanckDBeq}
\end{eqnarray}
i.e. the force $ \vec F(\vec x ) $ should be of the form
\begin{eqnarray}
   \vec F(\vec x ) = D (\vec x) \vec \nabla  \ln( P_{eq}(\vec x)  ) = - \beta D (\vec x) \vec \nabla U(\vec x)
\label{forcefokkerplanckDB}
\end{eqnarray}
Plugging this detailed-balance force into Eq. \ref{rate2.5diffodd},
one obtains after an integration by parts and using the stationarity constraint of Eq. \ref{i2.5diffusion}
that the odd contribution of Eq. \ref{rate2.5diffodd} vanishes
\begin{eqnarray}
 I^{Odd}_{2.5}[ \rho(.), \vec j(.)] && = \frac{1}{2} \int d^d \vec x \ 
  \vec j(\vec x) . \vec \nabla \left[    \beta   U(\vec x) +  \ln (\rho(\vec x) ) \right]
  \nonumber \\
&& = -   \frac{1}{2} \int d^d \vec x \ 
  \left[    \beta   U(\vec x) +  \ln (\rho(\vec x) ) \right]  \vec \nabla . \vec j(\vec x) =0
\label{rate2.5diffodd0}
\end{eqnarray}
As already mentioned after Eq. \ref{rate2.5masterajlinkodddbsum} concerning the analog property for Markov jump
processes, the physical meaning of this vanishing contribution is that a detailed-balance dynamics
 cannot have a global irreversible property.

So the rate function at Level 2.5 of Eq. \ref{rate2.5diffevenodd}
reduces to the even contribution of Eq. \ref{rate2.5diffeven} with the force of Eq. \ref{fokkerplanckDB}
\begin{eqnarray}
 I_{2.5}[ \rho(.), \vec j(.)] = I^{Even}_{2.5}[ \rho(.), \vec j(.)]  
 = \frac{1}{4} \int d^d \vec x \  D (\vec x) \rho(\vec x)  
 \left( \frac{ \vec j^2(\vec x) }{ D^2 (\vec x) \rho^2(\vec x)} 
 + \left[    \beta  \vec \nabla U(\vec x)+ \vec \nabla \ln( \rho(\vec x) ) \right]^2 \right)
\label{rate2.5diffevenDB}
\end{eqnarray}
As a consequence, when the empirical density $\rho(.) $ is given,
any configuration of the empirical current $\vec j(.)$ that satisfies the stationary constraint of Eq. \ref{i2.5diffusion}
has the same rate function as the configuration with the reversed empirical current $\vec j(.)$ that also satisfies the stationary constraint 
\begin{eqnarray}
I_{2.5}[ \rho(.), \vec j(.)]  =  I_{2.5}[ \rho(.), - \vec j(.)] 
\label{rate2.5diffevenREVERSED}
\end{eqnarray}

When the empirical density $\rho( \vec x)$ is given, the rate function of Eq. \ref{rate2.5diffevenDB}
is minimized 
when the empirical current vanishes everywhere 
\begin{eqnarray}
  \vec j_{opt} (\vec x) = \vec 0
 \label{joptdiff}
\end{eqnarray}
while the stationarity constraint of Eq. \ref{i2.5diffusion} is trivially satisfied.
As already mentioned after Eq. \ref{jopt0} concerning the analog property for Markov jump processes,
the physical meaning is that for any given empirical density $\rho(.) $ 
a detailed-balance diffusion process prefers to remain detailed-balance even at the empirical level via the 
vanishing of the empirical current everywhere (Eq. \ref{joptdiff}).

As a consequence, the contraction of the Level 2.5 of Eq. \ref{ld2.5diff}
over the empirical current $\vec j(\vec x) $ is explicit via the optimal solution of Eq. \ref{joptdiff}
that leads to the Level 2 for the empirical density $\rho( \vec x)$ alone
\begin{eqnarray}
 P^{[2]}_T[ \rho(.)]   \opsimeq_{T \to +\infty}  \delta \left(\int d^d \vec x \rho(\vec x) -1  \right)
e^{- \displaystyle I_{2}[ \rho(.)]
 }
\label{ld2diff}
\end{eqnarray}
with the rate function at Level 2 obtained from the rate function at Level 2.5 of Eq. \ref{rate2.5diffevenDB}
for vanishing current $\vec j_{opt} (\vec x) = \vec 0 $
\begin{eqnarray}
 I_{2}[ \rho(.)] =
   I_{2.5}[ \rho(.), \vec j(.)= \vec 0]
=\frac{1}{4} \int d^d \vec x \  D (\vec x) \rho(\vec x)  
  \left[    \beta  \vec \nabla U(\vec x)+ \vec \nabla \ln( \rho(\vec x) ) \right]^2 
\label{rate2diff}
\end{eqnarray}
So here the contraction of the Level 2.5 over the current gives directly the Level 2 for the empirical density,
since there are no 'activity degrees of freedom' in diffusion processes, in contrast to the Markov jump processes described previously.


\section{ Large deviations in the space of the $2^N$ configurations of the random soft East model  }

\label{app_configEast}

In this Appendix, the large deviations at various levels for Detailed-Balance Markov Jump processes summarized in the previous Appendix are applied in the space of the $2^N$ configurations of the random soft East model (Eq. \ref{weastrandomsoft}).

\subsection{ Application of Level 2.5 in the space of the $2^N$ configurations of the random soft East model}

For a trajectory $\{ S_1(t),...,S_N(t)\}$ of the $N$ spins over the large time-window $0 \leq t \leq T$,
the empirical time-averaged density of Eq. \ref{rhoc}
\begin{eqnarray}
 \rho(S_1,...,S_N) && \equiv \frac{1}{T} \int_0^T dt \  \prod_{n=1}^N \delta_{S_n(t),S_n}  
 \label{rhoconfig}
\end{eqnarray}
satisfies the normalization of Eq. \ref{rhocnorma}
\begin{eqnarray}
\sum_{S_1=\pm} ... \sum_{S_N=\pm}  \rho(S_1,...,S_N)  = 1
\label{rho1normaconfig}
\end{eqnarray}

The empirical flows of Eq. \ref{jumpempiricaldensity} associated to the flip rates $w_{i}^{S_i} (S_{i-1}) $ of the model
read 
\begin{eqnarray}
q_{i}^{S_i} (S_1,..,S_{i-1} ; S_{i+1} ,..,S_N)
\equiv  \frac{1}{T}    \sum_{t \in [0,T] : \substack{ S_i(t)=S_i   \\ S_i(t^+)= - S_i}} 
\left[ \prod_{n=1}^{i-1} \delta_{S_n(t),S_n}  \right] \left[ \prod_{p=i+1}^{N} \delta_{S_p(t),S_p}  \right] 
\label{qbulk}
\end{eqnarray}
The stationarity constraint of Eq. \ref{contrainteq} for the configuration $(S_1,...,S_N)$ reads
\begin{eqnarray}
 0  =  \sum_{i=1 }^{N} 
  \left[ q_{i}^{-S_i}  (S_1,..,S_{i-1} ; S_{i+1} ,..,S_N)
 - q_{i}^{S_i}  (S_1,..,S_{i-1} ; S_{i+1} ,..,S_N)  \right]
\label{contrainteqs}
\end{eqnarray}

The rate function of Eq. \ref{rate2.5master}
that governs the large deviations at Level 2.5 of Eq. \ref{level2.5master} 
reads 
\begin{footnotesize}
\begin{eqnarray}
&& I_{2.5}[ \rho(.) ; q_.^{\pm}(.) ]
 =  \sum_{S_1 = \pm}  \sum_{S_2 = \pm} ... \sum_{S_N = \pm}
\nonumber \\
&& \sum_{i=1}^{N}
\bigg[ q_{i}^{S_i} (..,S_{i-1} ; S_{i+1} ,..)  \ln \left( \frac{q_{i}^{S_i} (..,S_{i-1} ; S_{i+1} ,..) }
{  w_{i}^{S_i} (S_{i-1})   \rho(..S_{i-1},S_i,S_{i+1}..) }  \right) 
- q_{i}^{S_i} (..,S_{i-1} ; S_{i+1} ,..)  + w_{i}^{S_i} (S_{i-1})   \rho(..S_{i-1},S_i,S_{i+1}..)  \bigg]
\label{rate2.5}
\end{eqnarray}
\end{footnotesize}

The parametrizations of Eq. \ref{ajreci} for the empirical flows
\begin{eqnarray}
q_{i}^{+} (..,S_{i-1} ; S_{i+1} ,..) \equiv \frac{a_{i} (..,S_{i-1} ; S_{i+1} ,..)  +j_{i} (..,S_{i-1} ; S_{i+1} ,..) }{2} 
\nonumber \\
q_{i}^{-} (..,S_{i-1} ; S_{i+1} ,..) \equiv \frac{a_{i} (..,S_{i-1} ; S_{i+1} ,..)  - j_{i} (..,S_{i-1} ; S_{i+1} ,..) }{2} 
\label{ajreciEast}
\end{eqnarray}
in terms of the activities and the currents of Eq. \ref{aj}
\begin{eqnarray}
a_{i} (..,S_{i-1} ; S_{i+1} ,..)  \equiv q_{i}^{+} (..,S_{i-1} ; S_{i+1} ,..) + q_{i}^{-} (..,S_{i-1} ; S_{i+1} ,..)
\nonumber \\
j_{i} (..,S_{i-1} ; S_{i+1} ,..)  \equiv q_{i}^{+} (..,S_{i-1} ; S_{i+1} ,..) - q_{i}^{-} (..,S_{i-1} ; S_{i+1} ,..)
\label{ajEast}
\end{eqnarray}
allows to rewrite the stationarity constraints of Eq. \ref{contrainteqs} in terms of the currents only (Eq. \ref{contraintej})
 \begin{eqnarray}
 0  =  \sum_{i=1 }^{N}   j_{i}  (S_1,..,S_{i-1} ; S_{i+1} ,..,S_N)
\label{contraintejEast}
\end{eqnarray}


\subsection{ Application of Level 2.25 in the space of the $2^N$ configurations of the random soft East model}

As explained in detail in Appendix \ref{app_LargeDevDB}, the detailed-balance property satisfied by the rates
allows to make the explicit contraction of the Level 2.5 over the empirical currents
via the simple optimal solution where the empirical currents on all the links (Eq. \ref{jopt0}) 
\begin{eqnarray}
 j^{opt}_{i} (..,S_{i-1} ; S_{i+1} ,..)=0
\label{jopt0East}
\end{eqnarray} 
and leads to the Level 2.25 of Eq. \ref{level2.25master} for the empirical density and the empirical activities
\begin{eqnarray}
P^{[2.25]}_{T}[ \rho(.) ; a_.(.)  ]  \oppropto_{T \to +\infty} 
 \delta \left( \sum_{S_1=\pm} ... \sum_{S_N=\pm}  \rho(S_1,...,S_N) - 1 \right) 
 e^{- T I_{2.25}[ \rho(.) ; a(.)  ] }
\label{level2.25masterEast}
\end{eqnarray}
with the rate function
of Eq \ref{rate2.25master}
\begin{eqnarray}
&& I_{2.25}[ \rho(.) ; a_.(.)  ]   =  \sum_{i=1}^{N} \left[ \prod_{k \ne i}  \sum_{S_k = \pm}  \right]
 \nonumber \\ &&
\bigg[ \frac{a_{i} (..,S_{i-1} ; S_{i+1} ,..) }{2}   \ln \left( \frac{ a^2_{i} (..,S_{i-1} ; S_{i+1} ,..)   }
{  4 w_{i}^+ (S_{i-1})  \rho (..,S_{i-1},+, S_{i+1},..) 
w_{i}^{-} (S_{i-1})  \rho (..,S_{i-1},-, S_{i+1},..)}  \right) 
\nonumber \\
&& - a_{i} (..,S_{i-1} ; S_{i+1} ,..)     + w_{i}^{+} (S_{i-1})  \rho (..,S_{i-1},+, S_{i+1},..,)   
 + w_{i}^{-}(S_{i-1})   \rho(..,S_{i-1},-, S_{i+1},..) \bigg]
\label{rate2.25rhoaEast}
\end{eqnarray}


\subsection{ Application of Level 2 in the space of the $2^N$ configurations of the random soft East model}

Finally, the Level 2 for the empirical density alone of Eq. \ref{level2master}
\begin{eqnarray}
P^{[2]}_{T}[ \rho(.)   ]  \oppropto_{T \to +\infty} 
 \delta \left( \sum_{S_1=\pm} ... \sum_{S_N=\pm}  \rho(S_1,...,S_N) - 1 \right) 
 e^{- T I_{2}[ \rho(.)  ] }
\label{level2masterEast}
\end{eqnarray}
involves the rate function at Level 2 of Eq. \ref{rate2master}
\begin{eqnarray}
 I_{2}[ \rho(.)   ]  = \sum_{i=1}^{N} \left[ \prod_{k \ne i}  \sum_{S_k = \pm}  \right]
\left[  \sqrt{ w_{i}^{+} (S_{i-1})  \rho (..,S_{i-1},+, S_{i+1},..,)  }   - \sqrt{  w_{i}^{-}(S_{i-1})   \rho(..,S_{i-1},-, S_{i+1},..) }  \right]^2
\label{rate2masterEast}
\end{eqnarray}


\section{ Random Soft East Model : Contraction of the global Level 2.25 towards the local Level 2.25   }

\label{app_GlobalLocal}

In this Appendix, we describe the explicit contraction from the Level 2.25 of Eq. \ref{level2.25masterEast}
in the space of the $2^N$ configurations of the random soft East model
towards the Level 2.25 for the local densities and the local activities,
as given by Eq. \ref{level2.25e} of the main text.

\subsection{ Local empirical observables from global empirical observables in the space of the $2^N$ configurations   }

The empirical 2-spin density $\rho_{i-1,i}^{S_{i-1},S_{i}} $ 
of Eq. \ref{rho2}
can be obtained from the configuration empirical density $\rho(S_1,...,S_N) $ 
of Eq. \ref{rhoconfig} by summing over the $(N-2)$ other spins $n \ne (i-1,i)$
\begin{eqnarray}
\rho_{i-1,i}^{S_{i-1},S_{i}} 
= \left[ \prod_{n=1}^{i-2} \sum_{S_n=\pm}  \right] \left[ \prod_{p=i+1}^{N}\sum_{S_p=\pm}   \right]  \rho(S_1,...,S_N)
\label{rholocalfromglobal}
\end{eqnarray}

Similarly the local empirical activities $a_i(S_{i-1})$ of Eq. \ref{ajeast}
 can be obtained from the configuration activities $a_{i} (S_1,..,S_{i-1} ; S_{i+1} ,..,S_N) $ 
 of Eq. \ref{ajEast} by summing over the $(N-2)$ other spins $n \ne (i-1,i)$
\begin{eqnarray}
a_i(S_{i-1})  =  \left[ \prod_{n=1}^{i-2} \sum_{S_n=\pm}  \right] \left[ \prod_{p=i+1}^{N}\sum_{S_p=\pm}   \right]
a_{i} (..,S_{i-1} ; S_{i+1} ,..)
\label{alocalfromglobal}
\end{eqnarray}


\subsection{ Contraction of the global Level 2.25 with constraints fixing the local empirical observables    }

In order to optimize the rate function at Level 2.25 of Eq. \ref{rate2.25rhoaEast}
over the configuration empirical density $\rho(S_1,...,S_N) $
and over the configuration empirical activities $a_{i} (..,S_{i-1} ; S_{i+1} ,..) $
in the presence of the constraints of Eq. \ref{rholocalfromglobal}
and Eq. \ref{alocalfromglobal},
we consider the following Lagrangian involving the Lagrange multipliers $[\omega_{i-1,i}^{\pm,\pm}  ; \lambda_i(\pm)]$ for $i=1,..,N$
\begin{eqnarray}
 \Upsilon[ \rho(.) ; a_.(.)  ]  &&  \equiv I_{2.25}[ \rho(.) ; a_.(.)  ]   
 + \sum_{i=1}^N \sum_{S_{i-1}=\pm}\sum_{S_{i}=\pm} \omega_{i-1,i}^{S_{i-1},S_{i}} 
\left(
\rho_{i-1,i}^{S_{i-1},S_{i}} - \left[ \prod_{n=1}^{i-2} \sum_{S_n=\pm}  \right] \left[ \prod_{p=i+1}^{N}\sum_{S_p=\pm}   \right]  \rho(S_1,...,S_N)
\right)
  \nonumber \\
&& + \sum_{i=1}^N \sum_{S_{i-1}=\pm}\lambda_i(S_{i-1}) \left(  a_i(S_{i-1})  -  \left[ \prod_{n=1}^{i-2} \sum_{S_n=\pm}  \right] \left[ \prod_{p=i+1}^{N}\sum_{S_p=\pm}   \right] a_{i} (..,S_{i-1} ; S_{i+1} ,..) \right)
\label{lagrange2.25def}
\end{eqnarray}
The explicit rate function at Level 2.25 of Eq. \ref{rate2.25rhoaEast} yields
\begin{eqnarray}
&& \Upsilon[ \rho(.) ; a_.(.)  ]  =  \sum_{i=1}^{N} \left[ \prod_{k \ne i}  \sum_{S_k = \pm}  \right]
\bigg[ \frac{a_{i} (..,S_{i-1} ; S_{i+1} ,..) }{2}   \ln \left( \frac{ a^2_{i} (..,S_{i-1} ; S_{i+1} ,..)   }
{  4 w_{i}^+ (S_{i-1})  \rho (..,S_{i-1},+, S_{i+1},..) 
w_{i}^{-} (S_{i-1})  \rho (..,S_{i-1},-, S_{i+1},..)}  \right) 
\nonumber \\
&& - a_{i} (..,S_{i-1} ; S_{i+1} ,..)     + w_{i}^{+} (S_{i-1})  \rho (..,S_{i-1},+, S_{i+1},..,)   
 + w_{i}^{-}(S_{i-1})   \rho(..,S_{i-1},-, S_{i+1},..) \bigg]
 \nonumber \\
&& + \sum_{i=1}^N \sum_{S_{i-1}=\pm}\sum_{S_{i}=\pm} \omega_{i-1,i}^{S_{i-1},S_{i}} 
\left(
\rho_{i-1,i}^{S_{i-1},S_{i}} - \left[ \prod_{n=1}^{i-2} \sum_{S_n=\pm}  \right] \left[ \prod_{p=i+1}^{N}\sum_{S_p=\pm}   \right]  \rho(S_1,...,S_N)
\right)
  \nonumber \\
&& + \sum_{i=1}^N \sum_{S_{i-1}=\pm}\lambda_i(S_{i-1}) \left(  a_i(S_{i-1})  -  \left[ \prod_{n=1}^{i-2} \sum_{S_n=\pm}  \right] \left[ \prod_{p=i+1}^{N}\sum_{S_p=\pm}   \right] a_{i} (..,S_{i-1} ; S_{i+1} ,..) \right)
\label{lagrange2.25}
\end{eqnarray}

The optimization of this Lagrangian over the configuration activities $a_{i} (..,S_{i-1} ; S_{i+1} ,..)  $
\begin{eqnarray}
0= \frac{ \partial  \Upsilon[ \rho(.) ; a_.(.)  ] }{ \partial a_{i} (..,S_{i-1} ; S_{i+1} ,..)}  
=   \frac{1 }{2}   \ln \left( \frac{ a^2_{i} (..,S_{i-1} ; S_{i+1} ,..)   }
{  4 w_{i}^+ (S_{i-1})  \rho (..,S_{i-1},+, S_{i+1},..) 
w_{i}^{-} (S_{i-1})  \rho (..,S_{i-1},-, S_{i+1},..)}  \right) 
 - \lambda_i(S_{i-1}) 
\label{lagrange2.25deria}
\end{eqnarray}
leads to the optimal values
\begin{eqnarray}
  a^{opt}_{i} (..,S_{i-1} ; S_{i+1} ,..)   
= 2 e^{ \lambda_i(S_{i-1})} \sqrt{  w_{i}^+ (S_{i-1})  \rho (..,S_{i-1},+, S_{i+1},..) 
w_{i}^{-} (S_{i-1})  \rho (..,S_{i-1},-, S_{i+1},..) }
\label{aconfopt}
\end{eqnarray}
where the Lagrange multipliers $ \lambda_i(S_{i-1})$ have to be chosen to satisfy the corresponding constraints of Eq. \ref{alocalfromglobal}
\begin{eqnarray}
 a_i(S_{i-1}) && =  \left[ \prod_{n=1}^{i-2} \sum_{S_n=\pm}  \right] \left[ \prod_{p=i+1}^{N}\sum_{S_p=\pm}   \right]
a_{i}^{opt} (..,S_{i-1} ; S_{i+1} ,..)
\nonumber \\
&& = 2 e^{ \lambda_i(S_{i-1})}  \sqrt{  w_{i}^+ (S_{i-1}) w_{i}^{-} (S_{i-1})  } 
 D_i(S_{i-1}) 
\label{alocalopt}
\end{eqnarray}
where we have introduced the notation
\begin{eqnarray}
D_i(S_{i-1})  \equiv  \left[ \prod_{n=1}^{i-2} \sum_{S_n=\pm}  \right] \left[ \prod_{p=i+1}^{N}\sum_{S_p=\pm}   \right]
 \sqrt{   \rho (..,S_{i-1},+, S_{i+1},..) 
  \rho (..,S_{i-1},-, S_{i+1},..) }
\label{defd}
\end{eqnarray}
The optimal values of Eq. \ref{aconfopt} can be thus rewritten as
\begin{eqnarray}
  a^{opt}_{i} (..,S_{i-1} ; S_{i+1} ,..)   
= a_i(S_{i-1})  
\frac{ \sqrt{   \rho (..,S_{i-1},+, S_{i+1},..)   \rho (..,S_{i-1},-, S_{i+1},..) } }
{D_i(S_{i-1}) }
\label{aconfoptsanslambda}
\end{eqnarray}
The optimization of the Lagrangian of Eq. \ref{lagrange2.25}
over the configuration density $\rho(S_1,...,S_N) $
\begin{eqnarray}
0= \frac{ \partial  \Upsilon[ \rho(.) ; a_.(.)  ] }{ \partial \rho(S_1,...,S_N)}  
=  - \frac{1}{2  \rho(S_1,...,S_N)} \sum_{i=1}^{N} a_{i} (..,S_{i-1} ; S_{i+1} ,..)
 +  \sum_{i=1}^N \left( w_{i}^{S_i} (S_{i-1})  -  \omega_{i-1,i}^{S_{i-1},S_{i}} \right)
\label{lagrange2.25derirho}
\end{eqnarray}
yields that the optimal values $\rho^{opt}(S_1,...,S_N) $ have to satisfy together with
the optimal values $a^{opt}_{i} (..,S_{i-1} ; S_{i+1} ,..) $ of the activities of Eq. \ref{aconfoptsanslambda}
\begin{eqnarray}
0=   \sum_{i=1}^{N} \left[ - \frac{ a_{i}^{opt} (..,S_{i-1} ; S_{i+1} ,..) }{2} 
 +   \left( w_{i}^{S_i} (S_{i-1})  -  \omega_{i-1,i}^{S_{i-1},S_{i}} \right)\rho^{opt}(S_1,...,S_N) \right]
\label{rhoconfopt}
\end{eqnarray}
The simplest way to satisfy this equation for the sum of $N$ terms
is to impose the vanishing of each term for $i=1,..,N$ 
\begin{eqnarray}
0=   - \frac{ a_{i}^{opt} (..,S_{i-1} ; S_{i+1} ,..) }{2} 
 +   \left( w_{i}^{S_i} (S_{i-1})  -  \omega_{i-1,i}^{S_{i-1},S_{i}} \right)\rho^{opt}(S_1,...,S_N) 
\label{rhoconfopti}
\end{eqnarray}
The summation over the spins $S_k=\pm$ for $k =1,..,i-2$ and $k=i+1,..,N$ yields using the constraints of 
Eqs \ref{rholocalfromglobal} and \ref{alocalfromglobal}
\begin{eqnarray}
0=   - \frac{ a_{i} (S_{i-1}) }{2} 
 +   \left( w_{i}^{S_i} (S_{i-1})  -  \omega_{i-1,i}^{S_{i-1},S_{i}} \right)\rho_{i-1,i}^{S_{i-1},S_{i}} 
\label{rhoconfoptisum}
\end{eqnarray}
so the Lagrange multipliers $\omega_{i-1,i}^{S_{i-1},S_{i}} $ that modify the true rates $w_{i}^{S_i} (S_{i-1}) $
to produce the effective rates $\left( w_{i}^{S_i} (S_{i-1})  -  \omega_{i-1,i}^{S_{i-1},S_{i}} \right) $ can be computed from the ratios
\begin{eqnarray}
\left( w_{i}^{S_i} (S_{i-1})  -  \omega_{i-1,i}^{S_{i-1},S_{i}} \right)=    \frac{ a_{i} (S_{i-1}) }{2 \rho_{i-1,i}^{S_{i-1},S_{i}}}  
\label{omegaeff}
\end{eqnarray}
Eq. \ref{rhoconfopti} then yields the optimal density
\begin{eqnarray}
\rho^{opt}(S_1,...,S_N) =  \rho_{i-1,i}^{S_{i-1},S_{i}}  \ \frac{ a_{i}^{opt} (..,S_{i-1} ; S_{i+1} ,..) }{ a_{i} (S_{i-1})} 
\label{rhoconfoptifinal}
\end{eqnarray}
that can be plugged into Eq. \ref{defd} to obtain with the use of the constraint of Eq. \ref{alocalfromglobal}
\begin{eqnarray}
D_i(S_{i-1}) && =  \left[ \prod_{n=1}^{i-2} \sum_{S_n=\pm}  \right] \left[ \prod_{p=i+1}^{N}\sum_{S_p=\pm}   \right]
 \sqrt{   \rho^{opt} (..,S_{i-1},+, S_{i+1},..) 
  \rho^{opt} (..,S_{i-1},-, S_{i+1},..) }
  \nonumber \\
  && = \frac{ \sqrt{    \rho_{i-1,i}^{S_{i-1},+}    \rho_{i-1,i}^{S_{i-1},-}   } }{ a_{i} (S_{i-1}) }
     \left[ \prod_{n=1}^{i-2} \sum_{S_n=\pm}  \right] \left[ \prod_{p=i+1}^{N}\sum_{S_p=\pm}   \right]
   a_{i}^{opt} (..,S_{i-1} ; S_{i+1} ,..) 
  = \sqrt{    \rho_{i-1,i}^{S_{i-1},+}    \rho_{i-1,i}^{S_{i-1},-}   }
\label{defdopt}
\end{eqnarray}
so that
 the Lagrange multipliers $ \lambda_i(S_{i-1})$ of Eq. \ref{alocalopt}
 reduce to
\begin{eqnarray}
 \lambda_i(S_{i-1}) = \ln \left(  \frac{ a_i(S_{i-1}) } { 2  \sqrt{  w_{i}^+ (S_{i-1}) 
w_{i}^{-} (S_{i-1})  \rho_{i-1,i}^{S_{i-1},+}    \rho_{i-1,i}^{S_{i-1},-}  } } \right)
= \frac{1}{2}  \ln \left(  \frac{ a^2_i(S_{i-1}) } { 4 w_{i}^+ (S_{i-1}) 
w_{i}^{-} (S_{i-1})  \rho_{i-1,i}^{S_{i-1},+}    \rho_{i-1,i}^{S_{i-1},-}   } \right)
\label{lagrangesol}
\end{eqnarray}

The optimal value of the Lagrangian of Eq. \ref{lagrange2.25}
corresponding to the optimal solution $[ \rho^{opt}(.) ; a^{opt}_.(.)  ] $ satisfying the constraints
reads using Eq. \ref{lagrangesol}
\begin{eqnarray}
&&  \Upsilon^{opt} \equiv  \Upsilon[ \rho^{opt}(.) ; a^{opt}_.(.)  ] 
\label{lagrange2.25opt} \\
 &&
 =  \sum_{i=1}^{N} \left[ \prod_{k \ne i}  \sum_{S_k = \pm}  \right]
 \frac{a^{opt}_{i} (..,S_{i-1} ; S_{i+1} ,..) }{2}   \ln \left( \frac{ [ a^{opt}_{i} (..,S_{i-1} ; S_{i+1} ,..)]^2   }
{  4 w_{i}^+ (S_{i-1})  \rho^{opt} (..,S_{i-1},+, S_{i+1},..) 
w_{i}^{-} (S_{i-1})  \rho^{opt} (..,S_{i-1},-, S_{i+1},..)}  \right) 
\nonumber \\
&&
+ \sum_{i=1}^{N} \left[ \prod_{k \ne i}  \sum_{S_k = \pm}  \right]
\bigg[ - a^{opt}_{i} (..,S_{i-1} ; S_{i+1} ,..)     + w_{i}^{+} (S_{i-1})  \rho^{opt} (..,S_{i-1},+, S_{i+1},..,)   
 + w_{i}^{-}(S_{i-1})   \rho^{opt}(..,S_{i-1},-, S_{i+1},..) \bigg]
  \nonumber \\
 &&
 =  \sum_{i=1}^{N} \sum_{S_{i-1} = \pm}  
\left[ a_{i} (S_{i-1})
\lambda_i(S_{i-1})
 - a_{i} (S_{i-1})     + w_{i}^{+} (S_{i-1})  \rho_{i-1,i}^{S_{i-1},+}
 + w_{i}^{-}(S_{i-1})   \rho_{i-1,i}^{S_{i-1},-} \right]
   \nonumber \\
 &&
 =  \sum_{i=1}^{N} \sum_{S_{i-1} = \pm}  
\left[ 
\frac{a_{i} (S_{i-1})}{2}  \ln \left(  \frac{ a^2_i(S_{i-1}) } { 4 w_{i}^+ (S_{i-1}) 
w_{i}^{-} (S_{i-1})  \rho_{i-1,i}^{S_{i-1},+}    \rho_{i-1,i}^{S_{i-1},-}   } \right)
 - a_{i} (S_{i-1})     + w_{i}^{+} (S_{i-1})  \rho_{i-1,i}^{S_{i-1},+}
 + w_{i}^{-}(S_{i-1})   \rho_{i-1,i}^{S_{i-1},-} \right]
\nonumber 
\end{eqnarray}
This optimal value $ \Upsilon^{opt} $
corresponds to the rate function $I_{2.25} [  a_.(.) ;  \rho_{.,.}^{.,.}]   $
 at Level 2.25 for the local activities $a_{i} (S_{i-1})  $ 
and the local densities $\rho_{i-1,i}^{S_{i-1},S_i} $ as given in Eq. \ref{rate2.25e} of the main text.
\begin{eqnarray}
 \Upsilon^{opt} = I_{2.25} [  a_.(.) ;  \rho_{.,.}^{.,.}]  
 \label{upsilonopt}
 \end{eqnarray}


\section{ Pure Soft East Model : Contraction of the global Level 2.25 towards the local Level 2.25   }

\label{app_GlobalLocalPure}

In this Appendix, we describe the explicit contraction from the Level 2.25 of Eq. \ref{level2.25masterEast}
in the space of the $2^N$ configurations of the pure soft East model
towards the Level 2.25, for the empirical time-space-averaged densities and activities,
as given by Eq. \ref{level2.25pureEastsimplia} of the main text.
This contraction is very similar to the contraction described in the previous Appendix.

\subsection{ Empirical time-space-averaged observables from observables in the space of the $2^N$ configurations   }

The empirical time-space-averaged density $\rho^{S_L,S}$  of two consecutive spins $(S_L,S)$ 
of Eq. \ref{rho2pure}
can be obtained from the configuration empirical time-averaged density $\rho(S_1,...,S_N) $ 
of Eq. \ref{rhoconfig} via
\begin{eqnarray}
\rho^{S_L,S} 
= \frac{1}{N} \sum_{i=1}^N \left[ \prod_{n=1}^{N} \sum_{S_n=\pm}  \right] 
\delta_{S_{i-1},S_L} \delta_{S_{i},S}  \ \rho(S_1,...,S_N)
\label{rholocalfromglobalpure}
\end{eqnarray}

Similarly the empirical time-space-averaged activity $a(S_L) $ of Eq. \ref{ajeastpure}
 can be obtained from the configuration empirical activities $a_{i} (S_1,..,S_{i-1} ; S_{i+1} ,..,S_N) $ 
 of Eq. \ref{ajEast} via
\begin{eqnarray}
a(S_L)  =\frac{1}{N} \sum_{i=1}^N  \left[ \prod_{n=1}^{i-1} \sum_{S_n=\pm}  \right] \left[ \prod_{p=i+1}^{N}\sum_{S_p=\pm}   \right] \delta_{S_{i-1},S_L}  \ 
a_{i} (..,S_{i-1} ; S_{i+1} ,..)
\label{alocalfromglobalpure}
\end{eqnarray}


\subsection{ Contraction of the global Level 2.25 with constraints fixing the time-space-averaged observables    }

In order to optimize the rate function at Level 2.25 of Eq. \ref{rate2.25rhoaEast}
over the configuration empirical density $\rho(S_1,...,S_N) $
and over the configuration empirical activities $a_{i} (..,S_{i-1} ; S_{i+1} ,..) $
in the presence of the constraints of Eq. \ref{rholocalfromglobalpure}
and Eq. \ref{alocalfromglobalpure},
we consider the following Lagrangian involving the Lagrange multipliers $[\omega^{\pm,\pm}  ; \lambda(\pm)]$ 
\begin{eqnarray}
 \Upsilon[ \rho(.) ; a_.(.)  ]  &&  \equiv I_{2.25}[ \rho(.) ; a_.(.)  ]   
 +  \sum_{S_L=\pm}\sum_{S=\pm} \omega^{S_L,S} 
\left( N \rho^{S_L,S} 
-  \sum_{i=1}^N \left[ \prod_{n=1}^{N} \sum_{S_n=\pm}  \right] 
\delta_{S_{i-1},S_L} \delta_{S_{i},S}  \ \rho(S_1,...,S_N)
\right)
  \nonumber \\
&& +  \sum_{S_L=\pm}\lambda(S_L) \left(  
N a(S_L)  - \sum_{i=1}^N  \left[ \prod_{n=1}^{i-1} \sum_{S_n=\pm}  \right] \left[ \prod_{p=i+1}^{N}\sum_{S_p=\pm}   \right] \delta_{S_{i-1},S_L}  \ 
a_{i} (..,S_{i-1} ; S_{i+1} ,..)
 \right)
\label{lagrange2.25defpure}
\end{eqnarray}
The explicit rate function at Level 2.25 of Eq. \ref{rate2.25rhoaEast} yields
\begin{eqnarray}
&& \Upsilon[ \rho(.) ; a_.(.)  ]  =  \sum_{i=1}^{N} \left[ \prod_{k \ne i}  \sum_{S_k = \pm}  \right]
\bigg[ \frac{a_{i} (..,S_{i-1} ; S_{i+1} ,..) }{2}   \ln \left( \frac{ a^2_{i} (..,S_{i-1} ; S_{i+1} ,..)   }
{  4 w^+ (S_{i-1})  \rho (..,S_{i-1},+, S_{i+1},..) 
w^{-} (S_{i-1})  \rho (..,S_{i-1},-, S_{i+1},..)}  \right) 
\nonumber \\
&& - a_{i} (..,S_{i-1} ; S_{i+1} ,..)     + w^{+} (S_{i-1})  \rho (..,S_{i-1},+, S_{i+1},..,)   
 + w^{-}(S_{i-1})   \rho(..,S_{i-1},-, S_{i+1},..) \bigg]
 \nonumber \\
&& 
 +  \sum_{S_L=\pm}\sum_{S=\pm} \omega^{S_L,S} 
\left(N \rho^{S_L,S} 
- \sum_{i=1}^N \left[ \prod_{n=1}^{N} \sum_{S_n=\pm}  \right] 
\delta_{S_{i-1},S_L} \delta_{S_{i},S}  \ \rho(S_1,...,S_N)
\right)
  \nonumber \\
&& +  \sum_{S_L=\pm}\lambda(S_L) \left(  
N a(S_L)  -  \sum_{i=1}^N  \left[ \prod_{n=1}^{i-1} \sum_{S_n=\pm}  \right] \left[ \prod_{p=i+1}^{N}\sum_{S_p=\pm}   \right] \delta_{S_{i-1},S_L}  \ 
a_{i} (..,S_{i-1} ; S_{i+1} ,..)
 \right)
\label{lagrange2.25pure}
\end{eqnarray}

The optimization of this Lagrangian over the configuration activities $a_{i} (..,S_{i-1} ; S_{i+1} ,..)  $
\begin{eqnarray}
0= \frac{ \partial  \Upsilon[ \rho(.) ; a_.(.)  ] }{ \partial a_{i} (..,S_{i-1} ; S_{i+1} ,..)}  
=   \frac{1 }{2}   \ln \left( \frac{ a^2_{i} (..,S_{i-1} ; S_{i+1} ,..)   }
{  4 w^+ (S_{i-1})  \rho (..,S_{i-1},+, S_{i+1},..) 
w^{-} (S_{i-1})  \rho (..,S_{i-1},-, S_{i+1},..)}  \right) 
 - \lambda(S_{i-1}) 
\label{lagrange2.25deriapure}
\end{eqnarray}
leads to the optimal values
\begin{eqnarray}
  a^{opt}_{i} (..,S_{i-1} ; S_{i+1} ,..)   
= 2 e^{  \lambda(S_{i-1})} \sqrt{  w^+ (S_{i-1})  \rho (..,S_{i-1},+, S_{i+1},..) 
w^{-} (S_{i-1})  \rho (..,S_{i-1},-, S_{i+1},..) }
\label{aconfoptpure}
\end{eqnarray}
where the Lagrange multipliers $ \lambda(S_L)$ have to be chosen to satisfy the corresponding constraints of Eq. \ref{alocalfromglobalpure}
\begin{eqnarray}
 a(S_L) && =\frac{1}{N} \sum_{i=1}^N  \left[ \prod_{n=1}^{i-1} \sum_{S_n=\pm}  \right] \left[ \prod_{p=i+1}^{N}\sum_{S_p=\pm}   \right] \delta_{S_{i-1},S_L}  \ 
a_{i}^{opt} (..,S_{i-1} ; S_{i+1} ,..)
\nonumber \\
&& =
2 e^{  \lambda(S_L) } \sqrt{  w^+ (S_L)  
w^{-} (S_L)   }
D(S_L)
\label{alocaloptpure}
\end{eqnarray}
where we have introduced the notation
\begin{eqnarray}
D(S_L)  \equiv 
\frac{1}{N} \sum_{i=1}^N  \left[ \prod_{n=1}^{i-2} \sum_{S_n=\pm}  \right] \left[ \prod_{p=i+1}^{N}\sum_{S_p=\pm}   \right] 
 \sqrt{   \rho (..S_{i-2},S_L,+, S_{i+1},..) 
 \rho (..,S_{i-2},S_L,-, S_{i+1},..) }
\label{defdpure}
\end{eqnarray}
The optimal values of Eq. \ref{aconfoptpure} can be thus rewritten as
\begin{eqnarray}
  a^{opt}_{i} (..,S_{i-1} ; S_{i+1} ,..)   
= a(S_{i-1})  
\frac{ \sqrt{   \rho (..,S_{i-1},+, S_{i+1},..)   \rho (..,S_{i-1},-, S_{i+1},..) } }
{D(S_{i-1}) }
\label{aconfoptsanslambdapure}
\end{eqnarray}

The optimization of the Lagrangian of Eq. \ref{lagrange2.25}
over the configuration density $\rho(S_1,...,S_N) $
\begin{eqnarray}
0= \frac{ \partial  \Upsilon[ \rho(.) ; a_.(.)  ] }{ \partial \rho(S_1,...,S_N)}  
=  - \frac{1}{2  \rho(S_1,...,S_N)} \sum_{i=1}^{N} a_{i} (..,S_{i-1} ; S_{i+1} ,..)
 +  \sum_{i=1}^N  \left( w^{S_i} (S_{i-1})  -  \omega^{S_{i-1},S_{i}} \right)
 \label{lagrange2.25derirhopure}
\end{eqnarray}
yields that the optimal values $\rho^{opt}(S_1,...,S_N) $ have to satisfy together with
the optimal values $a^{opt}_{i} (..,S_{i-1} ; S_{i+1} ,..) $ of the activities of Eq. \ref{aconfoptsanslambdapure}
\begin{eqnarray}
0=   \sum_{i=1}^{N} \left[ - \frac{ a_{i}^{opt} (..,S_{i-1} ; S_{i+1} ,..) }{2} 
 +   \left( w^{S_i} (S_{i-1})  -  \omega^{S_{i-1},S_{i}} \right)\rho^{opt}(S_1,...,S_N) \right]
\label{rhoconfoptpure}
\end{eqnarray}
The simplest way to satisfy this equation for the sum of $N$ terms
is to impose the vanishing of each term for $i=1,..,N$ 
\begin{eqnarray}
0=   - \frac{ a_{i}^{opt} (..,S_{i-1} ; S_{i+1} ,..) }{2} 
 +   \left( w^{S_i} (S_{i-1})  -  \omega^{S_{i-1},S_{i}} \right)\rho^{opt}(S_1,...,S_N) 
\label{rhoconfoptipure}
\end{eqnarray}

Let us apply the sum $\left(  \frac{1}{N} \sum_{i=1}^N \left[ \prod_{n=1}^{N} \sum_{S_n=\pm}  \right] 
\delta_{S_{i-1},S_L} \delta_{S_{i},S} \right)$ to Eq. \ref{rhoconfoptipure}
in order to be able to use the constraints of Eqs \ref{rholocalfromglobalpure}
and \ref{alocalfromglobalpure}
\begin{eqnarray}
0 && = \frac{1}{N} \sum_{i=1}^N \left[ \prod_{n=1}^{N} \sum_{S_n=\pm}  \right] 
\delta_{S_{i-1},S_L} \delta_{S_{i},S} \left[ - \frac{ a_{i} (S_{i-1}) }{2} 
 +   \left( w^{S_i} (S_{i-1})  -  \omega^{S_{i-1},S_{i}} \right)\rho_{i-1,i}^{S_{i-1},S_{i}} \right]
 \nonumber \\
 && = - \frac{ a (S_L) }{2} 
 +   \left( w^{S} (S_L)  -  \omega^{S_L,S} \right)\rho^{S_L,S}
\label{rhoconfoptisumpure}
\end{eqnarray}
So the Lagrange multipliers $\omega^{S_L,S} $ that modify the true rates $w^{S} (S_L)$
to produce the effective rates $\left( w^{S} (S_L)  -  \omega^{S_L,S} \right) $ can be computed from the ratios
\begin{eqnarray}
\left( w^{S} (S_L)  -  \omega^{S_L,S} \right)=    \frac{ a (S_L) }{2 \rho^{S_L,S}}  
\label{omegaeffpure}
\end{eqnarray}
Eq. \ref{rhoconfoptipure} then yields the optimal density
\begin{eqnarray}
\rho^{opt}(S_1,...,S_N) =  \rho^{S_{i-1},S_{i}}  \ \frac{ a_{i}^{opt} (..,S_{i-1} ; S_{i+1} ,..) }{ a (S_{i-1})} 
\label{rhoconfoptifinalpure}
\end{eqnarray}
that can be plugged into Eq. \ref{defdpure} to obtain with the use of the constraint of Eq. \ref{alocalfromglobalpure}
\begin{eqnarray}
D(S_L) && = 
\frac{1}{N} \sum_{i=1}^N  \left[ \prod_{n=1}^{i-2} \sum_{S_n=\pm}  \right] \left[ \prod_{p=i+1}^{N}\sum_{S_p=\pm}   \right] 
 \sqrt{   \rho^{opt} (..S_{i-2},S_L,+, S_{i+1},..) 
 \rho^{opt} (..,S_{i-2},S_L,-, S_{i+1},..) }
 \nonumber \\
 && 
 =\frac{ \sqrt{ \rho^{S_L,+}   \rho^{S_L,-}   } }{ N a (S_L)}
 \sum_{i=1}^N  \left[ \prod_{n=1}^{i-2} \sum_{S_n=\pm}  \right] \left[ \prod_{p=i+1}^{N}\sum_{S_p=\pm}   \right] 
 a_{i}^{opt} (..,S_L ; S_{i+1} ,..)
 = \sqrt{ \rho^{S_L,+}   \rho^{S_L,-}   }
\label{defdoptpure}
\end{eqnarray}
So the Lagrange multipliers $ \lambda(S_L)$ of Eq. \ref{alocaloptpure}
 reduce to
\begin{eqnarray}
 \lambda(S_L) = \ln \left(  \frac{ a(S_L) } { 2  \sqrt{  w^+ (S_L) 
w^{-} (S_L)  \rho^{S_L,+}    \rho^{S_L,-}  } } \right)
= \frac{1}{2}  \ln \left(  \frac{ a^2(S_L) } { 4   w^+ (S_L) 
w^{-} (S_L)  \rho^{S_L,+}    \rho^{S_L,-}   } \right)
\label{lagrangesolpure}
\end{eqnarray}

The optimal value of the Lagrangian of Eq. \ref{lagrange2.25pure}
corresponding to the optimal solution $[ \rho^{opt}(.) ; a^{opt}_.(.)  ] $ satisfying the constraints
reads using Eq. \ref{lagrangesolpure}
\begin{small}
\begin{eqnarray}
&& \Upsilon^{opt} \equiv  \Upsilon[ \rho^{opt}(.) ; a^{opt}_.(.)  ] 
\nonumber \\
&&  =  \sum_{i=1}^{N} \left[ \prod_{k \ne i}  \sum_{S_k = \pm}  \right]
\bigg[ \frac{a^{opt}_{i} (..,S_{i-1} ; S_{i+1} ,..) }{2}   \ln \left( \frac{ (a^{opt}_{i} (..,S_{i-1} ; S_{i+1} ,..) )^2  }
{  4 w^+ (S_{i-1})  \rho^{opt} (..,S_{i-1},+, S_{i+1},..) 
w^{-} (S_{i-1})  \rho^{opt} (..,S_{i-1},-, S_{i+1},..)}  \right) 
\nonumber \\
&& - a^{opt}_{i} (..,S_{i-1} ; S_{i+1} ,..)     + w^{+} (S_{i-1})  \rho^{opt} (..,S_{i-1},+, S_{i+1},..,)   
 + w^{-}(S_{i-1})   \rho^{opt}(..,S_{i-1},-, S_{i+1},..) \bigg]
 \nonumber \\
 &&  = \sum_{S_L=\pm}  \lambda(S_L)   \sum_{i=1}^{N} \left[ \prod_{k \ne (i-1,i)}  \sum_{S_k = \pm}  \right]
 a^{opt}_{i} (..S_{i-2},S_L ; S_{i+1} ,..) 
 - \sum_{S_L=\pm}  \sum_{i=1}^{N} \left[ \prod_{k \ne (i-1,i)}  \sum_{S_k = \pm}  \right]
 a^{opt}_{i} (..,S_{i-2},S_L ; S_{i+1} ,..)
\nonumber \\
&& + \sum_{S_L=\pm}  \sum_{i=1}^{N} \left[ \prod_{k \ne (i-1,i)}  \sum_{S_k = \pm}  \right]
\bigg[  w^{+} (S_L)  \rho^{opt} (..,S_{i-2},S_L,+, S_{i+1},..,)   
 + w^{-}(S_L)   \rho^{opt}(..,S_{i-2},S_L,-, S_{i+1},..) \bigg]
  \nonumber \\
 &&  = N  \sum_{S_L=\pm} \left[     
  \frac{a(S_L)}{2}  \ln \left(  \frac{ a^2(S_L) } { 4   w^+ (S_L) 
w^{-} (S_L)  \rho^{S_L,+}    \rho^{S_L,-}   } \right)
  - a(S_L) 
+   w^{+} (S_L)  \rho^{S_L,+}   
 + w^{-}(S_L)   \rho^{opt}(S_L,-) \right]
\label{lagrange2.25optpure}
\end{eqnarray}
\end{small}

This optimal value $ \Upsilon^{opt} $
allows to recover the rate function $ {\cal I}_{2.25} [  a(.) ; \rho^{.,.}]   $ with respect to the space-time volume $(NT)$
 at Level 2.25 for the empirical time-space-averaged activities $a (\pm)  $ 
and the local time-space-averaged densities $\rho^{S_L,S} $ as given in Eq. \ref{rate2.25pureEastsimpli} of the main text
\begin{eqnarray}
 \Upsilon^{opt} = N {\cal I}_{2.25} [  a(.) ; \rho^{.,.}] 
 \label{upsilonoptpure}
 \end{eqnarray}


\end{document}